\documentclass[twocolumn]{aastex631}

\usepackage{graphicx}	
\usepackage{amsmath}	
\usepackage{amssymb}	
\usepackage{makecell}
\usepackage{comment}
\usepackage{multirow}

\begin{document}

\title{A dichotomy in the 1--24~GHz pc-scale radio spectra of radio-quiet quasars}


\correspondingauthor{Sina Chen}
\email{sina.chen@campus.technion.ac.il}

\author{Sina Chen}
\affiliation{Physics Department, Technion, Haifa 32000, Israel}

\author{Ari Laor}
\affiliation{Physics Department, Technion, Haifa 32000, Israel}

\author{Ehud Behar}
\affiliation{Physics Department, Technion, Haifa 32000, Israel}

\author{Ranieri D. Baldi}
\affiliation{INAF - Istituto di Radioastronomia, via Gobetti 101, 40129 Bologna, Italy}

\author{Joseph D. Gelfand}
\affiliation{NYU Abu Dhabi, PO Box 129188, Abu Dhabi, UAE}

\author{Amy E. Kimball}
\affiliation{National Radio Astronomy Observatory, 1011 Lopezville Road, Socorro, NM 87801, USA}


\begin{abstract}

We present the pc-scale radio spectra of a representative sample of 13 Palomar-Green radio-quiet quasars, based on our new Very Long Baseline Array (VLBA) observations at 8.4 and 23.6~GHz and our earlier VLBA studies at 1.5 and 5.0~GHz.
The radio core emission generally exhibits a flat spectrum at 1.5--5.0~GHz, which indicates a compact optically thick synchrotron source on a scale smaller than the broad-line region (BLR) radius $R_{\rm BLR} \sim 0.01-0.1$~pc.
The 8.4--23.6~GHz spectral slope remains flat in four objects indicating the inner radius of the radio source $R_{\rm in} < 0.1~R_{\rm BLR}$, and becomes steep in four other objects indicating $R_{\rm in} \sim 0.5~R_{\rm BLR}$.
The flat 8.4--23.6~GHz slope sources may be associated with a continuous ejection starting at the accretion disk corona.
The steep 8.4--23.6~GHz slope sources may be produced by an interaction of an AGN-driven wind with the BLR gas or a low-power jet extending to the BLR scale.
Seven of these eight objects, which have a flat or steep 8.4--23.6~GHz slope, reside at the Eddington ratios $L/L_{\rm Edd} < 0.3$, and four of the remaining five objects, where the 8.4--23.6~GHz fluxes are too faint to significantly constrain the slope, reside at $L/L_{\rm Edd} > 0.3$.
The 8.4--23.6~GHz radio emission in the high $L/L_{\rm Edd}$ objects may be weak due to more extended emission from a radiation pressure driven wind.
Future sub-millimeter observations can further constrain the inward radial extent of the radio emission down to the coronal scale.

\end{abstract}


\section{Introduction}

Active galactic nuclei (AGN) are typically classified as either radio-loud (RL) or radio-quiet (RQ) based on the radio 5~GHz to optical 4400~{\AA} flux ratio larger than 10 or smaller than 10 \citep{Kellermann1989}.
In RL AGN, the radio emission is primarily produced by a powerful relativistic jet, characterized by a radio to X-ray luminosity ratio of $\log L_{\rm R}/L_{\rm X} \simeq -2$ \citep{Panessa2007}.
However, the origin of the radio emission in RQ AGN, where $\log L_{\rm R}/L_{\rm X} \simeq -5$ \citep{Laor2008}, remains an open question.
Various mechanisms have been proposed, including a low-power jet, an AGN-driven wind, the accretion disk corona, star formation (SF), and free-free emission \citep[see][for a review]{Panessa2019}.
Differentiating between these potential radio emission mechanisms in RQ AGN is a key challenge.

\subsection{Spectral constraints on the synchrotron source \label{sec:sed}}

\begin{figure*}[ht!]
\centering
\includegraphics[width=1.8\columnwidth, trim={2cm, 0cm, 2cm, 0cm}, clip]{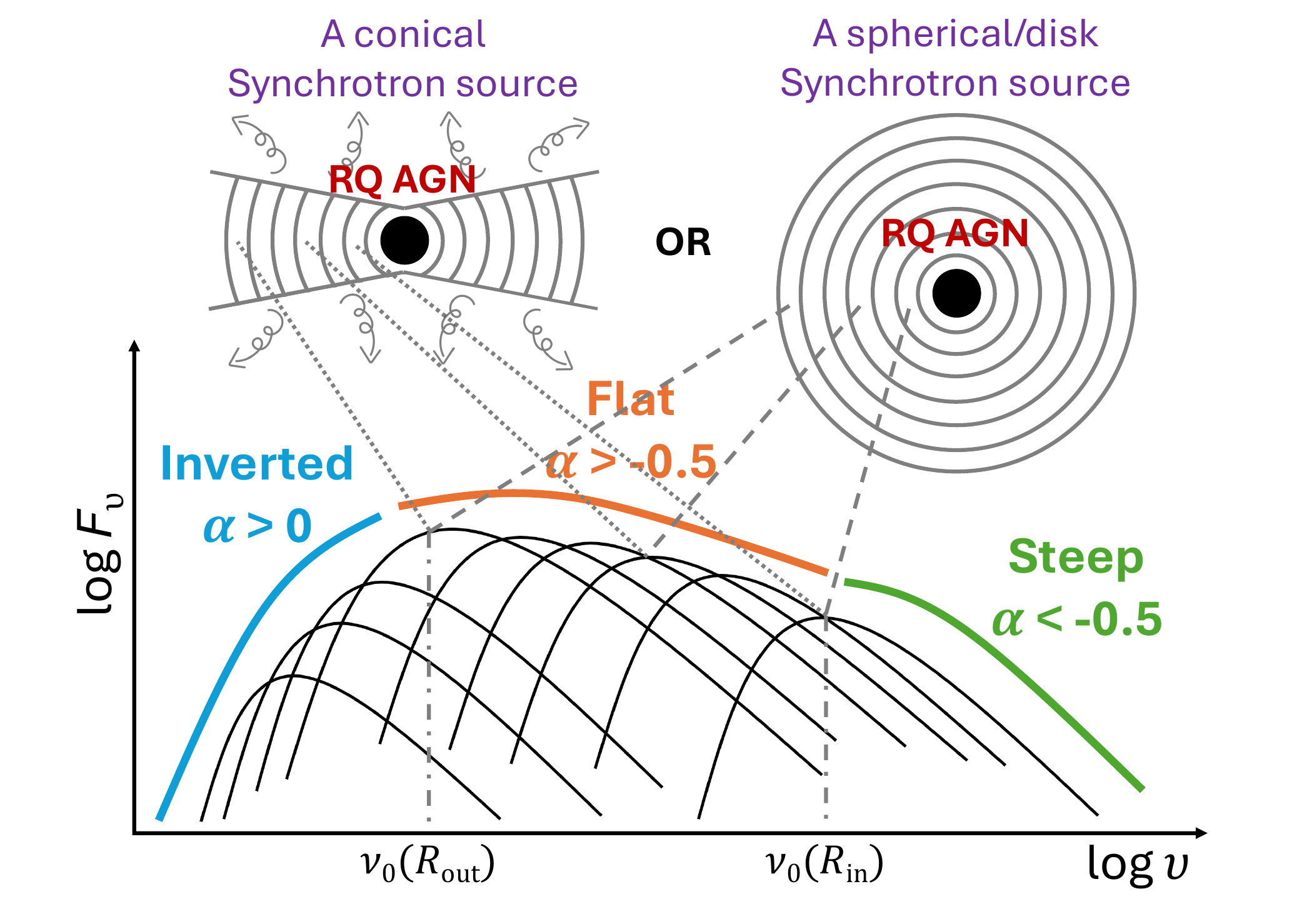}
\caption{A sketch of the radio spectrum produced by a conical, spherical, or disk-like synchrotron source in RQ AGN.
The synchrotron source is assumed to be azimuthally or spherically symmetric, and the emission is unbeamed and isotropically radiated, and thus depends on the radius only.
At each radius $R$, the spectrum is composed of self-absorbed emission below the peak frequency $\nu_0 (R)$, and optically thin emission above $\nu_0 (R)$.
An assumed equipartition $B(R)$ increasing inwards leads to an increasing $\nu_0 (R)$ inwards.
The observed radio spectrum is the superposition of all the components.
A flat spectrum indicates a continuous flow extending from an inner radius, which is set by a spectral break frequency from flat to steep slopes, to an outer radius, which is set by a spectral break frequency from inverted to flat slopes.}
\label{synchrotron}
\end{figure*}

Figure~\ref{synchrotron} presents a schematic description of the expected spectral shape from a generic synchrotron source extending over a certain range of radii $R$.
In contrast with RL AGN where the emission is produced by a relativistically beamed jet \citep[e.g.][]{Blandford1979}, in RQ AGN the emission is most likely unbeamed.
The geometry of the synchrotron source can be conical (e.g., a jet or an outflow), or spherical/disk-like (e.g., an accretion disk corona).
Each component of the emission volume is characterized by a magnetic field $B(R)$, which is a function of distance from the center, and produces a synchrotron spectrum.
The synchrotron emission is self-absorbed with a flux density $F_{\nu} \propto \nu^{2.5}$ at $\nu < \nu_0$, and is optically thin with $F_{\nu} \propto \nu^{\alpha}$ at $\nu > \nu_0$, where $\nu_0$ is the frequency at the peak $F_{\nu}$, $\alpha = -(p-1)/2$, and $p$ is the power-law slope of the relativistic electron energy distribution, typically $p \sim 2-3$.
Since $B(R)$ probably increases inwards, e.g., $B(R) \propto R^{-1}$ for equipartition, and $\nu_0 \propto B(R)$, we generically expect that the synchrotron spectrum from an inner region shifts to a higher frequency.
The overall spectral shape of a synchrotron source extending from an inner shell $R_{\rm in}$ to an outer shell $R_{\rm out}$ is depicted in Figure~\ref{synchrotron}.

This leads to a convenient result that the radiative transfer effect is not significant for the following reasons.
In the spherical geometry, the emission from a small $R$ transfers through the emitting shells at a larger $R$.
Since $\nu_0$ shifts to a lower value with an increasing $R$, the peak frequency of an inner shell $\nu_0 (R_{\rm in})$ is higher than that of outer shells, and thus the emission from the inner shell passes through the outer shells which are all optically thin at $\nu_0 (R_{\rm in})$.
Therefore, the total emission can be approximated by a sum of the emission from all shells.
The conical and disk-like geometries can be treated as a spherical geometry with a covering factor of $f_{\rm cover} = \Omega/(4\pi)$, where $\Omega$ is the solid angle of the total emitting area on both sides.
In a spherical source, $f_{\rm cover} = 1$.
A detailed radiative transfer calculation can be seen in \citet[][Figure 4 there]{Raginski2016}.

The spectrum at $\nu < \nu_0 (R_{\rm out})$ is self-absorbed, with $\alpha > 0$ typically.
The $\alpha = 2.5$ case is achieved when there is a sharp outer boundary of the emitting volume, as for example occurs in a supernova explosion \citep[e.g.][]{Soderberg2006}.
In AGN the outer boundary is likely not sharp, for example, due to an earlier outflow when the emission is intermittent, or a clumped outflow which gradually expands and cools.
This leads to a superposition of weaker emission shells, and a slope of $\alpha < 2.5$, as depicted in Figure~\ref{synchrotron}.
At $\nu > \nu_0 (R_{\rm in})$, the emission from all elements becomes optically thin and is characterized by $\alpha < -0.5$, and the emission is likely dominated by the innermost element.
In the intermediate regime $\nu_0 (R_{\rm out}) < \nu < \nu_0 (R_{\rm in})$, the slope is flat, i.e., $\alpha > -0.5$, and the exact value depends on the relative contribution of different elements.
The emission observed at a given $\nu$ is dominated by the component at a certain $R$ where $\nu_0 (R) = \nu$.
With an increasing $\nu$, the emission originates at a smaller $R$.

The amplitude of the optically thin emission depends on the emission per unit volume, that is $B(R)$, the total emitting volume, and the distribution function of the emitting electrons.
However, when the emission becomes optically thick, it is emitted mainly from a surface layer, and the emission per unit area is set by $B(R)$ only.
Thus, the total emitting surface area at a given $\nu$ can be derived directly from the observed luminosity density $L_\nu = F_\nu \cdot 4 \pi d^2$, where $d$ is the luminosity distance, once a $B(R)$ is given.

We consider a source of emission with a source function $S_\nu$, and an intensity $I_\nu$ emitted from its surface. When the emission is optically thick, $I_\nu = S_\nu$.
The flux density emitted from the source surface is \citep{Rybicki1986}
\begin{equation}
F_{\nu}^{*} = \int I_\nu \cos\theta d\Omega = \pi I_\nu = \pi S_\nu.
\end{equation}
In RQ AGN, the emission is unbeamed and isotropic, the source luminosity density is $L_\nu = F_{\nu}^{*} \cdot A$, where $A$ is the total emitting surface area.
For simplicity, we assume a spherical source with the AGN at its center, and $A = 4 \pi R_\nu^2 \cdot f_{\rm cover}$ to accommodate different geometries.
Therefore,
\begin{equation}
L_\nu = \pi S_\nu \cdot 4 \pi R_\nu^2 \cdot f_{\rm cover},
\end{equation}
and the radius or size of an optically thick synchrotron source is
\begin{equation}
R_\nu = \frac{1}{\sqrt{f_{\rm cover}}} \frac{1}{2\pi} \sqrt{\frac{L_\nu}{S_\nu}}.
\end{equation}
In the case of a spherical source, it is recovered to the Equation 18 in \citet{Laor2008}.

The size of the ``radio sphere'' is weakly dependent on $B(R)$, and is given by the Equation 19 in \citet{Laor2008} times a factor of $1/\sqrt{f_{\rm cover}}$,
\begin{equation}
R_\nu = 0.54 L_{30}^{1/2} \nu_{\rm GHz}^{-5/4} B^{1/4} f_{\rm cover}^{-1/2} ~ {\rm pc},
\end{equation}
where $L_{30}$ is the luminosity density in units of $10^{30}$~erg~s$^{-1}$~Hz$^{-1}$ and $\nu_{\rm GHz}$ is the observed frequency in GHz.
The magnetic field $B$ can be estimated in the case of an equipartition between the magnetic energy density and the photon energy density \citep[][Equation 21 there]{Laor2008},
\begin{equation}
B_{\rm eq} = 0.27 R_\nu^{-1} L_{46}^{1/2} ~ {\rm Gauss},
\end{equation}
where $L_{46}$ is the bolometric luminosity $L_{\rm bol}$ in units of $10^{46}$~erg~s$^{-1}$.
This gives
\begin{equation}
R_\nu = 0.47 L_{30}^{0.4} L_{46}^{0.1} \nu_{\rm GHz}^{-1} f_{\rm cover}^{-0.5} ~ {\rm pc},
\label{eq:radius}
\end{equation}
which is the Equation 22 in \citet{Laor2008} times a factor of $f_{\rm cover}^{-0.5}$.

In RQ Type 1 AGN, $L_{\rm R}/L_{\rm X} \sim 10^{-5}$ at 5~GHz and $L_{\rm X}/L_{\rm bol} \sim 0.1$ typically, and thus $L_{\rm R}/L_{\rm bol} \sim 10^{-6}$.
This gives $L_{30} \sim \frac{10^{-6} L_{\rm bol}}{10^{30} \nu} \sim L_{46}$, and therefore
\begin{equation}
R_{\rm 5GHz} \sim R_{\rm BLR} \sim 0.1 L_{46}^{0.5} ~ {\rm pc}.
\label{eq:blr}
\end{equation}
This happens to match the BLR size versus luminosity relation, which is converted from the Equation 1 in \citet{Kaspi2005} using the UV luminosity with a bolometric correction factor of 3 \citep{Richards2006}.

Since the radio emission in RQ AGN is most likely unbeamed and radiated isotropically, the observed radio spectra are independent of the geometry.
Therefore, the detection of the optically thick ($\alpha > -0.5$) synchrotron emission in RQ AGN at a given $\nu$ allows to derive $R_\nu$, that is the size of the region where the emission dominates at $\nu$.
A spectral break from inverted to flat provides $R_{\rm out}$, and a spectral break from flat to steep provides $R_{\rm in}$.
A steep spectrum throughout all $\nu$ provides only a lower limit on $R_{\rm in}$.
If a flat spectrum becomes steep at lower $\nu$, it implies the optically thin emission of an additional more extended source.
If a flat spectrum becomes inverted at higher $\nu$, it implies the self-absorbed emission of an additional more compact source.
Notably, the assumption of an equipartition $B(R)$, which underlies the quantitative derivation of $R_\nu$, may not hold, and only direct observational constraints on the size of the radio emission region, e.g., through variability, can test how well the equipartition assumption holds.

\subsection{The expected properties of the various radio emission mechanisms}

A low-power jet and an AGN-driven wind can be similar in terms of the radio morphology and the spectral shape.
Both mechanisms produce extended optically thin emission with a steep slope $\alpha < -0.5$, potentially extending from pc (mas) to kpc (arcsec) scales.
In principle, a jet is expected to be well-collimated and propagate at a higher velocity compared to a wind.
However, distinguishing between these two mechanisms based solely on radio observations is challenging.
The surface brightness temperature $T_{\rm B}$, which is likely enhanced in the jet emission, may provide a hint \citep{Chen2023}.
Additionally, proper motion studies of individual sources can reveal the bulk outflow velocity, allowing for the separation of a (mild) relativistic jet from a non-relativistic wind \citep[e.g.][]{Wang2023b}.

The accretion disk corona can produce compact optically thick emission characterized by a flat or inverted slope $\alpha > -0.5$ \citep{Laor2008}.
Since it originates on sub-pc scales, it remains unresolved in the Very Long Baseline Interferometry (VLBI) observations, such as those conducted with the Very Long Baseline Array (VLBA) and the European VLBI Network (EVN).
The corona can also produce extended optically thin emission, due to coronal mass ejections (CME) in a form of outflowing plasma, similar to the phenomena observed in the Sun and in coronally active stars.
The accretion disk corona is also suggested to form a base for the jet formation \citep{Merloni2002,King2017}.

SF produces diffuse optically thin emission with a steep slope $\alpha < -0.5$, extending across the host galaxy \citep{Kimball2011,Condon2013,Zakamska2016}.
Free-free emission from AGN photoionized gas is characterized by a specific flat slope $\alpha = -0.1$, and becomes self-absorbed with a slope of $\alpha = 2$ below a frequency determined by the distance from the AGN \citep{Baskin2021}.
Both mechanisms produce emission with a low surface brightness temperature, typically well below $T_{\rm B} \sim 10^4 - 10^5$~K, which is resolved out in mas-scale VLBI observations.

In this paper, we focus on the spectral properties of the pc-scale radio emission associated with the corona, a low-power jet, and an AGN-driven wind, exploring how these different mechanisms can be distinguished based on their radio properties.
We use the term ``core'' to refer to the unresolved radio emission, and the general term ``outflow'' to refer to both a wind and a jet when they are not specified.

\subsection{The observed core emission properties in RQ AGN}

Our earlier VLBA studies of 18 Palomar-Green (PG) RQ quasars (RQQ) \citep{Boroson1992} found a significant correlation between the VLBA 1.5--5~GHz spectral slopes and the VLBA to VLA flux ratios, indicating that the source becomes more compact as the spectral slope flattens \citep{Alhosani2022,Chen2023}.
This suggests that the radio emission in flat-spectrum objects is more compact, likely associated with the accretion disk corona, while the radio emission in steep-spectrum objects is more extended, possibly produced by an AGN-driven wind or a low-power jet.
Thus the spectral slope and the emission compactness provide important clues to the origin of the radio emission.

An intriguing result from our earlier VLBA studies \citep{Alhosani2022,Chen2023} is that the pc-scale radio emission in most RQQ consists of a compact core with a flat or inverted 1.5--5~GHz spectrum.
The origin of the radio emission of this compact flat-spectrum core is of particular interest.
If the core emission is associated with the X-ray corona \citep{Laor2008}, it originates from the inner part of the accretion disk.
The emission will remain optically thick with a flat or inverted spectrum up to a few hundreds GHz \citep{Raginski2016}.
Alternatively, if the core emission is produced by a compact jet or wind, it will become optically thin with a steep spectrum at higher frequencies.

The Eddington ratio $L/L_{\rm Edd}$ and the profiles of high-ionization UV emission lines, specifically the C\,IV line, can provide additional insights into distinguishing between a wind and a jet.
A Very Large Array (VLA) study of 25 PG RQQ \citep{Laor2019} found that high $L/L_{\rm Edd}$ ($>$0.3) objects generally have steep VLA spectra ($\alpha_{5-8.5}<-0.5$).
This trend is confirmed by a VLA study of 15 PG RQQ at 45~GHz \citep{Baldi2022}.
Additionally, a recent study of 19 PG RQQ combining VLBA and EVN observations \citep{Chen2024b} further found that the extended radio emission is typically observed in objects which have high $L/L_{\rm Edd}$ and strong excess blue wing in the C\,IV emission line profile.
Since the high $L/L_{\rm Edd}$ and the strong C\,IV excess blue wing emission are likely associated with a radiation pressure driven wind from the broad-line region (BLR) \citep[e.g.][]{Baskin2005}, the optically thin radio emission in RQQ with these characteristics may originate from shocks generated by an AGN-driven wind.

This study presents the results of our new VLBA observations at 8.4 and 23.6~GHz of 13 PG RQQ detected at 5~GHz in our earlier VLBA studies \citep{Alhosani2022,Chen2023}.
The advantage of higher frequencies is that they can probe the emission produced on smaller scales, as well as offer higher resolutions.
The combination of 1.5, 5.0, 8.4, and 23.6~GHz fluxes allows to construct a pc-scale radio spectrum at 1--24~GHz, providing constraints on the size of the radio core emission and insights into the underlying radio emission mechanisms.

The paper is organized as follows. In Section 2 we describe the sample selection, in Section 3 we detail the VLBA data reduction, in Section 4 we describe the VLBA data analysis. In Section 5 we present the results and discuss them in Section 6. Section 7 provides a summary.

\section{Sample selection}

The PG quasar sample \citep{Boroson1992} comprises $\sim$ 100 of the brightest AGN in one fourth of the sky, selected based on point-like morphology, blue colors, and the presence of broad emission lines \citep{Schmidt1983}.
These criteria result in a homogeneous and representative sample of Type 1 AGN, which are predominantly at the $L/L_{\rm Edd}$ range of $\sim$ 0.1--1 and are not significantly reddened.
The PG quasar sample is one of the most extensively studied samples of Type 1 AGN, including: the overall Spectral Energy Distribution (SED) \citep{Neugebauer1987,Sanders1989}, radio cm-band continuum and imaging \citep{Kellermann1989,Kellermann1994,Miller1993}, infrared photometry \citep{Haas2003,Shi2014,Petric2015}, optical spectroscopy \citep{Boroson1992}, optical polarization \citep{Berriman1990}, UV spectroscopy \citep{Baskin2005}, soft X-ray spectroscopy \citep{Brandt2000}, and many other studies.
A critical property of the PG quasar sample is that it is optically selected, and therefore not subject to a selection bias in terms of its radio properties.

The VLBA 8.4 and 23.6~GHz observations include 13 PG RQQ, which are part of the 71 $z<0.5$ PG RQQ.
They are selected from the 18 PG RQQ which are detected at 5.0~GHz in our earlier VLBA studies \citep{Alhosani2022,Chen2023}.
The 13 objects exhibit a wide range in the full width at half maximum (FWHM) of H$\beta$ line ($\sim$ 1000--7000\,km\,s$^{-1}$) and the $V$-band absolute magnitude ($-27 < M_{\rm V} < -21$).
These two parameters are used to determine the black hole (BH) mass $M_{\rm BH}$ and the Eddington ratio $L/L_{\rm Edd}$.
The sample spans a broad range of BH masses ($M_{\rm BH} \sim 10^{6.8}-10^{9.1}\,M_{\odot}$), bolometric luminosity ($L_{\rm bol} \sim 10^{44.5}-10^{46.6}$\,erg\,s$^{-1}$), and Eddington ratios ($L/L_{\rm Edd} \sim 0.03-2$).
All objects have a ratio of the VLBA 5~GHz luminosity to the X-ray 0.2--12~keV luminosity of $\log L_{\rm R}/L_{\rm X} \simeq -6$, which is typical for RQ AGN, except for two objects (PG\,1216+069 and PG\,1351+640) with $\log L_{\rm R}/L_{\rm X} \simeq -4$ \citep{Fischer2021,Chen2023}.
The radio emission is thus not contaminated by a powerful relativistic jet as observed in RL AGN.
Importantly, the sample is designed to have a consistent distribution with the parent sample of 71 $z<0.5$ PG RQQ, making it representative of the general properties of the parent sample \citep[see Figure 1 in][]{Chen2023}.

\section{Data reduction}

The observations were conducted with the VLBA in the X (3.6~cm = 8.4~GHz) and K (1.3~cm = 23.6~GHz) bands using the 8--10 main VLBA stations between July 2023 and June 2024 (Program ID: BC288).
We utilized the digital down conversion observing system with a 2-bit sampling at a data rate of 4~Gbps.
The setup includes four intermediate frequency (IF) bands, dual polarization, and a bandwidth of 128~MHz in each IF.
Phase-referencing continuum observations were carried out, with each session lasting approximately four hours.
The scans were interleaved across different frequencies to improve the $uv$-coverage.
A four-minute nodding cycle was used, with two minutes on a target and one minute on a phase calibrator before and after each target scan.
A ``Fringe Finder'' (3C84, 3C345, 3C454.3, or 4C39.25) was observed 2--3 times during each session when all the antennas were operational.
This strategy yields an integration time of approximately two hours on the target, with 60\% of the time in the X band and 40\% in the K band.

Data calibration was performed using the VLBA data calibration pipeline procedure VLBARUN \footnote{\url{http://www.aips.nrao.edu/vlbarun.shtml}} in the Astronomical Image Processing System \citep[AIPS;][]{Greisen2003}.
A step-by-step recipe of VLBARUN is described in Appendix C of the AIPS Cookbook \footnote{\url{http://www.aips.nrao.edu/cook.html}}.
We performed all calibrations to the phase calibrator, including corrections for ionosphere delay, Earth orientation, correlator sampler threshold errors, instrument delay, bandpass, amplitude, and parallactic angle.
Self-calibration was applied to the phase calibrator.
We then solved for complex amplitudes and phases for the phase calibrator and applied the solutions from the phase calibrator to the target by using a linear interpolation.
Upon the completion of the calibration procedures, the final calibrations were applied to the target using the AIPS task SPLIT.

The target imaging was performed using the DIFMAP \citep{Shepherd1994} through a number of iterations of model fitting with a point source function.
We inspected all baselines (antenna pairs) and spectral windows for radio frequency interference (RFI), flagging data affected by RFI.
The images were created with a size of 2048~mas $\times$ 2048~mas and a cell size of 0.1~mas, using a natural weighting to maximize sensitivity at the expense of angular resolution.
We consider a source to be present if the emission has a signal-to-noise (S/N) ratio of $\gtrsim$ 5.
In such cases, we place a model at the corresponding location and perform the MODELFIT in the DIFMAP.
If, after subtracting the initial model, there remains emission with a S/N ratio of $\gtrsim$ 5, we will place an additional model and run the MODELFIT again.
Self-calibration was not applied to the targets due to the insufficient S/N ratio.
To measure the spectral slope with a minimal bias from resolution differences between the X and K bands, we tapered the images to the same $uv$-range, typically $\sim$ 15--200~M$\lambda$ in both bands, corresponding to an angular resolution of $\sim$ 1--17~mas.
The exact $uv$-range, which is chosen to ensure data availability in both bands, varies with each observation.
The flux density of the fitted models was measured in the MODELFIT, and the background noise was measured over the entire image after subtracting the models.

\section{Data analysis}


\begin{figure*}[ht!]
\centering
\includegraphics[width=1.8\columnwidth, trim={0cm, 0cm, 2cm, 0cm}, clip]{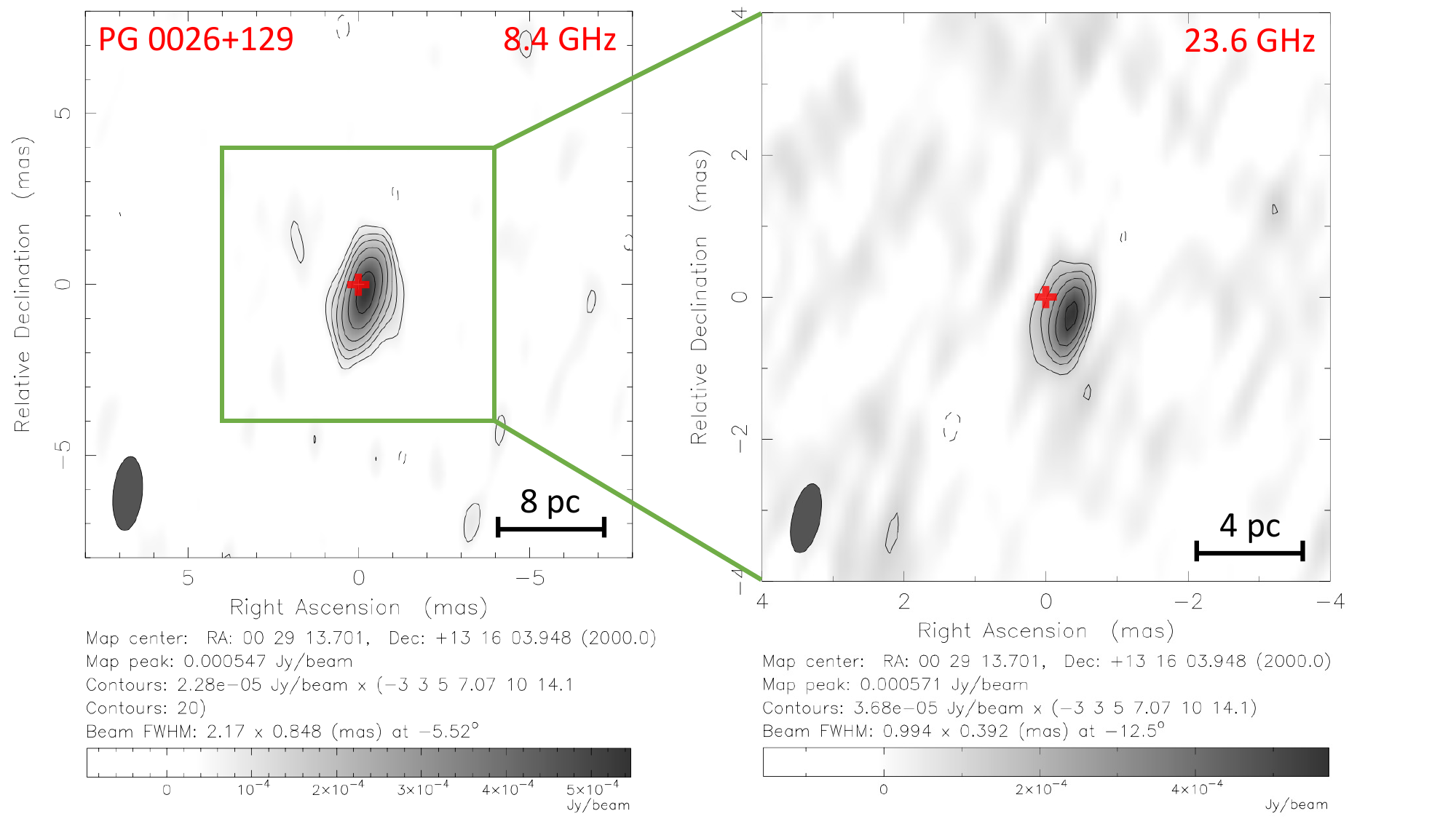}
\includegraphics[width=\columnwidth, trim={4cm, 0cm, 6cm, 1cm}, clip]{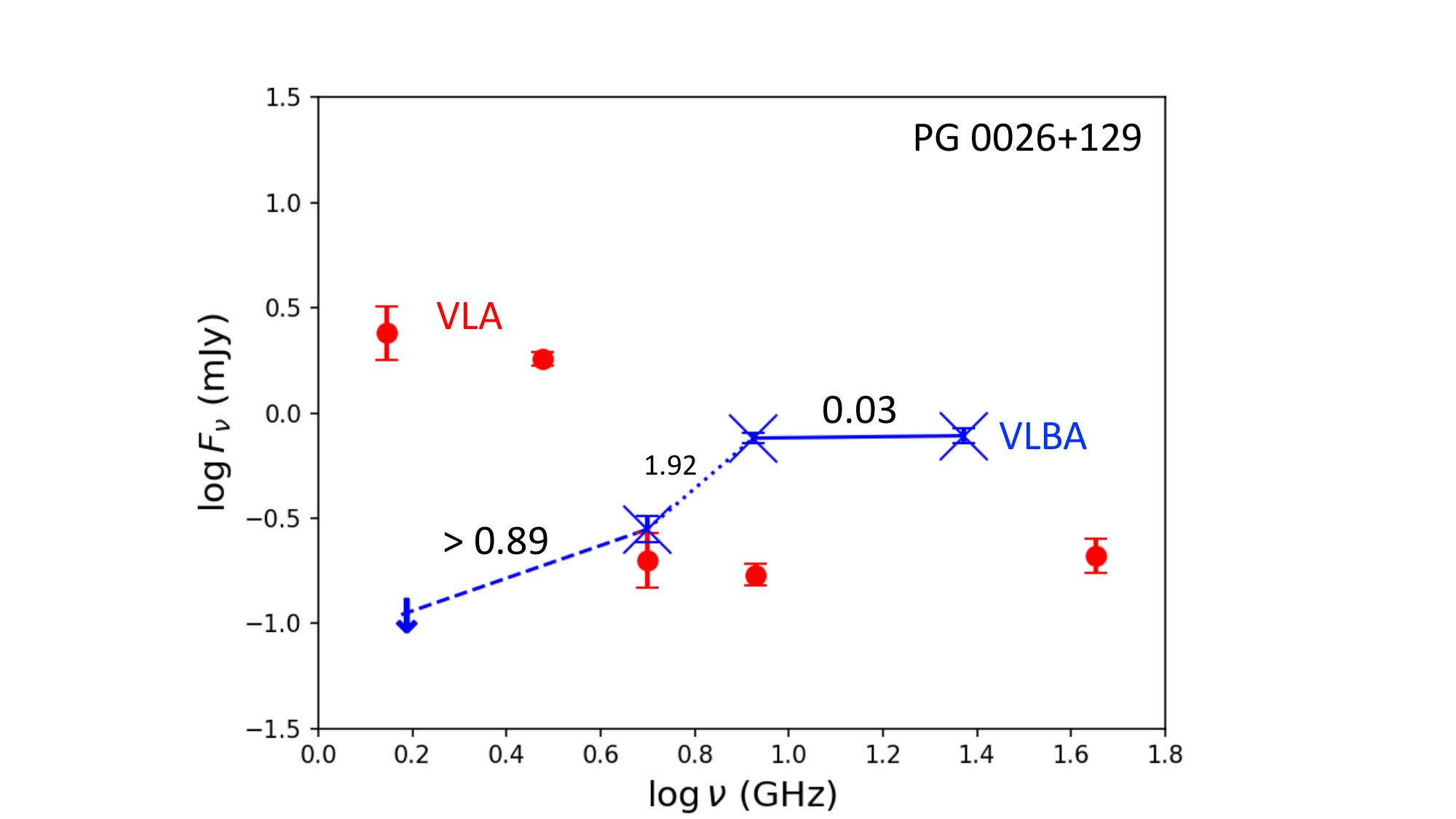}
\caption{PG\,0026+129: The radio maps at 8.4~GHz (upper-left panel) and 23.6~GHz (upper-right panel).
The contours are at ($-$3, 3, 5, 7.07, 10, 14.1, 20) $\times$ 0.0228~mJy/beam at 8.4~GHz, and ($-$3, 3, 5, 7.07, 10, 14.1) $\times$ 0.0368~mJy/beam at 23.6~GHz.
The synthesized beam sizes and orientations are 2.17~mas $\times$ 0.848~mas at $-$5.52$^{\circ}$ at 8.4~GHz, and 0.994~mas $\times$ 0.392~mas at $-$12.5$^{\circ}$ at 23.6~GHz, as shown in the lower-left corner of each map.
The images are centered at the {\it Gaia} position, which is marked as a red plus.
The 1--24~GHz radio spectrum (lower panel), where the VLBA tapered fluxes are used in all bands.
The core emission is flat at 1.5--5.0~GHz, and remains flat at 8.4--23.6~GHz.
The VLA detected fluxes are indicated by red circles.
The VLBA detected fluxes are indicated by blue crosses, and the 5$\sigma$ upper limits are marked with blue down-arrows.
The blue solid line represents the detected spectral slope, and the blue dashed line represents the limit on the spectral slope.
The values of the slope $\alpha$ are indicated near the lines.
The 1.5--5.0~GHz slope and the 8.4--23.6~GHz slope are derived from simultaneous observations, and thus are not affected by variability.
The blue dotted line represents the 5.0--8.4~GHz slope, which is affected by non-simultaneous observations and different resolutions.}
\label{0026}
\end{figure*}

\begin{figure*}[ht!]
\centering
\includegraphics[width=1.8\columnwidth, trim={0cm, 0cm, 2cm, 0cm}, clip]{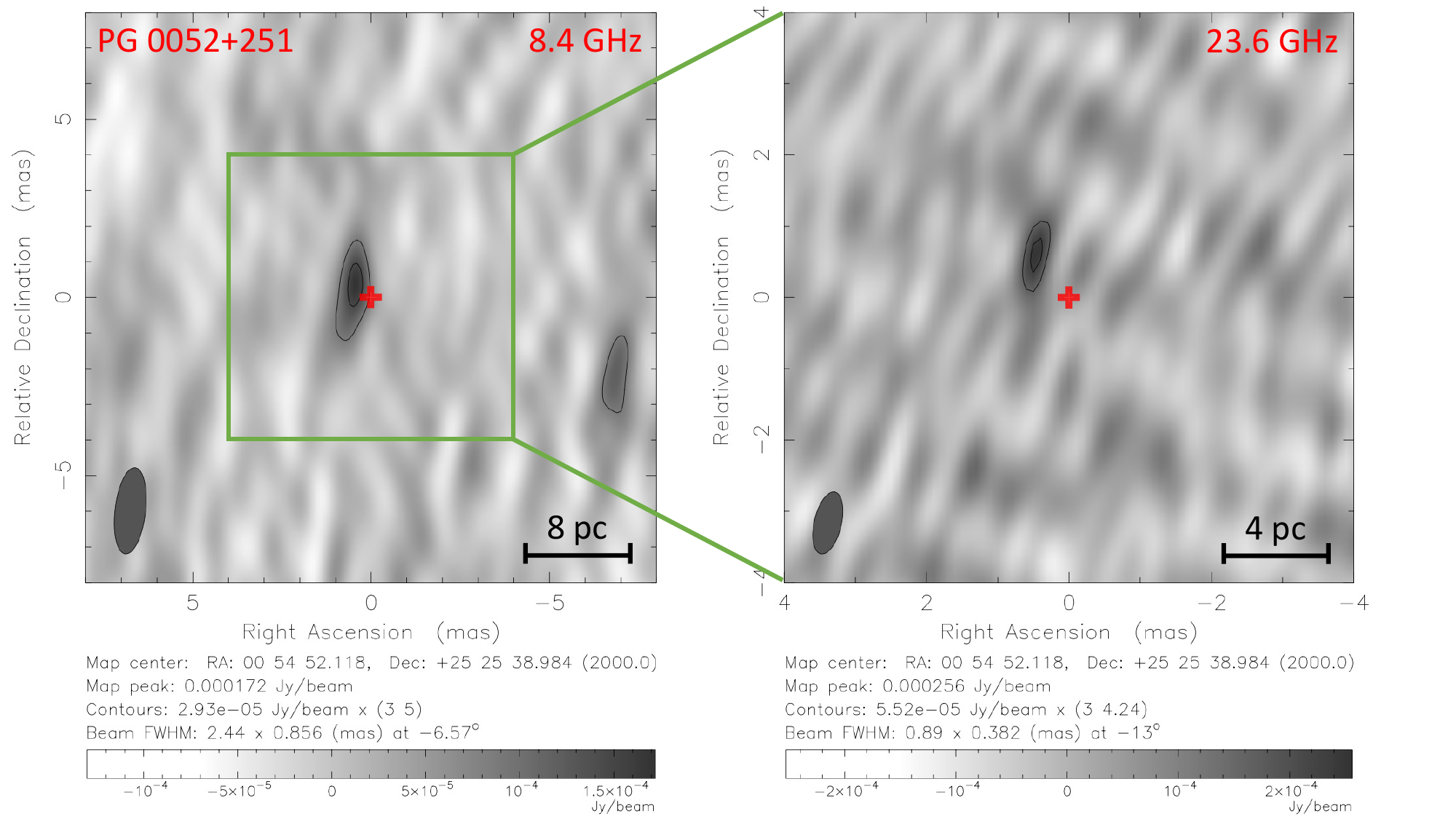}
\includegraphics[width=\columnwidth, trim={4cm, 0cm, 6cm, 1cm}, clip]{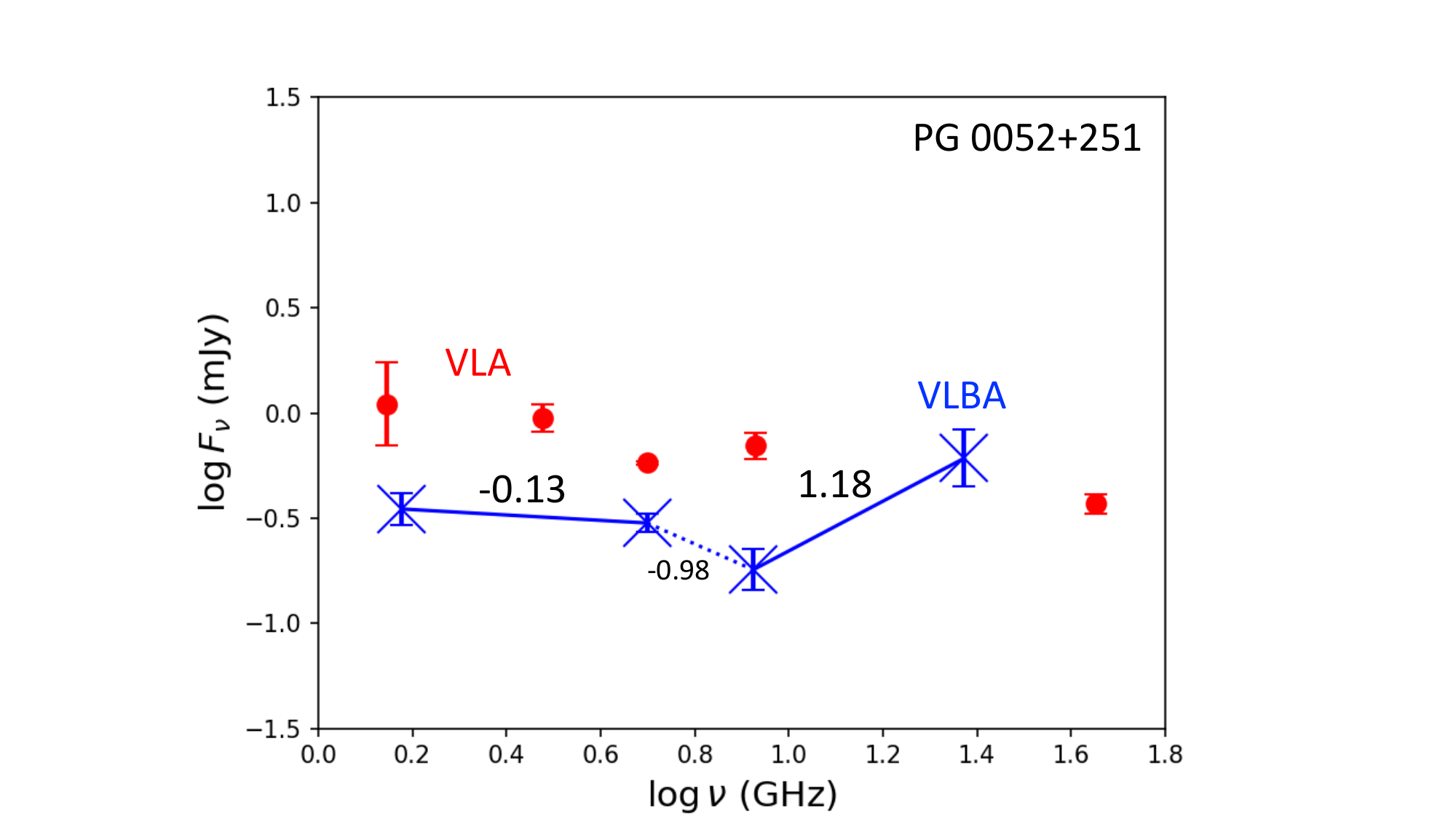}
\caption{PG\,0052+251: The radio maps at 8.4~GHz (upper-left panel) and 23.6~GHz (upper-right panel).
The contours are at (3, 5) $\times$ 0.0293~mJy/beam at 8.4~GHz, and (3, 4.24) $\times$ 0.0552~mJy/beam at 23.6~GHz.
The synthesized beam sizes and orientations are 2.44~mas $\times$ 0.856~mas at $-$6.57$^{\circ}$ at 8.4~GHz, and 0.890~mas $\times$ 0.382~mas at $-$13.0$^{\circ}$ at 23.6~GHz.
The 1--24~GHz radio spectrum (lower panel).
The core emission is flat at 1.5--5.0~GHz, and remains flat at 8.4--23.6~GHz.
The symbols and labels are the same as Figure~\ref{0026}.}
\label{0052}
\end{figure*}

\begin{figure*}[ht!]
\centering
\includegraphics[width=1.8\columnwidth, trim={0cm, 0cm, 2cm, 0cm}, clip]{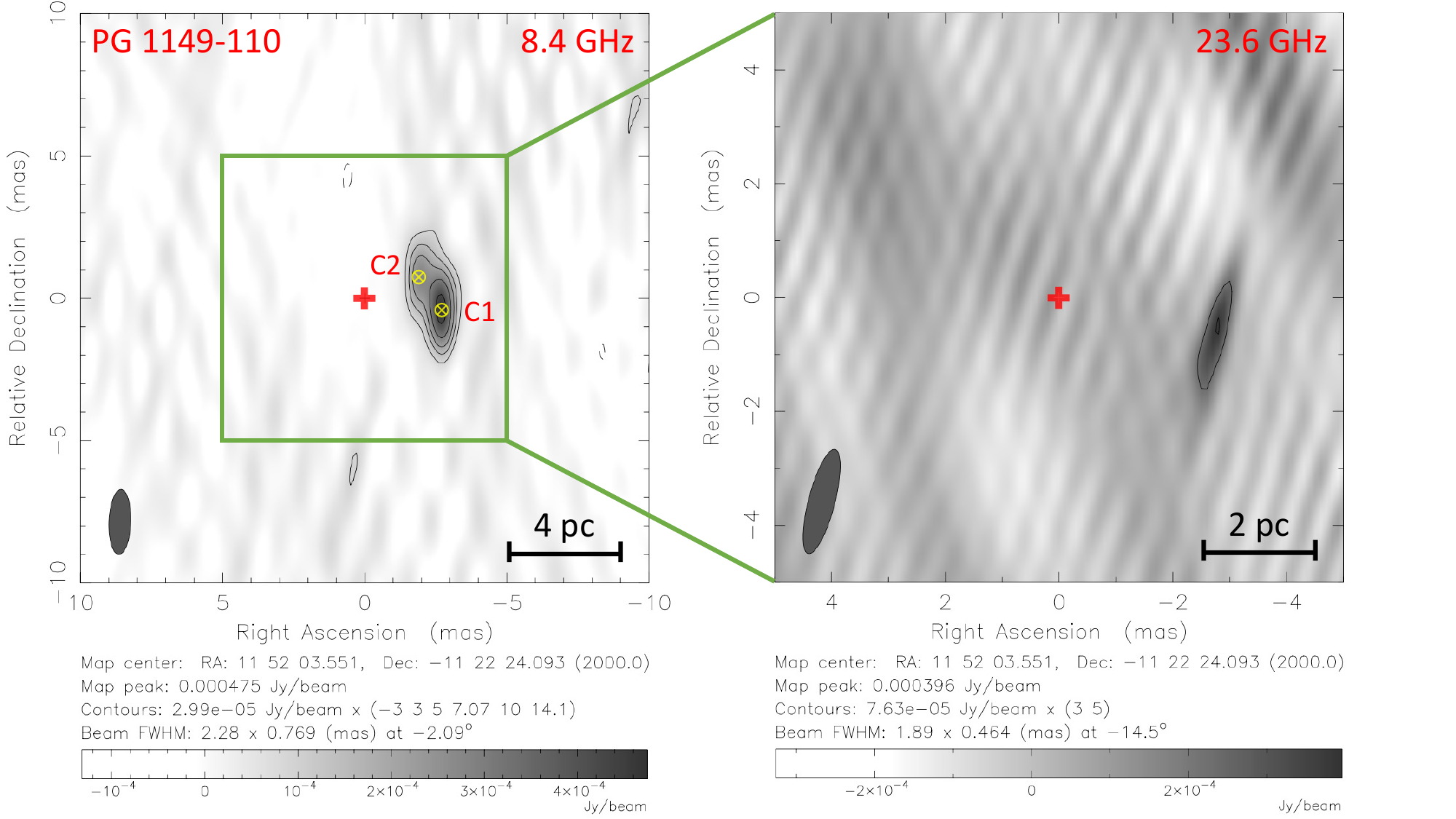}
\includegraphics[width=\columnwidth, trim={4cm, 0cm, 6cm, 1cm}, clip]{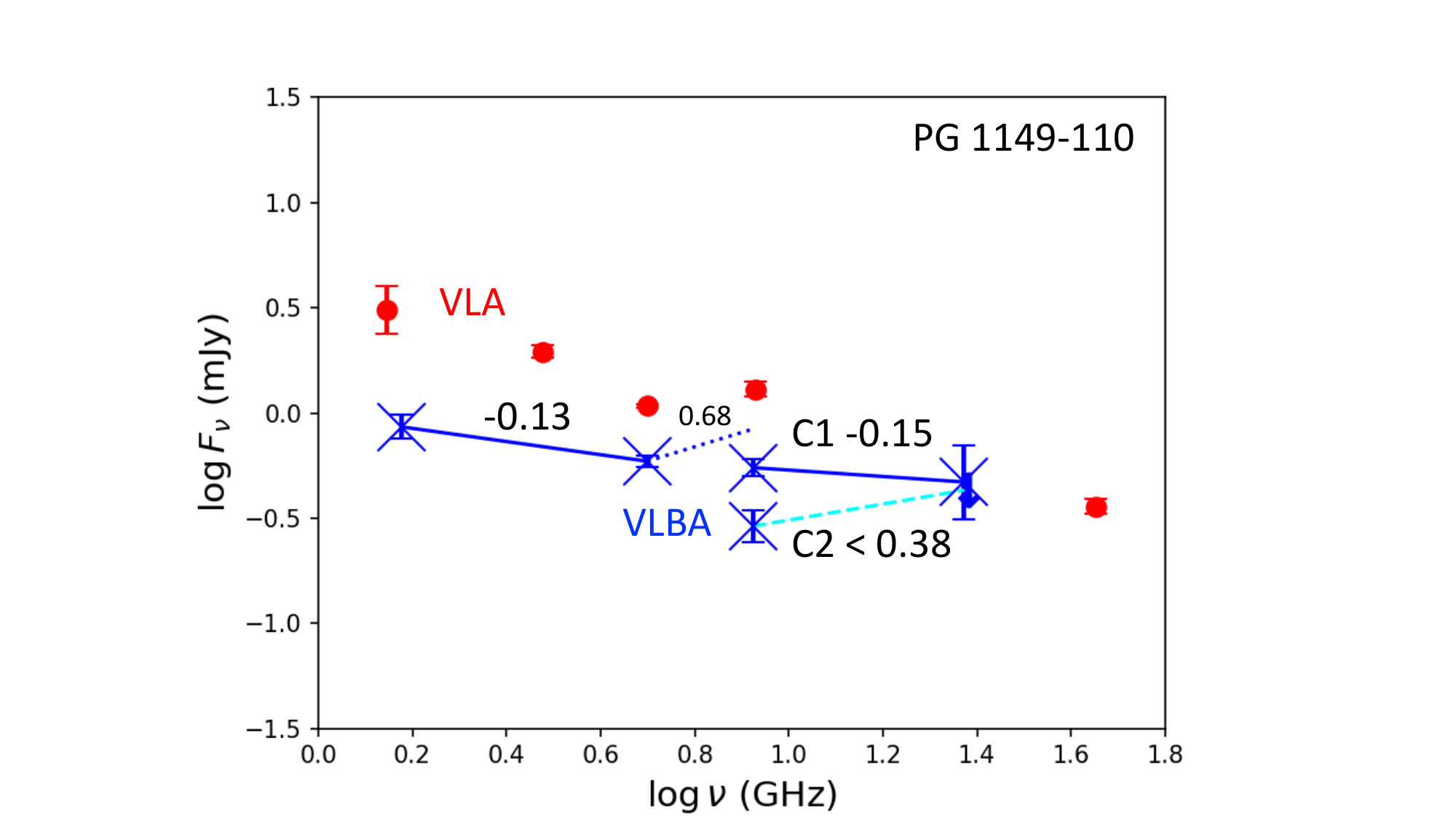}
\caption{PG\,1149$-$110: The radio maps at 8.4~GHz (upper-left panel) and 23.6~GHz (upper-right panel).
The contours are at ($-$3, 3, 5, 7.07, 10, 14.1) $\times$ 0.0299~mJy/beam at 8.4~GHz, and (3, 5) $\times$ 0.0763~mJy/beam at 23.6~GHz.
The synthesized beam sizes and orientations are 2.28~mas $\times$ 0.769~mas at $-$2.09$^{\circ}$ at 8.4~GHz, and 1.89~mas $\times$ 0.464~mas at $-$14.5$^{\circ}$ at 23.6~GHz.
The source has two components at 8.4~GHz, which are marked in yellow circles.
The 1--24~GHz radio spectrum (lower panel).
The core emission is flat at 1.5--5.0~GHz, and remains flat at 8.4--23.6~GHz.
The blue and cyan lines represent the slopes of different components.
The 5.0--8.4~GHz slope is derived from the 8.4~GHz total flux of two components which are unresolved at 5.0~GHz.
The symbols and labels are the same as Figure~\ref{0026}.}
\label{1149}
\end{figure*}

\begin{figure*}[ht!]
\centering
\includegraphics[width=1.8\columnwidth, trim={0cm, 0cm, 2cm, 0cm}, clip]{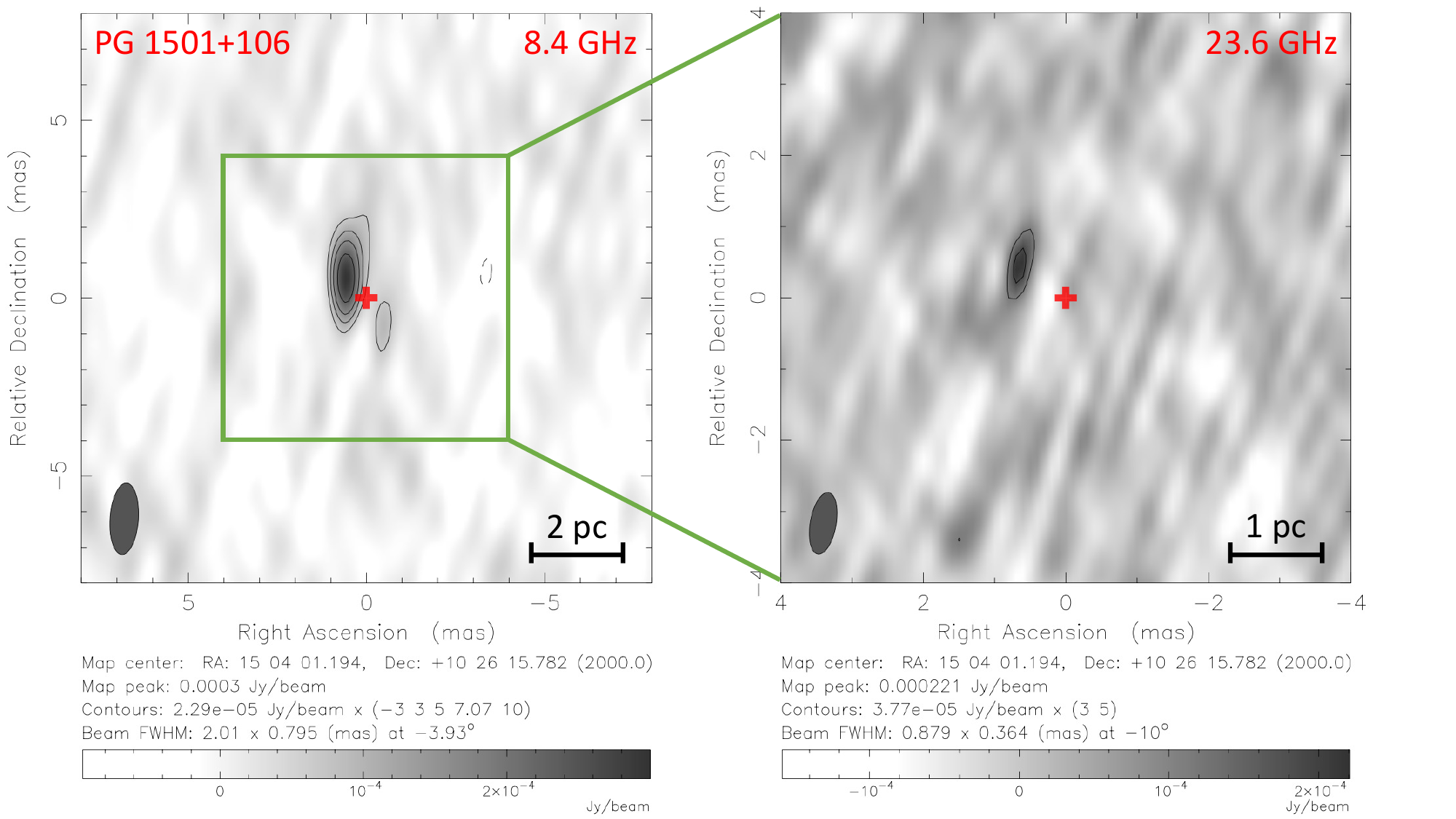}
\includegraphics[width=\columnwidth, trim={4cm, 0cm, 6cm, 1cm}, clip]{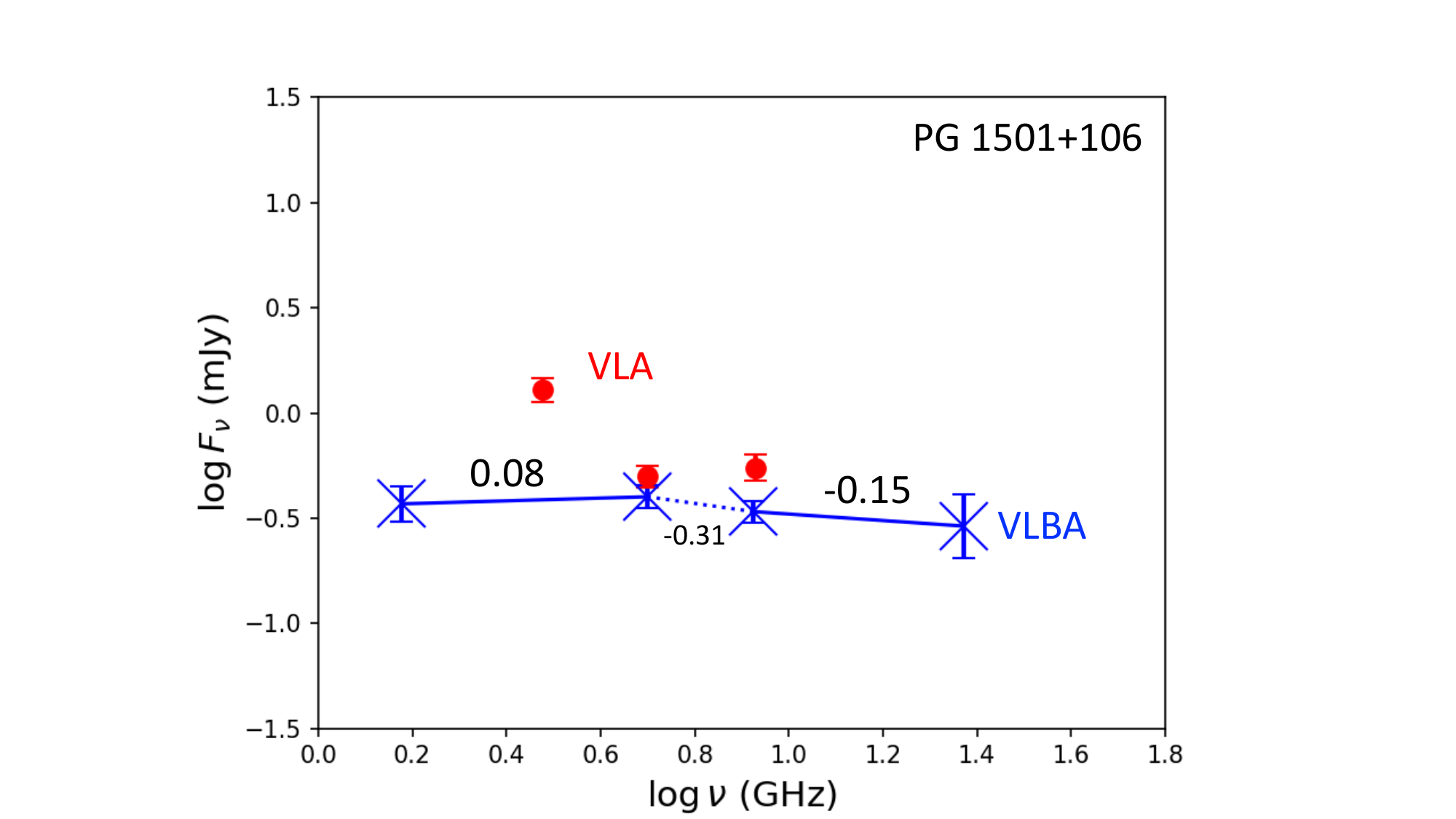}
\caption{PG\,1501+106: The radio maps at 8.4~GHz (upper-left panel) and 23.6~GHz (upper-right panel).
The contours are at ($-$3, 3, 5, 7.07, 10) $\times$ 0.0229~mJy/beam at 8.4~GHz, and (3, 5) $\times$ 0.0377~mJy/beam at 23.6~GHz.
The synthesized beam sizes and orientations are 2.01~mas $\times$ 0.795~mas at $-$3.93$^{\circ}$ at 8.4~GHz, and 0.879~mas $\times$ 0.364~mas at $-$10.0$^{\circ}$ at 23.6~GHz.
The 1--24~GHz radio spectrum (lower panel).
The core emission is flat at 1.5--5.0~GHz, and remains flat at 8.4--23.6~GHz.
The symbols and labels are the same as Figure~\ref{0026}.}
\label{1501}
\end{figure*}

\begin{figure*}[ht!]
\centering
\includegraphics[width=1.8\columnwidth, trim={0cm, 0cm, 2cm, 0cm}, clip]{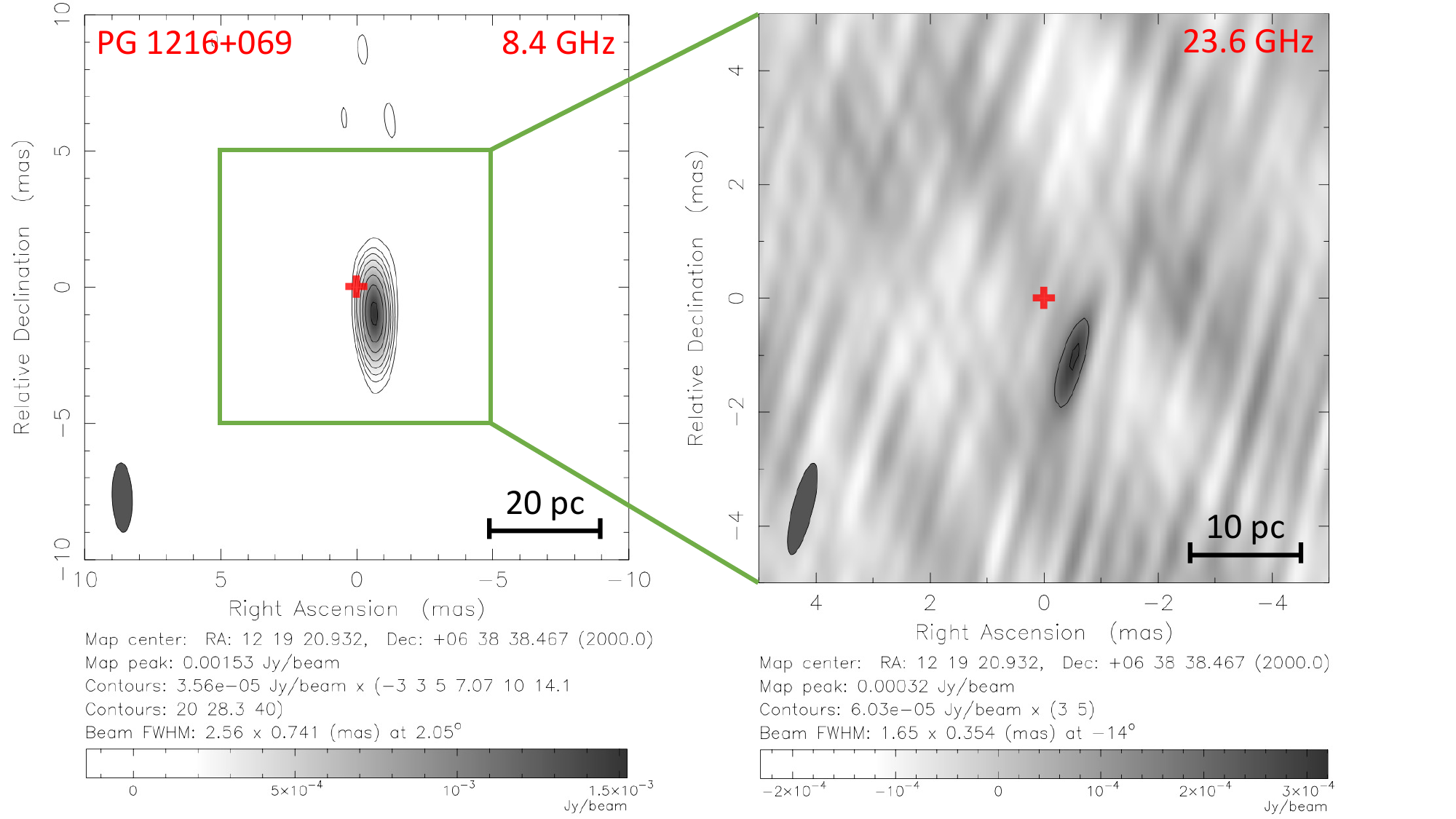}
\includegraphics[width=\columnwidth, trim={4cm, 0cm, 6cm, 1cm}, clip]{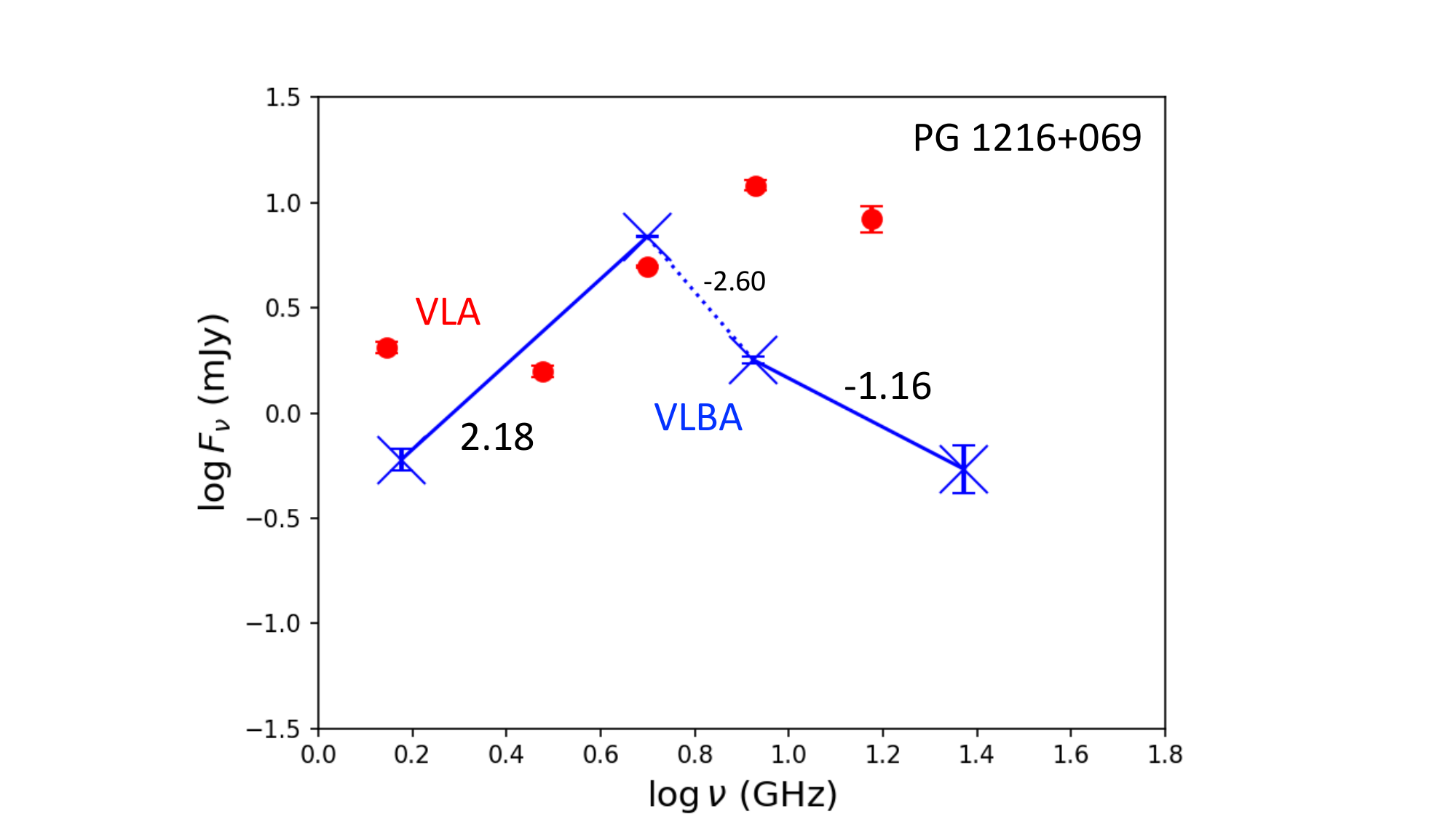}
\caption{PG\,1216+069: The radio maps at 8.4~GHz (upper-left panel) and 23.6~GHz (upper-right panel).
The contours are at ($-$3, 3, 5, 7.07, 10, 14.1, 20, 28.3, 40) $\times$ 0.0356~mJy/beam at 8.4~GHz, and (3, 5) $\times$ 0.0603~mJy/beam at 23.6~GHz.
The synthesized beam sizes and orientations are 2.56~mas $\times$ 0.741~mas at 2.05$^{\circ}$ at 8.4~GHz, and 1.65~mas $\times$ 0.354~mas at $-$14.0$^{\circ}$ at 23.6~GHz.
The 1--24~GHz radio spectrum (lower panel).
The core emission is flat at 1.5--5.0~GHz, and becomes steep at 8.4--23.6~GHz.
The symbols and labels are the same as Figure~\ref{0026}.}
\label{1216}
\end{figure*}

\begin{figure*}[ht!]
\centering
\includegraphics[width=.9\columnwidth, trim={0cm, 0cm, 18cm, 0cm}, clip]{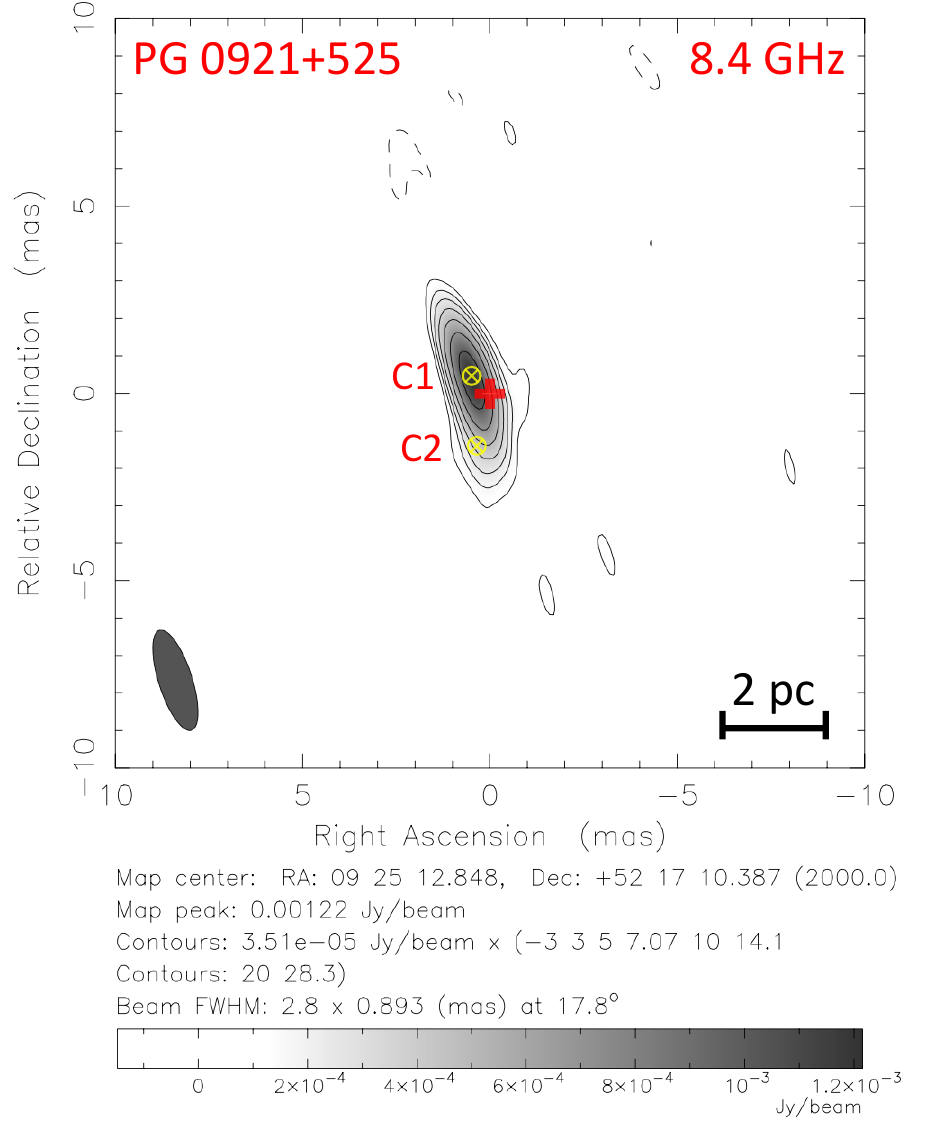}
\includegraphics[width=\columnwidth, trim={4cm, 0cm, 6cm, 1cm}, clip]{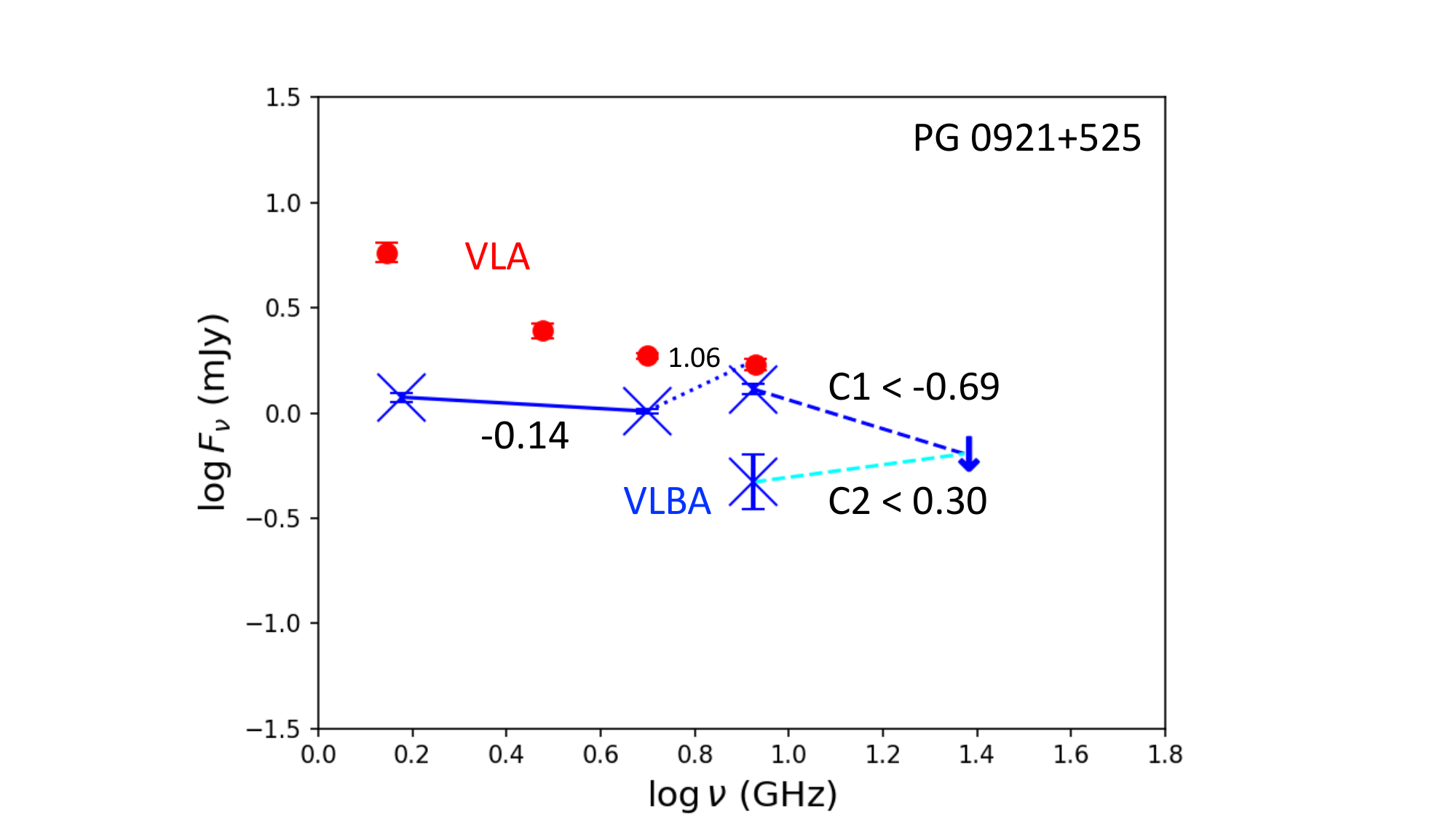}
\caption{PG\,0921+525: The radio maps at 8.4~GHz (left panel).
The contours are at ($-$3, 3, 5, 7.07, 10, 14.1, 20, 28.3) $\times$ 0.0351~mJy/beam at 8.4~GHz.
The synthesized beam sizes and orientations are 2.80~mas $\times$ 0.893~mas at 17.8$^{\circ}$ at 8.4~GHz.
The source has two components at 8.4~GHz, which are marked in yellow circles.
The 1--24~GHz radio spectrum (right panel).
The core emission is flat at 1.5--5.0~GHz, and becomes steep at 8.4--23.6~GHz.
The blue and cyan lines represent the slopes of different components.
The 5.0--8.4~GHz slope is derived from the 8.4~GHz total flux of two components which are unresolved at 5.0~GHz.
The symbols and labels are the same as Figure~\ref{0026}.}
\label{0921}
\end{figure*}

\begin{figure*}[ht!]
\centering
\includegraphics[width=.9\columnwidth, trim={0cm, 0cm, 18cm, 0cm}, clip]{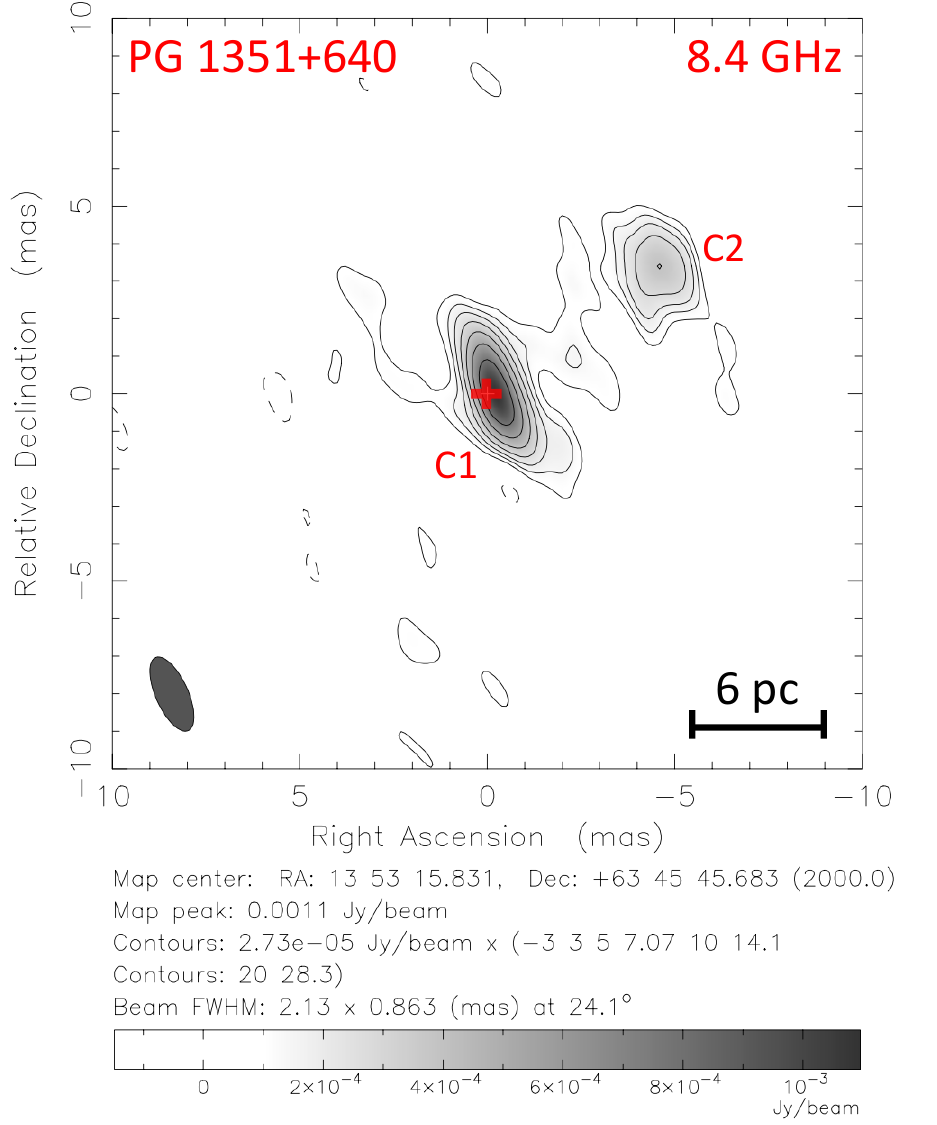}
\includegraphics[width=\columnwidth, trim={4cm, 0cm, 6cm, 1cm}, clip]{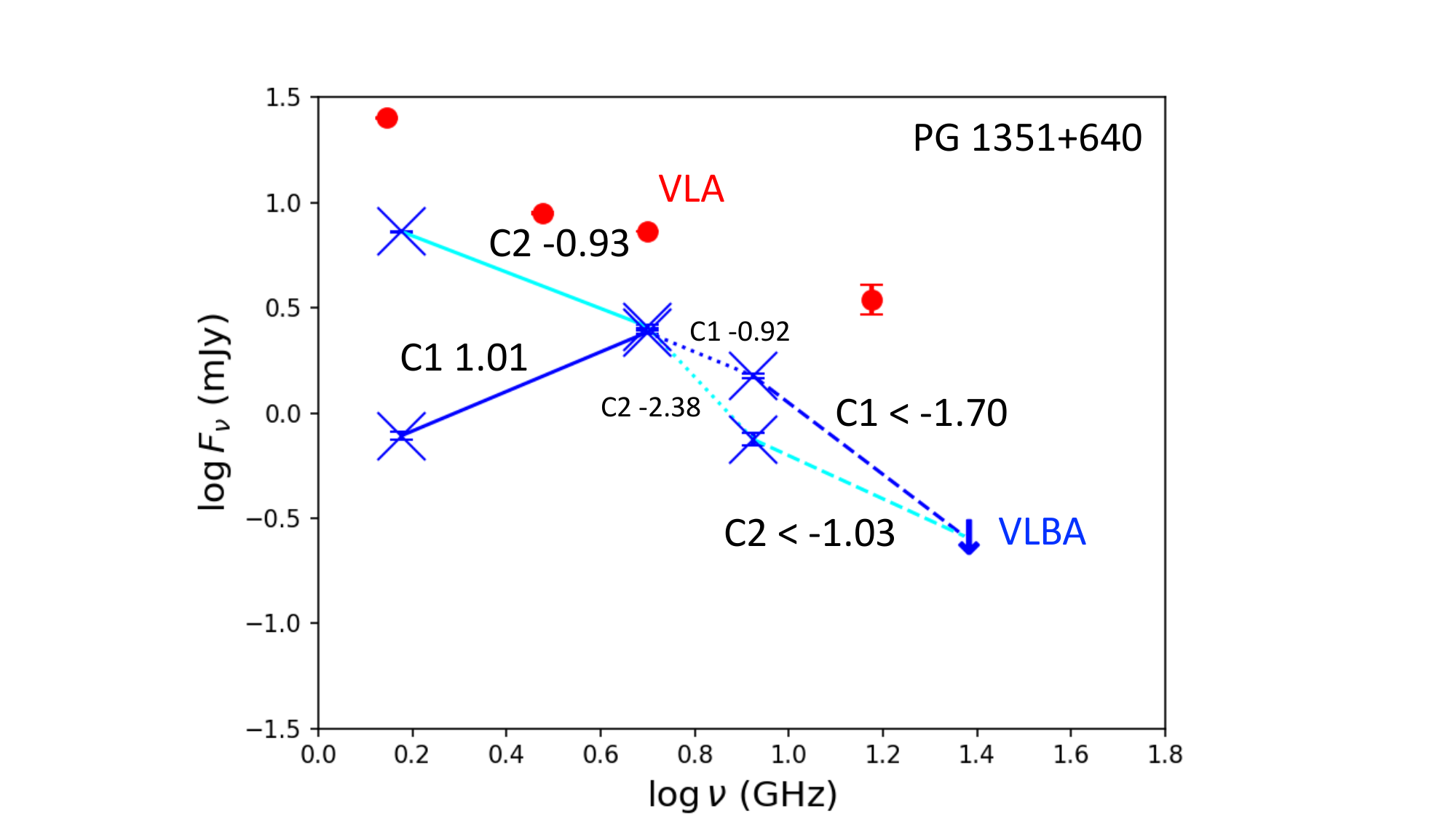}
\caption{PG\,1351+640: The radio maps at 8.4~GHz (left panel).
The contours are at ($-$3, 3, 5, 7.07, 10, 14.1, 20, 28.3) $\times$ 0.0273~mJy/beam at 8.4~GHz.
The synthesized beam sizes and orientations are 2.13~mas $\times$ 0.863~mas at 24.1$^{\circ}$ at 8.4~GHz.
The source has two components at 8.4~GHz.
The 1--24~GHz radio spectrum (right panel).
The core emission is flat at 1.5--5.0~GHz, and becomes steep at 8.4--23.6~GHz.
The blue and cyan lines represent the slopes of different components.
The symbols and labels are the same as Figure~\ref{0026}.}
\label{1351}
\end{figure*}

\begin{figure*}[ht!]
\centering
\includegraphics[width=.9\columnwidth, trim={0cm, 0cm, 18cm, 0cm}, clip]{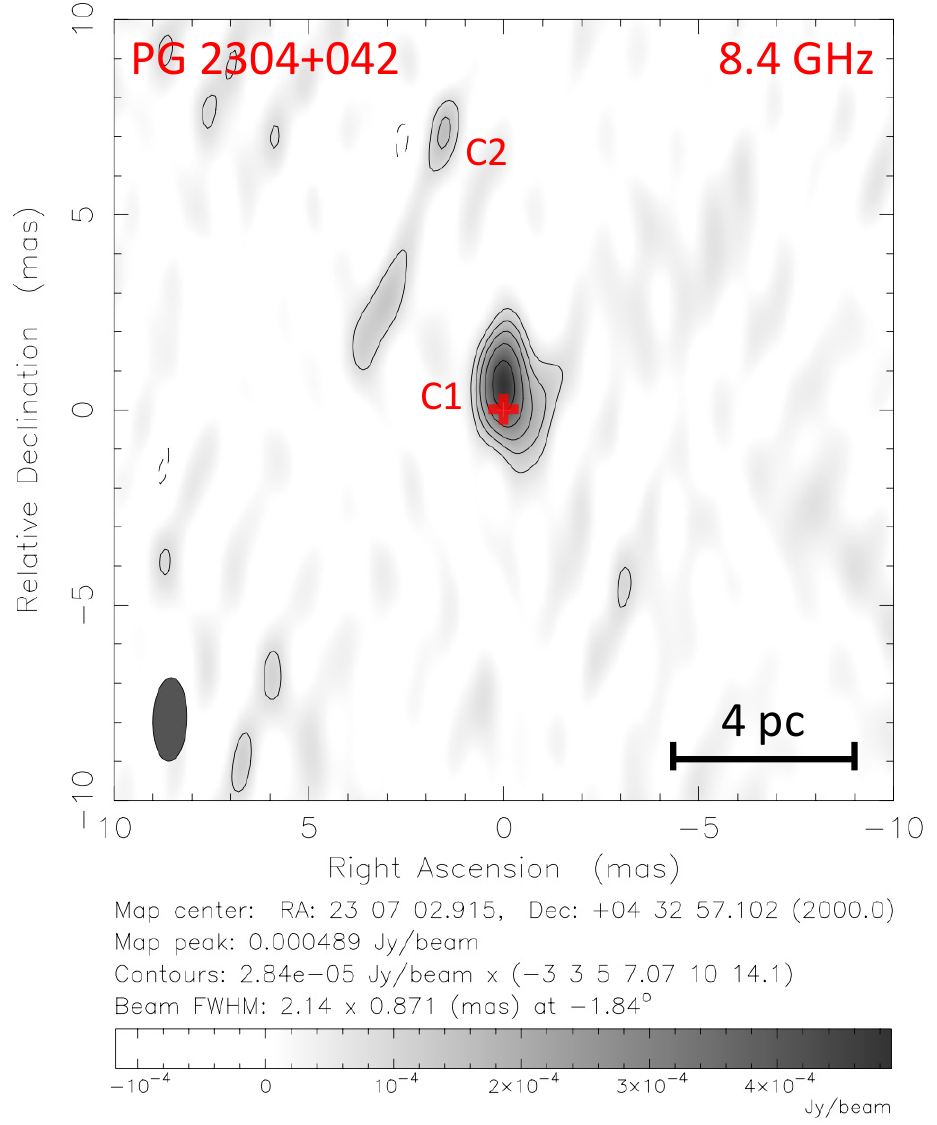}
\includegraphics[width=\columnwidth, trim={4cm, 0cm, 6cm, 1cm}, clip]{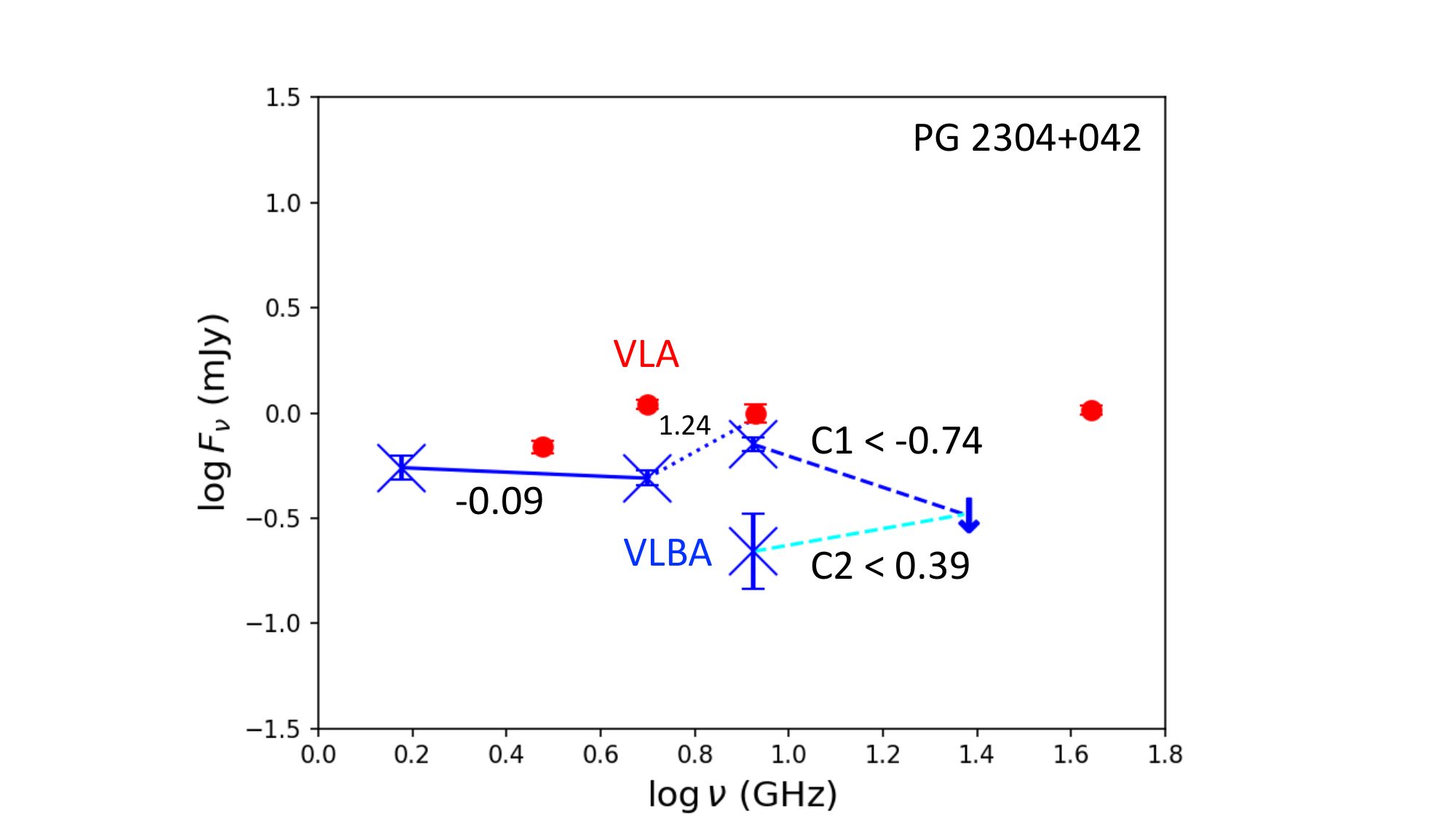}
\caption{PG\,2304+042: The radio maps at 8.4~GHz (left panel).
The contours are at ($-$3, 3, 5, 7.07, 10, 14.1) $\times$ 0.0284~mJy/beam at 8.4~GHz.
The synthesized beam sizes and orientations are 2.14~mas $\times$ 0.871~mas at $-$1.84$^{\circ}$ at 8.4~GHz.
The source has two components at 8.4~GHz.
The 1--24~GHz radio spectrum (right panel).
The core emission is flat at 1.5--5.0~GHz, and becomes steep at 8.4--23.6~GHz.
The blue and cyan lines represent the slopes of different components.
The 5.0--8.4~GHz slope is derived from the 8.4~GHz total flux of two components which are unresolved at 5.0~GHz.
The symbols and labels are the same as Figure~\ref{0026}.}
\label{2304}
\end{figure*}

\begin{figure*}[ht!]
\centering
\includegraphics[width=.9\columnwidth, trim={0cm, 0cm, 18cm, 0cm}, clip]{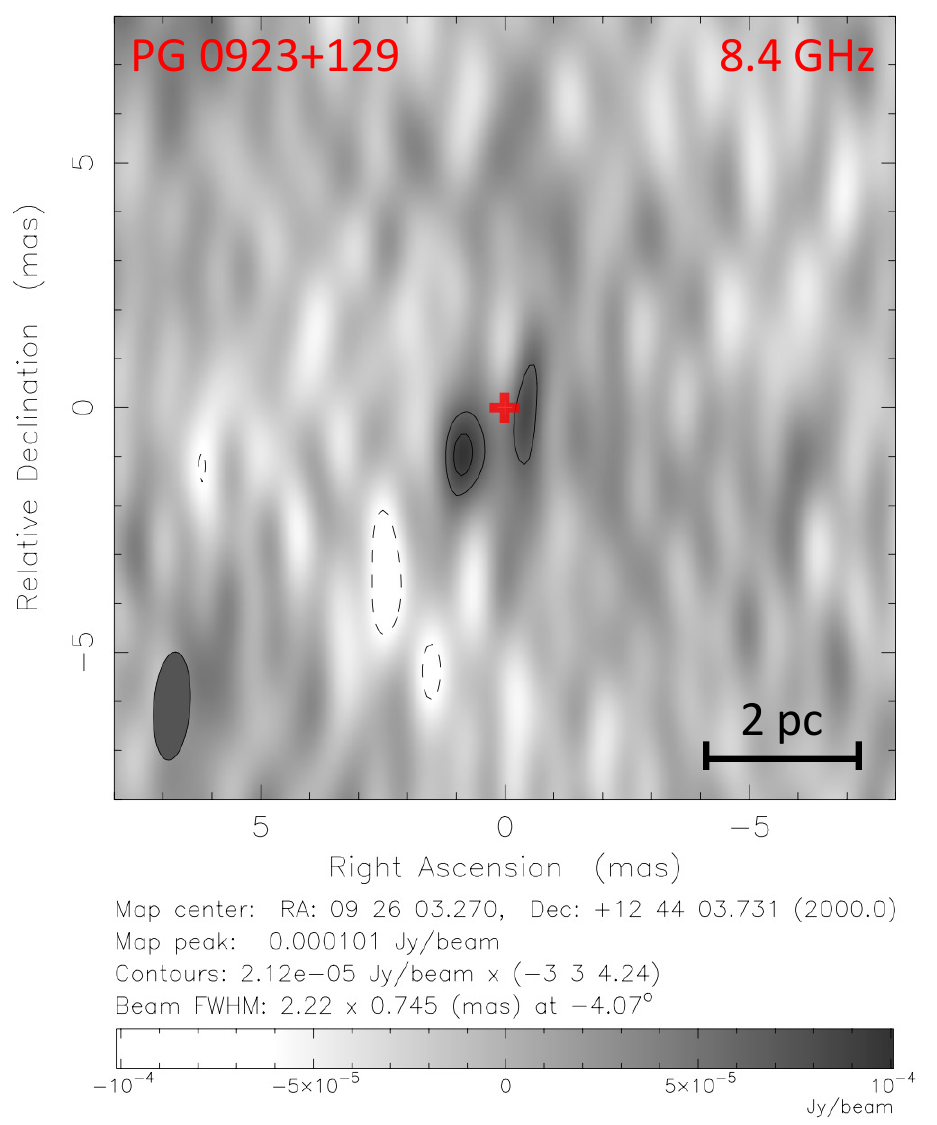}
\includegraphics[width=\columnwidth, trim={4cm, 0cm, 6cm, 1cm}, clip]{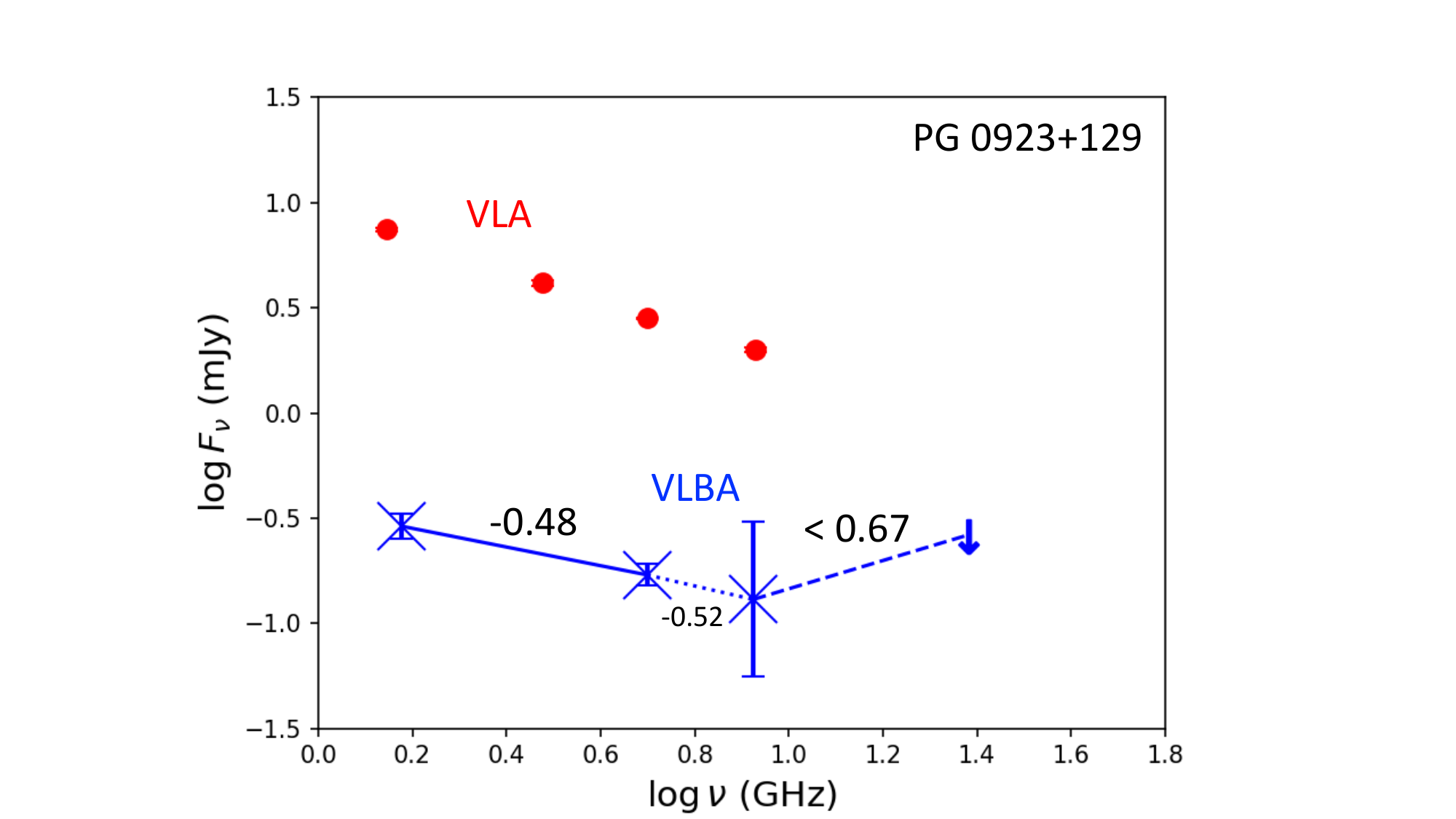}
\caption{PG\,0923+129: The radio maps at 8.4~GHz (left panel).
The contours are at ($-$3, 3, 4.24) $\times$ 0.0212~mJy/beam at 8.4~GHz.
The synthesized beam sizes and orientations are 2.22~mas $\times$ 0.745~mas at $-$4.07$^{\circ}$ at 8.4~GHz.
The 1--24~GHz radio spectrum (right panel).
The core emission is flat at 1.5--5.0~GHz, but uncertain at 8.4--23.6~GHz.
The symbols and labels are the same as Figure~\ref{0026}.}
\label{0923}
\end{figure*}

\begin{figure*}[ht!]
\centering
\includegraphics[width=.9\columnwidth, trim={0cm, 0cm, 18cm, 0cm}, clip]{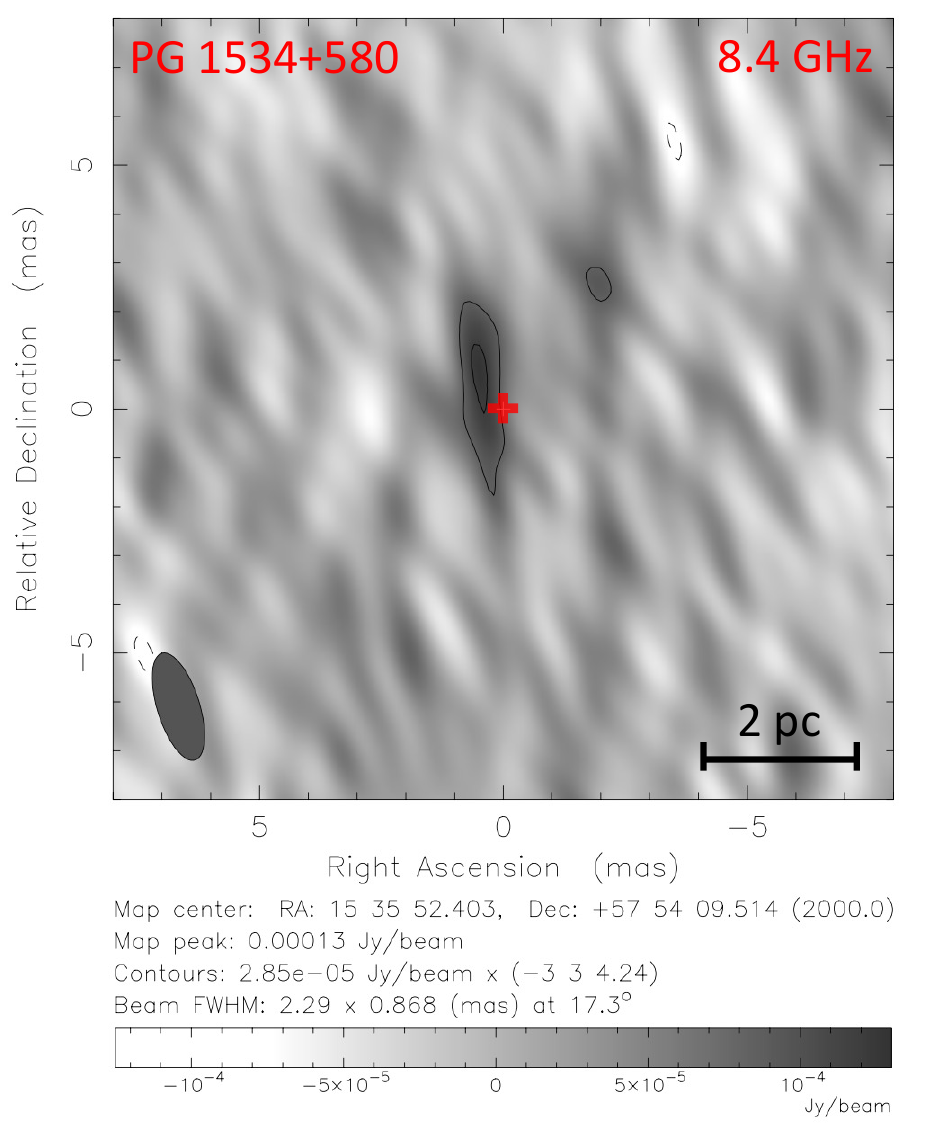}
\includegraphics[width=\columnwidth, trim={4cm, 0cm, 6cm, 1cm}, clip]{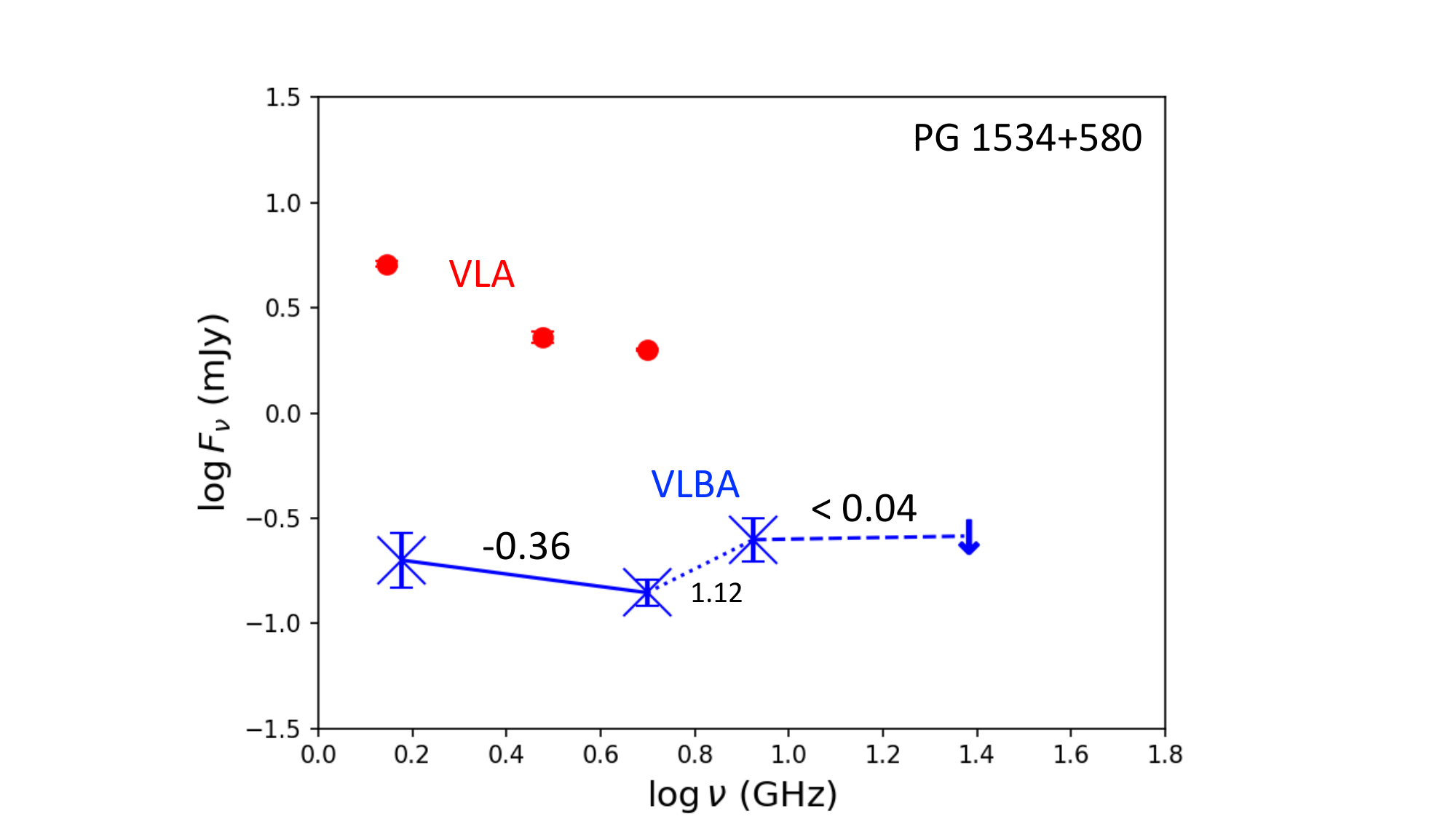}
\caption{PG\,1534+580: The radio maps at 8.4~GHz (left panel).
The contours are at ($-$3, 3, 4.24) $\times$ 0.0285~mJy/beam at 8.4~GHz.
The synthesized beam sizes and orientations are 2.29~mas $\times$ 0.868~mas at 17.3$^{\circ}$ at 8.4~GHz.
The 1--24~GHz radio spectrum (right panel).
The core emission is flat at 1.5--5.0~GHz, but uncertain at 8.4--23.6~GHz.
The symbols and labels are the same as Figure~\ref{0026}.}
\label{1534}
\end{figure*}

\begin{figure*}[ht!]
\centering
\includegraphics[width=.9\columnwidth, trim={0cm, 0cm, 18cm, 0cm}, clip]{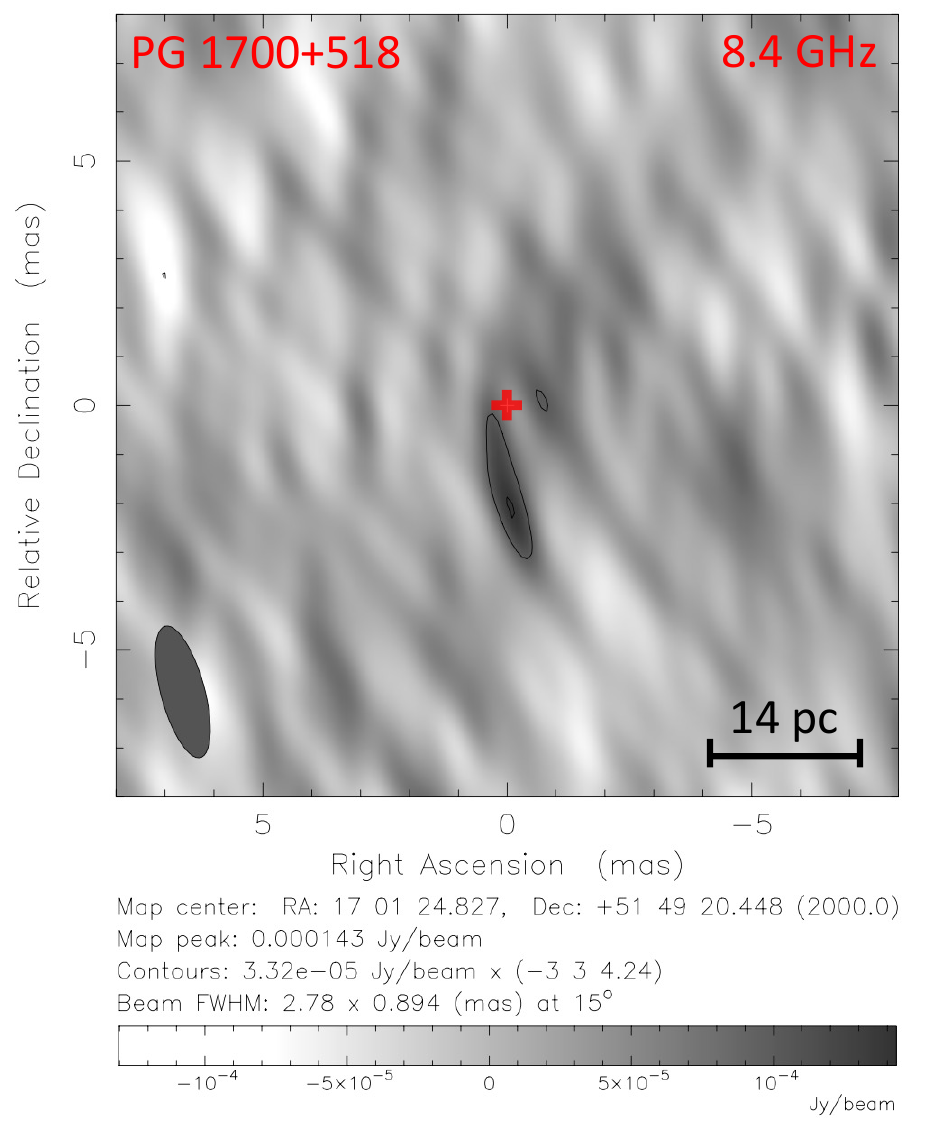}
\includegraphics[width=\columnwidth, trim={4cm, 0cm, 6cm, 1cm}, clip]{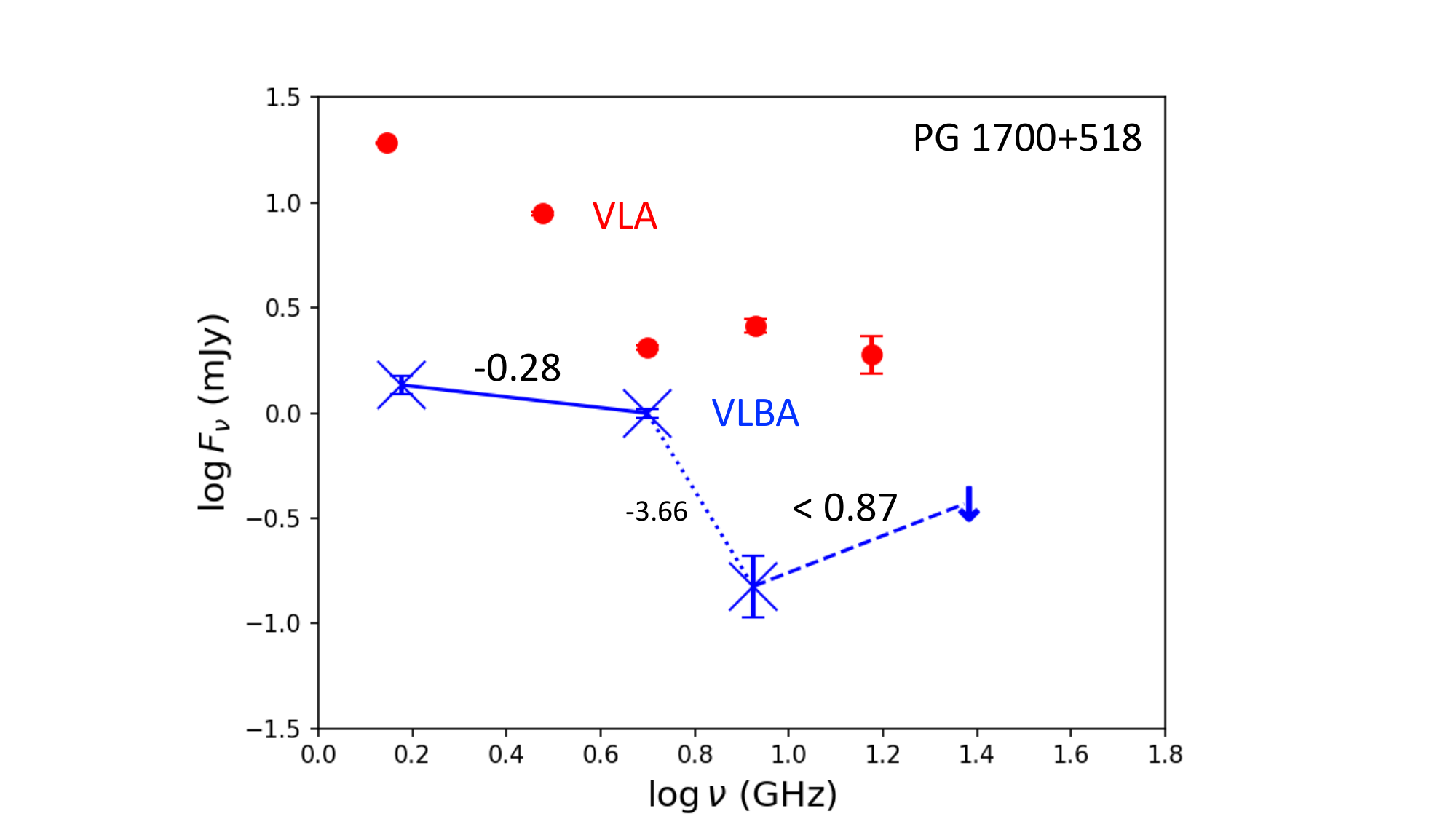}
\caption{PG\,1700+518: The radio maps at 8.4~GHz (left panel).
The contours are at ($-$3, 3, 4.24) $\times$ 0.0332~mJy/beam at 8.4~GHz.
The synthesized beam sizes and orientations are 2.78~mas $\times$ 0.894~mas at 15.0$^{\circ}$ at 8.4~GHz.
The 1--24~GHz radio spectrum (right panel).
The core emission is flat at 1.5--5.0~GHz, but uncertain at 8.4--23.6~GHz.
The symbols and labels are the same as Figure~\ref{0026}.}
\label{1700}
\end{figure*}

\begin{figure*}[ht!]
\centering
\includegraphics[width=.95\columnwidth, trim={4cm, 0cm, 6cm, 1cm}, clip]{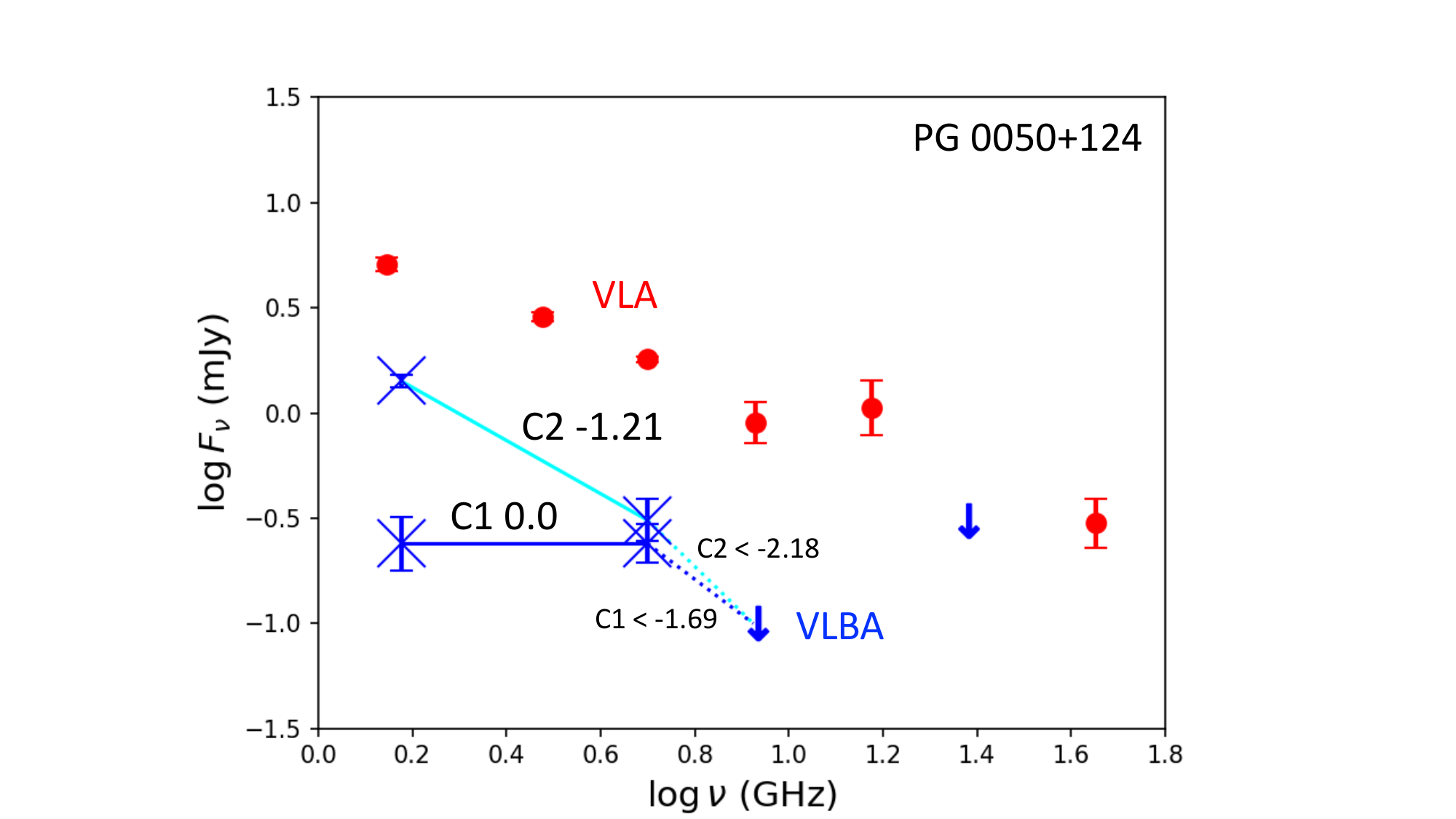}
\includegraphics[width=.95\columnwidth, trim={4cm, 0cm, 6cm, 1cm}, clip]{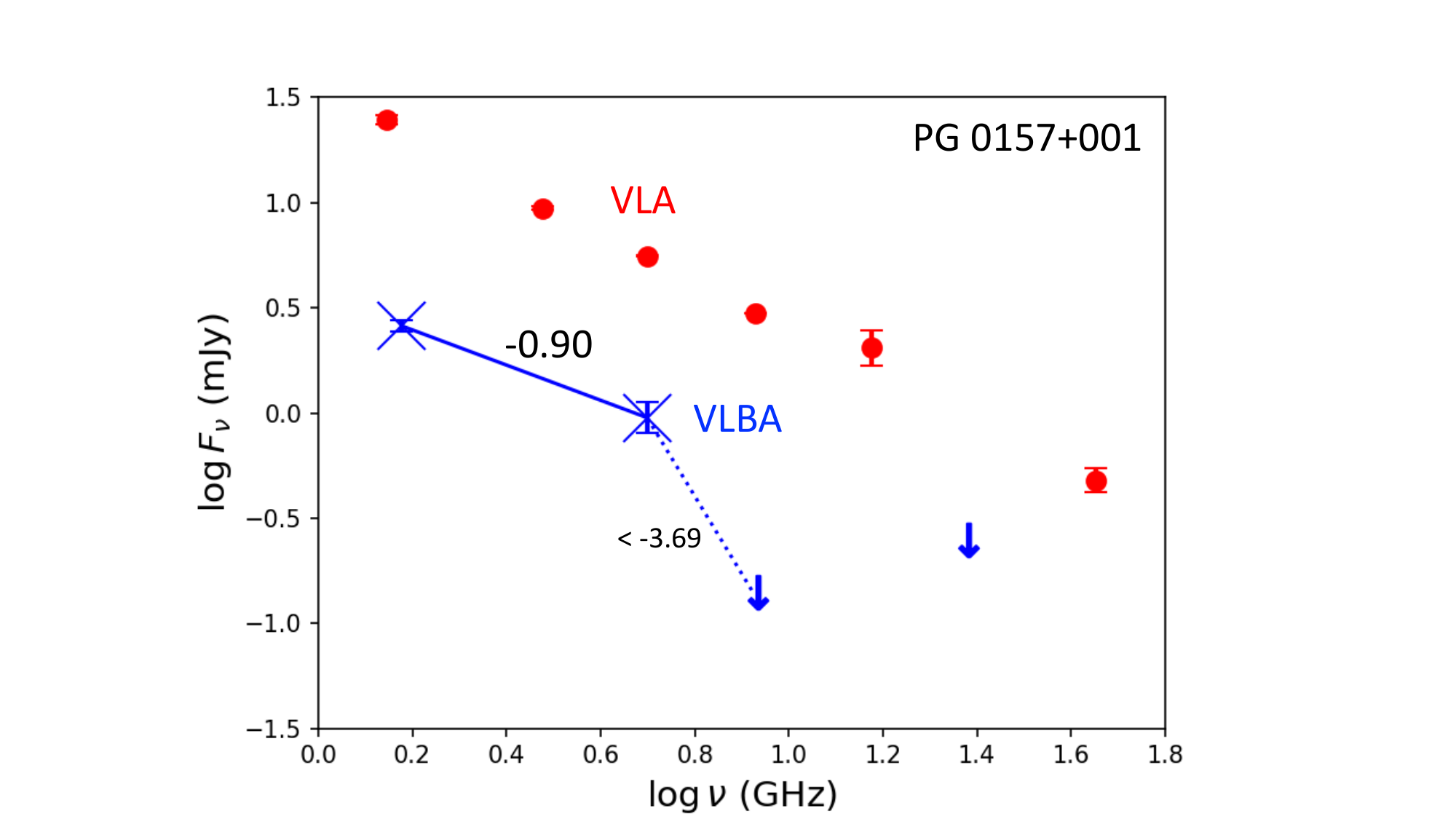}
\caption{The 1--24~GHz radio spectrum of PG\,0050+124 (left panel) and PG\,0157+001 (right panel).
The 8.4--23.6~GHz slopes are uncertain.
The blue and cyan lines represent the slopes of different components.
The symbols and labels are the same as Figure~\ref{0026}.}
\label{0050+0157}
\end{figure*}

We consider a $\ge 5 \sigma$ level, where $\sigma$ is the background noise, as a reliable detection, and also include a $\sim 4 \sigma$ level as a potential detection.
If the source is not detected, a 5$\sigma$ upper limit on the flux density is adopted.
Figures~\ref{0026}--\ref{1700} present the radio maps of the 11 PG RQQ detected in the VLBA observations at 8.4 and 23.6~GHz centered at the {\it Gaia} position \citep{Gaia2016,Gaia2023}.
We label the component dominated by the core emission as ``C1'' and the component dominated by the extended emission as ``C2''.
Five objects (PG\,0026+129, PG\,0052+251, PG\,1149$-$110, PG\,1216+069, and PG\,1501+106) are detected at both 8.4 and 23.6~GHz.
Six objects are only detected at 8.4~GHz, with three at $\ge 5 \sigma$ (PG\,0921+525, PG\,1351+640, and PG\,2304+042) and three at $\sim 4 \sigma$ (PG\,0923+129, PG\,1534+580, and PG\,1700+518).
Two objects (PG\,0050+124 and PG\,0157+001) are not detected at either 8.4 or 23.6~GHz.

Both PG\,1351+640 and PG\,2304+042 exhibit two spatially separated components at 8.4~GHz.
The C1 component is much closer to the {\it Gaia} position than C2, suggesting that C1 and C2 are likely associated with the core emission and the extended emission, respectively.
PG\,1149$-$110 also shows two components at 8.4~GHz, with only C1 detected at 23.6~GHz, indicating that C1 is the core emission and C2 is the extended emission.
In PG\,0921+525 which shows two spatially unresolved components at 8.4~GHz, we consider C1 with a smaller source size as the core emission, and C2 with a larger source size as the extended emission.

Table~\ref{size} lists the VLBA positions, the offsets from the {\it Gaia} positions \citep{Gaia2016,Gaia2023}, and the deconvolved source sizes in the full-array maps.
The offsets between the {\it Gaia} and the VLBI positions typically range from 0.1~mas to 10~mas in about 90\% of the AGN population with a median value of $\sim$ 2~mas \citep{Petrov2017}.
In our sample, the X and K band coordinates are consistent with the {\it Gaia} positions to within $\sim$ 2~mas, except for the C2 component in PG\,1351+640 and PG\,2304+042, which show offsets of about 5--7~mas from the {\it Gaia} positions.
If the deconvolved source size is smaller than half of the beam size, we adopt half of the beam size as an upper limit on the source size.

Table~\ref{flux_XK} reports the total flux density $F_{\rm total}$, the core flux density $F_{\rm core}$, the core to total flux ratio $F_{\rm core}/F_{\rm total}$, the background noise RMS, and the $uv$-range of the full-array and the tapered maps in the VLBA 8.4 and 23.6~GHz observations.
If a source is not detected, a 5$\sigma$ upper limit on $F_{\rm core}$ is provided.
For objects with two components, the $F_{\rm core}/F_{\rm total}$ is the ratio of the core flux to the total flux of both components.
Table~\ref{flux_LC} lists the VLBA 1.5 and 5.0~GHz total flux density in the full-array and the tapered maps from \citet{Alhosani2022} and \citet{Chen2023}.
For objects with more than one component, the total flux density of the C1 component there, which is dominated by the core emission, is provided.
We focus on the 1.5 and 5.0~GHz core emission, because the extended emission detected at these frequencies is generally resolved out on higher resolutions and/or becomes too faint to be detected at higher frequencies, except for PG\,1351+640 in which both the C1 and C2 components are detected at 5.0 and 8.4~GHz.
Table~\ref{vla} includes the VLA 1.4, 3.0, 5.0, 8.5, 15, and 45~GHz core flux densities observed with the VLA A/B/C configurations from literature.

The VLBA spectral slope at 8.4--23.6~GHz $\alpha_{\rm 8.4-23.6}$ is measured based on the $F_{\rm total}$ in the tapered maps, which have comparable resolutions and cover emission on similar scales at both frequencies.
For objects detected only at 8.4~GHz, an upper limit on the slope is provided.
The slope can not be determined for objects not detected at either frequencies.
The brightness temperature of all components is calculated using the formula \citep[e.g.][]{Ulvestad2005}
\begin{equation}
T_{\rm B} = 1.8 \times 10^{9} (1+z) \frac{F_\nu}{\nu^2 \theta_{\rm max} \theta_{\rm min}}
\end{equation}
where $F_\nu$ is the flux density in mJy, $\nu$ is the observing frequency in GHz, and $\theta_{\rm max}$ and $\theta_{\rm min}$ are the major and minor axes of the source size in mas.
We calculate the $T_{\rm B}$ at 8.4~GHz, where most of the objects are detected.
For objects where $\theta_{\rm max}$ and/or $\theta_{\rm min}$ are smaller than half of the beam size, the derived $T_{\rm B}$ is a lower limit.
Table~\ref{slope} presents the brightness temperature $T_{\rm B}$ calculated at 8.4~GHz, the VLBA spectral slopes at 8.4--23.6~GHz $\alpha_{\rm 8.4-23.6}$, and at 1.5--5.0~GHz $\alpha_{\rm 1.5-5.0}$ adapted from \citet{Alhosani2022} and \citet{Chen2023}.

\section{Results}

Figures~\ref{0026}--\ref{0050+0157} display the VLBA and VLA spectra at 1--24~GHz of the 13 objects observed in the VLBA observations.
We focus on the 1.5 and 5.0~GHz core emission, and how its spectral slope behaves differently at 8.4 and 23.6~GHz.
In four objects, PG\,0026+129, PG\,0052+251, PG\,1149$-$110, and PG\,1501+106, the spectral slope of the core emission (C1) is flat at 1.5--5.0~GHz, and remains flat at 8.4--23.6~GHz, which we refer to as the ``flat-flat slope'' objects hereafter.
In four other objects, PG\,1216+069, PG\,0921+525, PG\,1351+640, and PG\,2304+042, the spectral slope of the core emission (C1) is flat at 1.5--5.0~GHz, and becomes steep at 8.4--23.6~GHz, which we refer to as the ``flat-steep slope'' objects hereafter.
In the remaining five objects, PG\,0923+129, PG\,1534+580, PG\,1700+518, PG\,0050+124, and PG\,0157+001, the 8.4--23.6~GHz slope can not be determined or significantly constrained due to the upper limits.

\subsection{Flat at 1.5--5.0~GHz remains Flat at 8.4--23.6~GHz \label{sec:flat-flat}}

\subsubsection{PG\,0026+129}

The object shows a single component at both 8.4 and 23.6~GHz.
The source is dominated by the core emission as indicated by the $F_{\rm core}/F_{\rm total}$ = 0.7--0.8 at both frequencies.
The slope is inverted at 1.5--5.0~GHz, $\alpha_{1.5-5.0} > 0.89$ \citep{Chen2023}, and remains flat at 8.4--23.6~GHz, $\alpha_{8.4-23.6} = 0.03$, indicating that the emission is self-absorbed at 1.5--5.0~GHz and optically thick at 8.4--23.6~GHz.
The flat VLA 5--45~GHz slope \citep[0.02,][]{Baldi2022} suggests that the emission remains optically thick up to 45~GHz.
Additionally, the VLBA fluxes at 8.4--23.6~GHz are higher than the VLA fluxes, indicating significant variability.

\subsubsection{PG\,0052+251}

The object exhibits a single component at both 8.4 and 23.6~GHz.
The emission is compact at 8.4~GHz with the $F_{\rm core}/F_{\rm total}$ = 0.9, while half of the emission is resolved at 23.6~GHz with the $F_{\rm core}/F_{\rm total}$ = 0.5.
The flat 1.5--5.0~GHz slope $\alpha_{1.5-5.0} = -0.13$ \citep{Chen2023} indicates optically thick emission.
The inverted 8.4--23.6~GHz slope $\alpha_{8.4-23.6} = 1.18$ suggests that an additional component dominates at higher frequencies and is self-absorbed at lower frequencies.
The flat VLA 5--45~GHz slope \citep[$-$0.20,][]{Baldi2022} manifests that the emission is optically thick up to 45~GHz.

\subsubsection{PG\,1149$-$110}

The object displays two components at 8.4~GHz, which are unresolved at 1.5 and 5.0~GHz \citep{Alhosani2022,Wang2023c}.
A fraction of extended emission (C2) is seen at 8.4~GHz, as indicated by the $F_{\rm core}/F_{\rm total}$ = 0.6.
Only the core emission (C1) is detected at 23.6~GHz, and it is compact with the $F_{\rm core}/F_{\rm total}$ = 0.9.
The 1.5--5.0~GHz slope is flat $\alpha_{1.5-5.0} = -0.13$ \citep{Chen2023}.
The C1 slope remains flat at 8.4--23.6~GHz $\alpha_{8.4-23.6} = -0.15$, indicating that the core emission is optically thick at 1--24~GHz.
The C2 slope can not be determined due to the upper limit $\alpha_{8.4-23.6} < 0.38$.
The VLA 5--45~GHz slope \citep[$-$0.50,][]{Baldi2022}, which is at the transition between optically thin and optically thick emission, suggests that the core emission dominates at higher frequencies while the extended emission dominates at lower frequencies.

\subsubsection{PG\,1501+106}

The object shows a single component at both 8.4 and 23.6~GHz.
The emission is compact with the $F_{\rm core}/F_{\rm total}$ = 0.8--0.9 at both frequencies.
The slope is flat at 1.5--5.0~GHz $\alpha_{1.5-5.0} = 0.08$ \citep{Chen2023}, and remains flat at 8.4--23.6~GHz $\alpha_{8.4-23.6} = -0.15$, indicating optically thick emission at 1--24~GHz.
The 5--8~GHz fluxes on the VLBA scale are comparable to those on the VLA scale, suggesting that the core emission is dominant from pc to kpc scales.

\subsection{Flat at 1.5--5.0~GHz becomes Steep at 8.4--23.6~GHz \label{sec:flat-steep}}

\subsubsection{PG\,1216+069}

The object exhibits a single component at both 8.4 and 23.6~GHz, while a slightly extended structure is seen at 5.0~GHz \citep{Chen2023}.
The emission is compact with the $F_{\rm core}/F_{\rm total}$ = 0.8--0.9 at both frequencies.
The inverted 1.5--5.0~GHz slope $\alpha_{1.5-5.0} = 2.18$ \citep{Chen2023} indicates that the emission is self-absorbed at lower frequencies.
The emission becomes optically thin at higher frequencies as indicated by the steep 8.4--23.6~GHz slope $\alpha_{8.4-23.6} = -1.16$.
The VLBA fluxes at 5.0 and 8.4~GHz differ by a factor of $\sim$ 4, demonstrating significant variability on a time scale of $\sim$ 2 years.
The fluxes vary by a factor of $\sim$ 5 are also seen by comparing other VLBA observations \citep{Chen2023,Wang2023a,Wang2023c}.
The VLA spectrum at 1--15~GHz also shows a turnover around 8~GHz, despite of the data being collected over a few decades.

\subsubsection{PG\,0921+525}

The object is only detected at 8.4~GHz, showing two spatially unresolved components, and additional components are seen at 1.5 and 5.0~GHz \citep{Chen2023}.
A fraction of extended emission is present as indicated by the $F_{\rm core}/F_{\rm total}$ = 0.7.
The 1.5--5.0~GHz core emission is optically thick with a flat slope $\alpha_{1.5-5.0} = -0.14$ \citep{Chen2023}.
The 8.4--23.6~GHz core emission (C1) becomes optically thin with a steep slope $\alpha_{8.4-23.6} < -0.69$.
The slope of the extended emission (C2) can not be determined due to the upper limit $\alpha_{8.4-23.6} < 0.30$.
The steep VLA 1--8~GHz spectrum suggests the presence of extended emission on larger scales.
Radio variability was detected on both the VLBA and the VLA scales \citep{Panessa2022a,Chen2022b}.

\subsubsection{PG\,1351+640}

The object displays two components at 8.4~GHz with a separation of $\sim$ 5~mas, but is not detected at 23.6~GHz.
The two components are also detected at 5.0~GHz, but unresolved at 1.5~GHz \citep{Chen2023,Wang2023c}.
A large fraction of the emission is extended, as indicated by the $F_{\rm core}/F_{\rm total}$ = 0.4.
The core emission (C1) is self-absorbed at 1.5--5.0~GHz with an inverted slope $\alpha_{1.5-5.0} = 1.01$ \citep{Chen2023}, and becomes optically thin at 8.4--23.6~GHz with a steep slope $\alpha_{8.4-23.6} < -1.70$.
The extended emission (C2) is optically thin at 1--24~GHz with steep slopes $\alpha_{1.5-5.0} = -0.93$ \citep{Chen2023} and $\alpha_{8.4-23.6} < -1.03$.
A factor of $\sim$ 2 variability was observed \citep{Wang2023b}.
The steep VLA 1--15~GHz spectrum suggests the presence of extended emission on larger scales.

\subsubsection{PG\,2304+042}

The object is only detected at 8.4~GHz, showing two components with a separation of $\sim$ 7~mas.
The two components are unresolved at 1.5 and 5.0~GHz \citep{Alhosani2022,Wang2023c}.
A fraction of extended emission is present, as indicated by the $F_{\rm core}/F_{\rm total}$ = 0.6.
The 1.5--5.0~GHz core emission is optically thick with a flat slope $\alpha_{1.5-5.0} = -0.09$ \citep{Chen2023}.
The 8.4--23.6~GHz core emission (C1) becomes optically thin with a steep slope $\alpha_{8.4-23.6} < -0.74$.
The slope of the extended emission (C2) can not be determined due to the upper limit $\alpha_{8.4-23.6} < 0.39$.
The flat VLA 5--45~GHz slope ($-$0.03) suggests that the core emission dominates on larger scales and an additional component may contribute to the emission at 45~GHz.

\subsection{Indeterminate at 8.4--23.6~GHz \label{sec:indeterminate}}

\subsubsection{PG\,0923+129, PG\,1534+580, PG\,1700+518}

The three objects are potentially detected at 8.4~GHz at a 4$\sigma$ level and are not detected at 23.6~GHz.
They all show extended emission at 1.5~GHz, but the extended emission is probably resolved out at 5.0~GHz \citep{Yang2012,Chen2023}.
The core emission at 1.5--5.0~GHz is optically thick with flat slopes $\alpha_{1.5-5.0} = -0.48$ for PG\,0923+129, $\alpha_{1.5-5.0} = -0.36$ for PG\,1534+580, and $\alpha_{1.5-5.0} = -0.28$ for PG\,1700+518 \citep{Chen2023}.
The 8.4--23.6~GHz slopes are upper limits only, with $\alpha_{8.4-23.6} < 0.67$ for PG\,0923+129, $\alpha_{8.4-23.6} < 0.04$ for PG\,1534+580, and $\alpha_{8.4-23.6} < 0.87$ for PG\,1700+518, and thus can not be determined.
The VLA slopes are generally steeper at lower frequencies and become flatter at higher frequencies, suggesting that the extended emission dominates at lower frequencies and the core emission dominates at higher frequencies.
Additionally, the VLA fluxes are approximately an order of magnitude higher than the VLBA fluxes, demonstrating that significant extended emission is present on larger scales.

\subsubsection{PG\,0050+124, PG\,0157+001}

The two objects are not detected at either 8.4 or 23.6~GHz, likely because the emission is optically thin, and thus falls below the detection limits at higher frequencies and/or is resolved out at higher resolutions.
At 1.5 and 5.0~GHz, PG\,0050+124 shows two components with a flat slope ($\alpha_{1.5-5.0} = 0.0$) in one and a steep slope ($\alpha_{1.5-5.0} = -1.21$) in the other, and PG\,0157+001 shows a single component with a steep slope ($\alpha_{1.5-5.0} = -0.90$) \citep{Alhosani2022,Chen2023,Wang2023c}.
The 8.4--23.6~GHz slopes thus can not be determined.
The VLA fluxes are approximately an order of magnitude higher than the VLBA fluxes, suggesting that significant extended emission dominates on larger scales.
Additionally, the steep VLA 5--45~GHz slopes \citep[$-$0.8 for PG\,0050+124 and $-$1.1 for PG\,0157+001,][]{Baldi2022} manifest that the emission is optically thin up to 45~GHz.

\subsection{The $L/L_{\rm Edd}$ distribution}

\begin{figure}[ht!]
\centering
\includegraphics[width=\columnwidth, trim={7cm, 0cm, 9cm, 2cm}, clip]{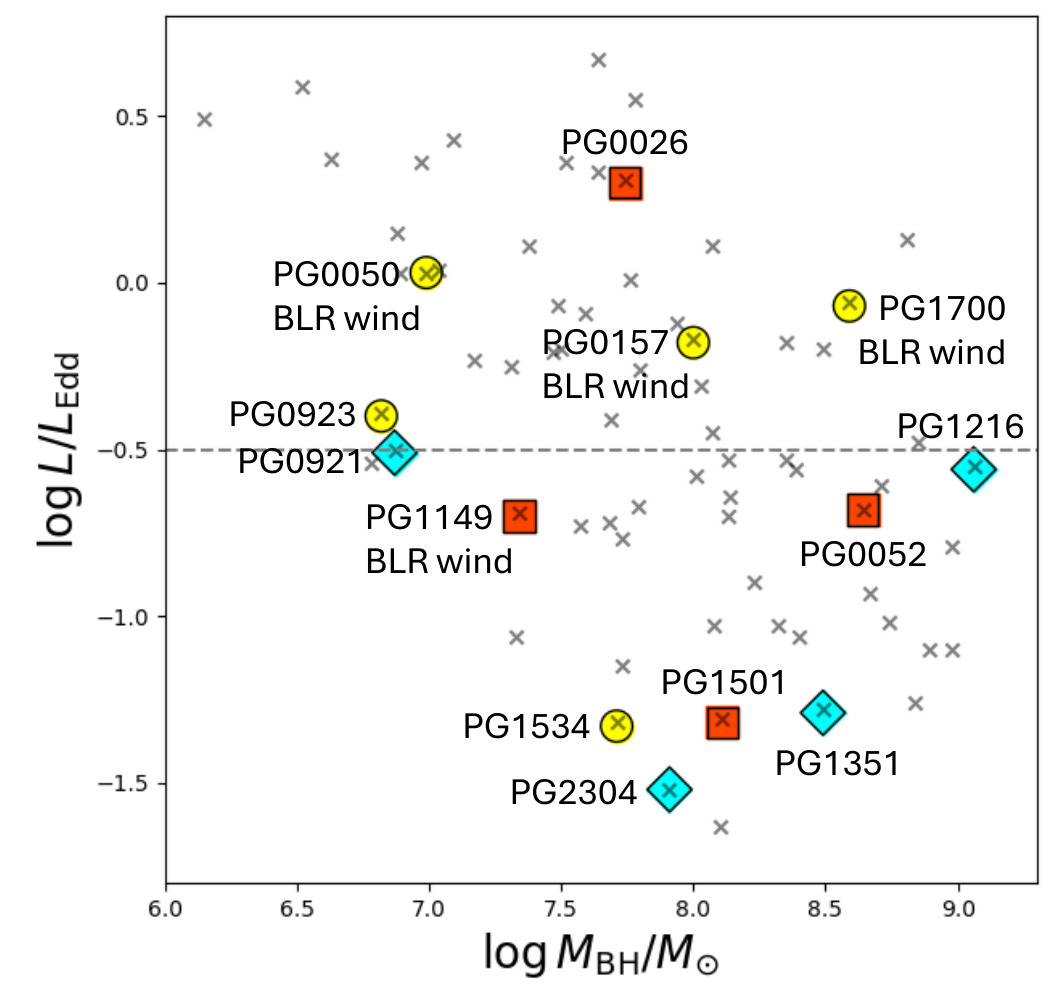}
\caption{The $L/L_{\rm Edd}$ and $M_{\rm BH}$ distributions of the sample.
The 71 $z<0.5$ PG RQQ are marked in grey crosses.
The four flat-flat slope objects are marked in red squares, and the four flat-steep slope objects are marked in cyan diamonds.
Seven of these eight objects have $L/L_{\rm Edd} < 0.3$.
The five objects detected at $\sim 4 \sigma$ or undetected are marked in yellow circles.
Four of these five objects have $L/L_{\rm Edd} > 0.3$.
The short name and the BLR wind if present are indicated nearby the objects.
The grey dashed line marks $L/L_{\rm Edd} = 0.3$ ($\log L/L_{\rm Edd} \simeq -0.5$).}
\label{Edd+Mbh}
\end{figure}

Figure~\ref{Edd+Mbh} presents the $L/L_{\rm Edd}$ and $M_{\rm BH}$ distributions of the 13 objects in the sample, which is consistent with the distribution of the parent sample of 71 $z<0.5$ PG RQQ, making it representative of the optically selected RQ Type 1 AGN.
Interestingly, seven of the eight objects with $\ge 5 \sigma$ detections, i.e., the flat-flat and flat-steep slope objects, have $L/L_{\rm Edd} < 0.3$, while four of the five objects detected at $\sim 4 \sigma$ or undetected, i.e., the indeterminate slope objects, have $L/L_{\rm Edd} > 0.3$.
However, the four flat-flat slope objects and the four flat-steep slope objects do not show differences in their $L/L_{\rm Edd}$ distribution.

In addition, a BLR wind, as indicated by the blue wing or blueshift in the broad C\,IV emission or absorption line, is found in four objects, including PG\,0050+124, PG\,0157+001, and PG\,1149$-$110, where the BLR wind is observed in emission \citep{Baskin2005,Chen2024b}, and PG\,1700+518, where the BLR wind is observed in absorption \citep{Young2007,Runnoe2018}.
These four BLR wind objects reside at $L/L_{\rm Edd} \ge 0.2$, of which three are not detected or only detected at $\sim 4 \sigma$.

The extended radio emission in high $L/L_{\rm Edd}$ objects strongly suggests that it is likely associated with a radiation pressure driven wind \citep{Laor2019,Chen2024b}.
The specific value of $L/L_{\rm Edd} \simeq 0.3$ ($\log L/L_{\rm Edd} \simeq -0.5$), which separates the objects that tend to exhibit winds from those that do not, is the boundary between the geometrically thin and thick accretion disks \citep{Laor1989}.
A transition from a thin to a thick disk may be related to the wind launching.

\section{Discussion}

\subsection{The radio source size}

The primary goal of our study is to construct the 1--24~GHz radio spectra to constrain the inner and outer boundaries of the radio core emission regions $R_{\rm in}$ and $R_{\rm out}$.
The value of $R_{\rm in}$ is estimated by the turnover frequency where the flat slope becomes steep $\nu_0 (R_{\rm in})$, and the value of $R_{\rm out}$ is estimated by the turnover frequency where the flat slope becomes self-absorbed (inverted) $\nu_0 (R_{\rm out})$, using Equation~\ref{eq:radius} as discussed in Section~\ref{sec:sed}.

We assume a spherical synchrotron source with $f_{\rm cover} = 1$, which gives the minimal possible size or radius $R_\nu$ of the sphere.
A conical or disk-like geometry with $f_{\rm cover} < 1$ would imply a larger $R_\nu$.
The BLR size is estimated using Equation~\ref{eq:blr} \citep{Kaspi2005}.
The estimated values or limits on $R_{\rm in}$ and $R_{\rm out}$ of the radio core emission, and their scales to $R_{\rm BLR}$ are presented in Table~\ref{region}.

Overall, the emission typically extends from $< 0.1 R_{\rm BLR}$ in the four flat-flat slope objects, and from $\sim 0.5 R_{\rm BLR}$ in the four flat-steep slope objects, out to at least a few $\times R_{\rm BLR}$ in most of the objects.
This suggests that the emission in the flat-flat slope objects originates from a radius at least a factor of 5 smaller than that in the flat-steep slope objects.

\subsubsection{The four flat-flat slope objects}

The spectra are flat or inverted at 8.4--23.6~GHz in the four flat-flat slope objects discussed in Section~\ref{sec:flat-flat}.
We can set a lower limit on $\nu_0 (R_{\rm in}) > 23.6$~GHz, which results in an upper limit on $R_{\rm in}/R_{\rm BLR} < 0.1$.
In PG\,0026+129, the spectrum is inverted at 1.5--5.0~GHz, which gives $\nu_0 (R_{\rm out})$ = 5.0--8.4~GHz and $R_{\rm out}/R_{\rm BLR}$ = 0.3--0.4.
In PG\,0052+251, PG\,1149$-$110, and PG\,1501+106, the spectra are flat (but not inverted) at 1.5--5.0~GHz, which gives $\nu_0 (R_{\rm out}) < 1.5$~GHz and $R_{\rm out}/R_{\rm BLR} > 1.4-3.2$.
Notably, in PG\,0052+251, a transition from a flat 1.5--5.0~GHz slope to an inverted 8.4--23.6~GHz slope implies that a more compact source produces enhanced emission at higher frequencies.

\subsubsection{The four flat-steep slope objects}

The spectra are steep at 8.4--23.6~GHz in the four flat-steep slope objects discussed in Section~\ref{sec:flat-steep}.
This allows to estimate $\nu_0 (R_{\rm in})$ = 5.0--8.4~GHz and $R_{\rm in}/R_{\rm BLR}$ = 0.6--1.9.
In PG\,1216+069 and PG\,1351+640, the spectra are inverted at 1.5--5.0~GHz, which gives $\nu_0 (R_{\rm out})$ = 5.0--8.4~GHz and $R_{\rm out}/R_{\rm BLR}$ = 0.7--1.9.
The emission likely originates in a source with a small radial extent, as it switches from self-absorbed below 5~GHz to optically thin above 8.4~GHz, which implies $R_{\rm out} \sim R_{\rm in}$.
In PG\,0921+525 and PG\,2304+042, the spectra are flat at 1.5--5.0~GHz, which gives $\nu_0 (R_{\rm out}) < 1.5$~GHz and $R_{\rm out}/R_{\rm BLR} > 2.8-3.4$.

\subsubsection{The five indeterminate slope objects}

The sizes can not be significantly constrained and only limits can be set in the five indeterminate slope objects discussed in Section~\ref{sec:indeterminate}.
In PG\,0923+129, PG\,1534+580, PG\,1700+518, and PG\,0050+124, the 1.5--5.0~GHz slope is flat and the 8.4--23.6~GHz slope is uncertain.
This gives limits on $\nu_0 (R_{\rm out}) < 1.5$~GHz and $\nu_0 (R_{\rm in}) > 5.0$~GHz, which result in $R_{\rm out}/R_{\rm BLR} > 2.2-3.0$ and $R_{\rm in}/R_{\rm BLR} < 0.4-0.8$.
In PG\,0157+001, the steep 1.5--5~GHz slope gives $\nu_0 (R_{\rm in}) < 1.5$~GHz and $R_{\rm in}/R_{\rm BLR} > 6.1$.

\subsection{The origin of the radio emission}

The brightness temperature measured at 8.4~GHz is $\log T_{\rm B} \gtrsim 6.6 - 8.1$, which is about one or two orders of magnitude higher than that is expected for SF and free-free emission \citep[$\ll 10^6$~K,][]{Njeri2024}.
We therefore focus on the accretion disk corona, a low-power jet, and an AGN-driven wind, as the possible mechanisms of the pc-scale radio emission.

\subsubsection{The four flat-flat slope objects}

The accretion disk corona is expected to be extremely compact, on a scale of $\sim$ 10--100 gravitational radii or $\sim 10^{-3}-10^{-2}~R_{\rm BLR}$.
The synchrotron emission from such a compact source is highly self-absorbed in the few GHz regime.
Analytic approximations \citep[Equation~\ref{eq:radius},][]{Laor2008} and radiative transfer solutions \citep{Raginski2016} indicate that the coronal-scale synchrotron emission is expected to become detectable above 100~GHz and peak at the sub-millimeter (sub-mm) regime $\sim$ 300--1000~GHz.
In stars, the coronal activity is associated with the CME, where magnetized plasma blobs escape the system.
If such events occur in the accretion disk corona as well, one would expect a stream of magnetized plasma blobs extending from the corona to larger scales.
This leads to optically thick synchrotron emission with a flat spectrum from a few hundred GHz down to possibly a few GHz, depending on the outer extent of the synchrotron source.
The coronal emission scenario naturally explains the very compact radio source sizes of $R_{\rm in} < 0.08-0.15~R_{\rm BLR}$ derived in the four flat-flat slope objects.

The radio emission is generally compact, with three objects exhibiting a single component and one object showing two components at 8.4~GHz in the VLBA images, which is consistent with a compact coronal source.
In PG\,1149$-$110, the extended emission (C2), which is offset by a projected distance of $\sim$ 2~pc from the core emission (C1), is likely associated with a radiation pressure driven wind, which is possible for $L/L_{\rm Edd} = 0.2$.
Additionally, a BLR wind is observed, as indicated by an excess blue wing in the C\,IV emission line \citep{Baskin2005,Chen2024b}.
The wind associated radio emission is not detected in PG\,0026+129, PG\,0052+251, and PG\,1501+106.
a radiation pressure driven wind may be possible but is not necessary to be present in PG\,0052+251 with $L/L_{\rm Edd} = 0.2$.
Such a wind is less likely in PG\,1501+106 with a low $L/L_{\rm Edd}$ (0.05).
The high $L/L_{\rm Edd}$ (2.0) in PG\,0026+129 is possibly an overestimate due to a nearly face-on view \citep[see discussion in][]{Chen2024b}.

\subsubsection{The four flat-steep slope objects}

The radio core emission in the four flat-steep slope objects starts at $R_{\rm in} \sim 0.6-1.0~R_{\rm BLR}$, which suggests that it originates on a scale comparable in size to the BLR.
In PG\,1216+069 and PG\,1351+640, $R_{\rm out} \sim 1.3-1.9~R_{\rm BLR}$ indicates that the core emission is confined to the BLR scale.
In PG\,0921+525 and PG\,2304+042, the core emission extends further outwards to $R_{\rm out} > 2.8-3.4~R_{\rm BLR}$.
The emission mechanism is therefore likely related to the BLR, possibly an interaction of an AGN-driven outflow (a wind or a jet) with the BLR gas, leading to shocks and the associated radio emission.
Alternatively, a low-power jet or an outflowing CME blob, which extend to the BLR, are also possible mechanisms.

The VLBA images reveal that all the four flat-steep slope objects display extended emission on pc scales.
The projected offsets between C1 and C2 at 8.4~GHz are approximately 1.5~pc in PG\,0921+525, 9.5~pc in PG\,1351+640, and 6.0~pc in PG\,2304+042.
In PG\,1216+069, although only a single component is detected at 8.4~GHz, a slightly extended structure is seen at 5.0~GHz \citep{Chen2023}.

PG\,1216+069 and PG\,1351+640 have a larger $M_{\rm BH} (10^{8.5}-10^{9.1}~M_{\odot})$ which is characteristic of RL AGN \citep{Laor2000,Laor2008}, and a higher $L_{\rm R}/L_{\rm X} (10^{-4})$ which is $\sim$ 100 times higher than the typical $L_{\rm R}/L_{\rm X}$ in our sample \citep{Chen2023}.
The radio to optical flux ratio is in an intermediate (1--10) regime between RL and RQ AGN \citep[4.3 for PG\,1351+640 and 1.65 for PG\,1216+069,][]{Kellermann1989}.
Notably, PG\,1351+640 is found to have a mildly relativistic jet where the C2 is moving away from the C1 at a speed of $\sim 0.5c$ \citep{Wang2023b}.
A radiation pressure driven wind is less likely as suggested by its $L/L_{\rm Edd} = 0.05$.
In contrast, the evidence for a jet in PG\,1216+069 is only circumstantial, and a radiation pressure driven wind is also possible given its $L/L_{\rm Edd} = 0.3$.

PG\,0921+525 and PG\,2304+042 have a smaller $M_{\rm BH} (10^{6.9}-10^{7.9}~M_{\odot})$ and a lower $L_{\rm R}/L_{\rm X} (10^{-6})$, which are characteristic of RQ AGN.
The $L/L_{\rm Edd}$ in PG\,0921+525 and PG\,2304+042 is 0.3 and 0.03, respectively, thus a wind appears more likely in PG\,0921+525 and a jet may be more likely in PG\,2304+042.
Notably, in PG\,2304+042, the flat VLA 5--45~GHz slope implies an additional compact source extending inwards to $R_{\rm in} < 0.1~R_{\rm BLR}$, possibly a new CME blob or a jet.

\subsubsection{The five indeterminate slope objects}

Interestingly, in the five potentially detected or not detected objects, four (PG\,0050+124, PG\,0157+001, PG\,0923+129, and PG\,1700+518) have a high $L/L_{\rm Edd}$ (0.4--1), and three (PG\,0050+124, PG\,0157+001, and PG\,1700+518) also show a BLR wind, specifically the blueshifted C\,IV emission or absorption line.
A high $L/L_{\rm Edd}$ is likely associated with a radiation pressure driven wind \citep[e.g.][]{Baskin2005,Chen2024b}.
Indeed, VLBA or EVN images of these five objects display extended emission at 1.5 and/or 5.0~GHz \citep{Yang2012,Alhosani2022,Chen2023,Wang2023c}.
The extended emission is optically thin with a steep spectrum, and thus is probably too faint or resolved out at 8.4 and 23.6~GHz.
PG\,1534+580 has a low $L/L_{\rm Edd}$ (0.05), making a radiatively driven wind less likely, and leaving a weak jet or a CME blob as possible options.

Physically, a radiation pressure driven wind may originate from the accretion disk, and a strong wind can deplete the coronal gas and thus reduce the coronal radio and X-ray emission.
A high $L/L_{\rm Edd}$ indeed suppresses the X-ray emission, as indicated by a steep optical to X-ray slope $\alpha_{\rm OX}$ \citep[e.g.][]{Vasudevan2007,Lusso2012}.
It is interesting that $L_{\rm R}/L_{\rm X}$ remains $\sim 10^{-6}$ in these high $L/L_{\rm Edd}$ objects, so both the radio and the X-ray emission are similarly suppressed.
This is in contrast with the weak jet RQ AGN (e.g., PG\,1351+640) with $L_{\rm R}/L_{\rm X} \sim 10^{-4}$, which implies that the weak jet enhances the radio emission by a factor of $\sim$ 100, but apparently does not enhance the X-ray emission.
The lack of X-ray enhancement by the jet emission is also generally observed in unbeamed RL AGN \citep{Miller2011}.

\subsection{Future studies}

The unique prediction of the corona scenario is that the synchrotron emission originates on the smallest possible scales.
The coronal scale of a hundred gravitational radii or less, cannot be spatially resolved in RQ AGN due to the low $\sim$ mJy level fluxes, in contrast with the bright $\sim$ Jy level RL AGN, which can be resolved with the Event Horizon Telescope.
However, the emission scales can be inferred from the SED, which may reveal compact optically thick emission up to a few hundred GHz if the synchrotron source extends down to the horizon scale.
Nevertheless, even the most compact coronal-scale emission is expected to become optically thin at $\nu > 500$~GHz \citep{Raginski2016}.
ALMA observations can go up to $\nu > 500$~GHz and measure $R_{\rm in}$ of RQQ down to a scale of a few gravitational radii.
Therefore, the detection of the sub-mm flat-to-steep spectral turnover will provide a robust confirmation to the coronal origin of the radio core emission in RQQ.
Furthermore, the detection of rapid variability at sub-mm bands \citep[e.g.][]{Shablovinskaya2024} will allow to monitor directly the production of relativistic electrons and the ensuing coronal heating close to the BH event horizon.

Repeated VLBA observations will allow to measure or set limits on the proper motion of the spatially resolved extended emission (e.g., PG\,0921+525, PG\,1149$-$110, PG\,1351+640, and PG\,2304+042), and determine if the extended emission is associated with a sub-relativistic jet or a non-relativistic wind.
In addition, the repeated observations will allow to measure the time dependence of $R_{\rm in}$ and $R_{\rm out}$.
For instance, if the core emission is produced by a series of outflowing blobs, we expect that the turnover frequencies $\nu_0 (R_{\rm in})$ and $\nu_0 (R_{\rm out})$ will shift to lower values as $R_{\rm in}$ and $R_{\rm out}$ increase.
This behavior is observed in supernova radio monitoring, where an expanding radio sphere is detected \citep[e.g.][]{Soderberg2006}.

The flat-spectrum optically thick radio emission may originate from the jet base, that is the jet launching region, which may overlap with the accretion disk corona, and the jet may be associated with the CME phenomenon.
Future radio polarization observations can probe the magnetic field structure in the radio-emitting region, helping to distinguish between the low polarization expected for the coronal emission, which is possibly associated with a turbulent magnetic field, and the high polarization observed in the jet emission, which is due to an ordered large-scale magnetic field.
Similarly, the steep-spectrum optically thin emission of a jet versus a wind may be distinguishable based on the polarization levels \citep[e.g.][]{Silpa2022}.
Another possible discrimination between a jet and a wind is the observed emission intensity, that is $T_{\rm B}$.
A spatially resolved jet is likely to be highly confined, and thus it has a higher $T_{\rm B}$ than a wind \citep[as observed in PG\,1351+640][]{Wang2023b,Chen2023}.
In contrast, a wind probably expands laterally to larger scales, and thus its surface brightness drops.

\section{Summary}

In this study, we construct the pc-scale 1--24~GHz radio spectra of 13 PG RQQ based on our earlier VLBA studies at 1.5 and 5.0~GHz \citep{Alhosani2022,Chen2023} and our new VLBA observations at 8.4 and 23.6~GHz.
The radio core emission typically exhibits a flat spectrum at 1.5--5.0~GHz.
In contrast, the spectral behavior at 8.4--23.6~GHz shows a large variety, from inverted (1.18) to steep ($< -1.70$).
The spectral slopes allow to constrain the size of the radio emission regions.
The key findings are summarized as follows.

(1) In four objects, the radio core emission exhibits a flat or inverted spectral slope at both 1.5--5.0~GHz and 8.4--23.6~GHz.
In four other objects, the spectral slope is flat or inverted at 1.5--5.0~GHz but becomes steep at 8.4--23.6~GHz.
In the remaining five objects, which are detected at $< 5 \sigma$ or not detected at 8.4~GHz, the 8.4--23.6~GHz slope cannot be significantly constrained.

(2) The radio source in the four flat-flat slope objects has an inner radius of $R_{\rm in} < 0.1~R_{\rm BLR}$, indicating a continuous radio emission may originating from the accretion disk corona and expanding outwards.
The radio source in the four flat-steep slope objects has $R_{\rm in} \sim 0.5~R_{\rm BLR}$, suggesting that the radio emission may be related to the BLR, possibly an interaction of an AGN-driven wind with the BLR gas, a low-power jet on the BLR scales, or an outflowing CME blob expanding to the BLR.

(3) In the eight objects with a $\ge 5 \sigma$ detection, seven have $L/L_{\rm Edd} < 0.3$, while in the five objects detected at $\sim 4 \sigma$ or not detected, four have $L/L_{\rm Edd} > 0.3$ and three also show a BLR wind.
The 8.4--23.6~GHz radio emission in the high $L/L_{\rm Edd}$ objects may be weaker due to more extended emission arising from a radiation pressure driven wind, compared to that in the lower $L/L_{\rm Edd}$ objects.

Our recent and upcoming observations of the same sample with the VLA A-configuration at 1--15~GHz and the enhanced Multi Element Remotely Linked Interferometer Network (e-MERLIN) at 1.5--5.0~GHz will allow us to study the various radio emission mechanisms in RQ AGN extending out to kpc scales.

\begin{acknowledgments}

We thank the anonymous referee for suggestions leading to the improvement of this work.
S.C. is supported in part at the Technion by a fellowship from the Lady Davis Foundation.
A.L. acknowledges support by the Israel Science Foundation (grant no.1008/18).
The Technion team is partially supported by a grant from the U.S.-Israel Binational Science Foundation (BSF) and the U.S. National Science Foundation (NSF).
The National Radio Astronomy Observatory is a facility of the National Science Foundation operated under cooperative agreement by Associated Universities, Inc.

\end{acknowledgments}

\vspace{5mm}
\facilities{VLBA}

\software{AIPS \citep{Greisen2003}, DIFMAP \citep{Shepherd1994}, astropy \citep{Astropy2022}}

\bibliography{main.bbl}

\begin{thebibliography}{}
\expandafter\ifx\csname natexlab\endcsname\relax\def\natexlab#1{#1}\fi
\providecommand{\url}[1]{\href{#1}{#1}}
\providecommand{\dodoi}[1]{doi:~\href{http://doi.org/#1}{\nolinkurl{#1}}}
\providecommand{\doeprint}[1]{\href{http://ascl.net/#1}{\nolinkurl{http://ascl.net/#1}}}
\providecommand{\doarXiv}[1]{\href{https://arxiv.org/abs/#1}{\nolinkurl{https://arxiv.org/abs/#1}}}

\bibitem[{{Alhosani} {et~al.}(2022){Alhosani}, {Gelfand}, {Zaw}, {Laor},
  {Behar}, {Chen}, \& {Wrzosek}}]{Alhosani2022}
{Alhosani}, A., {Gelfand}, J.~D., {Zaw}, I., {et~al.} 2022, \apj, 936, 73,
  \dodoi{10.3847/1538-4357/ac8665}

\bibitem[{{Astropy Collaboration} {et~al.}(2022){Astropy Collaboration},
  {Price-Whelan}, {Lim}, {Earl}, {Starkman}, {Bradley}, {Shupe}, {Patil},
  {Corrales}, {Brasseur}, {N{\"o}the}, {Donath}, {Tollerud}, {Morris},
  {Ginsburg}, {Vaher}, {Weaver}, {Tocknell}, {Jamieson}, {van Kerkwijk},
  {Robitaille}, {Merry}, {Bachetti}, {G{\"u}nther}, {Aldcroft},
  {Alvarado-Montes}, {Archibald}, {B{\'o}di}, {Bapat}, {Barentsen},
  {Baz{\'a}n}, {Biswas}, {Boquien}, {Burke}, {Cara}, {Cara}, {Conroy},
  {Conseil}, {Craig}, {Cross}, {Cruz}, {D'Eugenio}, {Dencheva}, {Devillepoix},
  {Dietrich}, {Eigenbrot}, {Erben}, {Ferreira}, {Foreman-Mackey}, {Fox},
  {Freij}, {Garg}, {Geda}, {Glattly}, {Gondhalekar}, {Gordon}, {Grant},
  {Greenfield}, {Groener}, {Guest}, {Gurovich}, {Handberg}, {Hart},
  {Hatfield-Dodds}, {Homeier}, {Hosseinzadeh}, {Jenness}, {Jones}, {Joseph},
  {Kalmbach}, {Karamehmetoglu}, {Ka{\l}uszy{\'n}ski}, {Kelley}, {Kern},
  {Kerzendorf}, {Koch}, {Kulumani}, {Lee}, {Ly}, {Ma}, {MacBride}, {Maljaars},
  {Muna}, {Murphy}, {Norman}, {O'Steen}, {Oman}, {Pacifici}, {Pascual},
  {Pascual-Granado}, {Patil}, {Perren}, {Pickering}, {Rastogi}, {Roulston},
  {Ryan}, {Rykoff}, {Sabater}, {Sakurikar}, {Salgado}, {Sanghi}, {Saunders},
  {Savchenko}, {Schwardt}, {Seifert-Eckert}, {Shih}, {Jain}, {Shukla}, {Sick},
  {Simpson}, {Singanamalla}, {Singer}, {Singhal}, {Sinha}, {Sip{\H{o}}cz},
  {Spitler}, {Stansby}, {Streicher}, {{\v{S}}umak}, {Swinbank}, {Taranu},
  {Tewary}, {Tremblay}, {de Val-Borro}, {Van Kooten}, {Vasovi{\'c}}, {Verma},
  {de Miranda Cardoso}, {Williams}, {Wilson}, {Winkel}, {Wood-Vasey}, {Xue},
  {Yoachim}, {Zhang}, {Zonca}, \& {Astropy Project Contributors}}]{Astropy2022}
{Astropy Collaboration}, {Price-Whelan}, A.~M., {Lim}, P.~L., {et~al.} 2022,
  \apj, 935, 167, \dodoi{10.3847/1538-4357/ac7c74}

\bibitem[{{Baldi} {et~al.}(2022){Baldi}, {Laor}, {Behar}, {Horesh}, {Panessa},
  {McHardy}, \& {Kimball}}]{Baldi2022}
{Baldi}, R.~D., {Laor}, A., {Behar}, E., {et~al.} 2022, \mnras, 510, 1043,
  \dodoi{10.1093/mnras/stab3445}

\bibitem[{{Barvainis} {et~al.}(2005){Barvainis}, {Leh{\'a}r}, {Birkinshaw},
  {Falcke}, \& {Blundell}}]{Barvainis2005}
{Barvainis}, R., {Leh{\'a}r}, J., {Birkinshaw}, M., {Falcke}, H., \&
  {Blundell}, K.~M. 2005, \apj, 618, 108, \dodoi{10.1086/425859}

\bibitem[{{Barvainis} {et~al.}(1996){Barvainis}, {Lonsdale}, \&
  {Antonucci}}]{Barvainis1996}
{Barvainis}, R., {Lonsdale}, C., \& {Antonucci}, R. 1996, \aj, 111, 1431,
  \dodoi{10.1086/117888}

\bibitem[{{Baskin} \& {Laor}(2005)}]{Baskin2005}
{Baskin}, A., \& {Laor}, A. 2005, \mnras, 356, 1029,
  \dodoi{10.1111/j.1365-2966.2004.08525.x}

\bibitem[{{Baskin} \& {Laor}(2021)}]{Baskin2021}
---. 2021, \mnras, 508, 680, \dodoi{10.1093/mnras/stab2555}

\bibitem[{{Berriman} {et~al.}(1990){Berriman}, {Schmidt}, {West}, \&
  {Stockman}}]{Berriman1990}
{Berriman}, G., {Schmidt}, G.~D., {West}, S.~C., \& {Stockman}, H.~S. 1990,
  \apjs, 74, 869, \dodoi{10.1086/191523}

\bibitem[{{Berton} {et~al.}(2018){Berton}, {Congiu}, {J{\"a}rvel{\"a}},
  {Antonucci}, {Kharb}, {Lister}, {Tarchi}, {Caccianiga}, {Chen}, {Foschini},
  {L{\"a}hteenm{\"a}ki}, {Richards}, {Ciroi}, {Cracco}, {Frezzato}, {La Mura},
  \& {Rafanelli}}]{Berton2018}
{Berton}, M., {Congiu}, E., {J{\"a}rvel{\"a}}, E., {et~al.} 2018, \aap, 614,
  A87, \dodoi{10.1051/0004-6361/201832612}

\bibitem[{{Blandford} \& {K{\"o}nigl}(1979)}]{Blandford1979}
{Blandford}, R.~D., \& {K{\"o}nigl}, A. 1979, \apj, 232, 34,
  \dodoi{10.1086/157262}

\bibitem[{{Boroson} \& {Green}(1992)}]{Boroson1992}
{Boroson}, T.~A., \& {Green}, R.~F. 1992, \apjs, 80, 109,
  \dodoi{10.1086/191661}

\bibitem[{{Brandt} {et~al.}(2000){Brandt}, {Laor}, \& {Wills}}]{Brandt2000}
{Brandt}, W.~N., {Laor}, A., \& {Wills}, B.~J. 2000, \apj, 528, 637,
  \dodoi{10.1086/308207}

\bibitem[{{Chen} {et~al.}(2022){Chen}, {Laor}, \& {Behar}}]{Chen2022b}
{Chen}, S., {Laor}, A., \& {Behar}, E. 2022, \mnras, 515, 1723,
  \dodoi{10.1093/mnras/stac1891}

\bibitem[{{Chen} {et~al.}(2023){Chen}, {Laor}, {Behar}, {Baldi}, \&
  {Gelfand}}]{Chen2023}
{Chen}, S., {Laor}, A., {Behar}, E., {Baldi}, R.~D., \& {Gelfand}, J.~D. 2023,
  \mnras, 525, 164, \dodoi{10.1093/mnras/stad2289}

\bibitem[{{Chen} {et~al.}(2024){Chen}, {Laor}, {Behar}, {Baldi}, {Gelfand},
  {Kimball}, {McHardy}, {Orosz}, \& {Paragi}}]{Chen2024b}
{Chen}, S., {Laor}, A., {Behar}, E., {et~al.} 2024, \apj, 975, 35,
  \dodoi{10.3847/1538-4357/ad74fc}

\bibitem[{{Condon} {et~al.}(2013){Condon}, {Kellermann}, {Kimball},
  {Ivezi{\'c}}, \& {Perley}}]{Condon2013}
{Condon}, J.~J., {Kellermann}, K.~I., {Kimball}, A.~E., {Ivezi{\'c}}, {\v{Z}}.,
  \& {Perley}, R.~A. 2013, \apj, 768, 37, \dodoi{10.1088/0004-637X/768/1/37}

\bibitem[{{Davis} \& {Laor}(2011)}]{Davis2011}
{Davis}, S.~W., \& {Laor}, A. 2011, \apj, 728, 98,
  \dodoi{10.1088/0004-637X/728/2/98}

\bibitem[{{Fischer} {et~al.}(2021){Fischer}, {Secrest}, {Johnson}, {Dorland},
  {Cigan}, {Fernandez}, {Hunt}, {Koss}, {Schmitt}, \&
  {Zacharias}}]{Fischer2021}
{Fischer}, T.~C., {Secrest}, N.~J., {Johnson}, M.~C., {et~al.} 2021, \apj, 906,
  88, \dodoi{10.3847/1538-4357/abca3c}

\bibitem[{{Gaia Collaboration} {et~al.}(2016){Gaia Collaboration}, {Prusti},
  {de Bruijne}, {Brown}, {Vallenari}, {Babusiaux}, {Bailer-Jones}, {Bastian},
  {Biermann}, {Evans}, {Eyer}, {Jansen}, {Jordi}, {Klioner}, {Lammers},
  {Lindegren}, {Luri}, {Mignard}, {Milligan}, {Panem}, {Poinsignon},
  {Pourbaix}, {Randich}, {Sarri}, {Sartoretti}, {Siddiqui}, {Soubiran},
  {Valette}, {van Leeuwen}, {Walton}, {Aerts}, {Arenou}, {Cropper}, {Drimmel},
  {H{\o}g}, {Katz}, {Lattanzi}, {O'Mullane}, {Grebel}, {Holland}, {Huc},
  {Passot}, {Bramante}, {Cacciari}, {Casta{\~n}eda}, {Chaoul}, {Cheek}, {De
  Angeli}, {Fabricius}, {Guerra}, {Hern{\'a}ndez}, {Jean-Antoine-Piccolo},
  {Masana}, {Messineo}, {Mowlavi}, {Nienartowicz}, {Ord{\'o}{\~n}ez-Blanco},
  {Panuzzo}, {Portell}, {Richards}, {Riello}, {Seabroke}, {Tanga},
  {Th{\'e}venin}, {Torra}, {Els}, {Gracia-Abril}, {Comoretto},
  {Garcia-Reinaldos}, {Lock}, {Mercier}, {Altmann}, {Andrae}, {Astraatmadja},
  {Bellas-Velidis}, {Benson}, {Berthier}, {Blomme}, {Busso}, {Carry},
  {Cellino}, {Clementini}, {Cowell}, {Creevey}, {Cuypers}, {Davidson}, {De
  Ridder}, {de Torres}, {Delchambre}, {Dell'Oro}, {Ducourant}, {Fr{\'e}mat},
  {Garc{\'\i}a-Torres}, {Gosset}, {Halbwachs}, {Hambly}, {Harrison}, {Hauser},
  {Hestroffer}, {Hodgkin}, {Huckle}, {Hutton}, {Jasniewicz}, {Jordan},
  {Kontizas}, {Korn}, {Lanzafame}, {Manteiga}, {Moitinho}, {Muinonen},
  {Osinde}, {Pancino}, {Pauwels}, {Petit}, {Recio-Blanco}, {Robin}, {Sarro},
  {Siopis}, {Smith}, {Smith}, {Sozzetti}, {Thuillot}, {van Reeven}, {Viala},
  {Abbas}, {Abreu Aramburu}, {Accart}, {Aguado}, {Allan}, {Allasia},
  {Altavilla}, {{\'A}lvarez}, {Alves}, {Anderson}, {Andrei}, {Anglada Varela},
  {Antiche}, {Antoja}, {Ant{\'o}n}, {Arcay}, {Atzei}, {Ayache}, {Bach},
  {Baker}, {Balaguer-N{\'u}{\~n}ez}, {Barache}, {Barata}, {Barbier}, {Barblan},
  {Baroni}, {Barrado y Navascu{\'e}s}, {Barros}, {Barstow}, {Becciani},
  {Bellazzini}, {Bellei}, {Bello Garc{\'\i}a}, {Belokurov}, {Bendjoya},
  {Berihuete}, {Bianchi}, {Bienaym{\'e}}, {Billebaud}, {Blagorodnova},
  {Blanco-Cuaresma}, {Boch}, {Bombrun}, {Borrachero}, {Bouquillon}, {Bourda},
  {Bouy}, {Bragaglia}, {Breddels}, {Brouillet}, {Br{\"u}semeister},
  {Bucciarelli}, {Budnik}, {Burgess}, {Burgon}, {Burlacu}, {Busonero}, {Buzzi},
  {Caffau}, {Cambras}, {Campbell}, {Cancelliere}, {Cantat-Gaudin}, {Carlucci},
  {Carrasco}, {Castellani}, {Charlot}, {Charnas}, {Charvet}, {Chassat},
  {Chiavassa}, {Clotet}, {Cocozza}, {Collins}, {Collins}, {Costigan}, {Crifo},
  {Cross}, {Crosta}, {Crowley}, {Dafonte}, {Damerdji}, {Dapergolas}, {David},
  {David}, {De Cat}, {de Felice}, {de Laverny}, {De Luise}, {De March}, {de
  Martino}, {de Souza}, {Debosscher}, {del Pozo}, {Delbo}, {Delgado},
  {Delgado}, {di Marco}, {Di Matteo}, {Diakite}, {Distefano}, {Dolding}, {Dos
  Anjos}, {Drazinos}, {Dur{\'a}n}, {Dzigan}, {Ecale}, {Edvardsson}, {Enke},
  {Erdmann}, {Escolar}, {Espina}, {Evans}, {Eynard Bontemps}, {Fabre},
  {Fabrizio}, {Faigler}, {Falc{\~a}o}, {Farr{\`a}s Casas}, {Faye}, {Federici},
  {Fedorets}, {Fern{\'a}ndez-Hern{\'a}ndez}, {Fernique}, {Fienga}, {Figueras},
  {Filippi}, {Findeisen}, {Fonti}, {Fouesneau}, {Fraile}, {Fraser}, {Fuchs},
  {Furnell}, {Gai}, {Galleti}, {Galluccio}, {Garabato}, {Garc{\'\i}a-Sedano},
  {Gar{\'e}}, {Garofalo}, {Garralda}, {Gavras}, {Gerssen}, {Geyer}, {Gilmore},
  {Girona}, {Giuffrida}, {Gomes}, {Gonz{\'a}lez-Marcos},
  {Gonz{\'a}lez-N{\'u}{\~n}ez}, {Gonz{\'a}lez-Vidal}, {Granvik}, {Guerrier},
  {Guillout}, {Guiraud}, {G{\'u}rpide}, {Guti{\'e}rrez-S{\'a}nchez}, {Guy},
  {Haigron}, {Hatzidimitriou}, {Haywood}, {Heiter}, {Helmi}, {Hobbs},
  {Hofmann}, {Holl}, {Holland}, {Hunt}, {Hypki}, {Icardi}, {Irwin}, {Jevardat
  de Fombelle}, {Jofr{\'e}}, {Jonker}, {Jorissen}, {Julbe}, {Karampelas},
  {Kochoska}, {Kohley}, {Kolenberg}, {Kontizas}, {Koposov}, {Kordopatis},
  {Koubsky}, {Kowalczyk}, {Krone-Martins}, {Kudryashova}, {Kull}, {Bachchan},
  {Lacoste-Seris}, {Lanza}, {Lavigne}, {Le Poncin-Lafitte}, {Lebreton},
  {Lebzelter}, {Leccia}, {Leclerc}, {Lecoeur-Taibi}, {Lemaitre}, {Lenhardt},
  {Leroux}, {Liao}, {Licata}, {Lindstr{\o}m}, {Lister}, {Livanou}, {Lobel},
  {L{\"o}ffler}, {L{\'o}pez}, {Lopez-Lozano}, {Lorenz}, {Loureiro},
  {MacDonald}, {Magalh{\~a}es Fernandes}, {Managau}, {Mann}, {Mantelet},
  {Marchal}, {Marchant}, {Marconi}, {Marie}, {Marinoni}, {Marrese},
  {Marschalk{\'o}}, {Marshall}, {Mart{\'\i}n-Fleitas}, {Martino}, {Mary},
  {Matijevi{\v{c}}}, {Mazeh}, {McMillan}, {Messina}, {Mestre}, {Michalik},
  {Millar}, {Miranda}, {Molina}, {Molinaro}, {Molinaro}, {Moln{\'a}r},
  {Moniez}, {Montegriffo}, {Monteiro}, {Mor}, {Mora}, {Morbidelli}, {Morel},
  {Morgenthaler}, {Morley}, {Morris}, {Mulone}, {Muraveva}, {Musella},
  {Narbonne}, {Nelemans}, {Nicastro}, {Noval}, {Ord{\'e}novic},
  {Ordieres-Mer{\'e}}, {Osborne}, {Pagani}, {Pagano}, {Pailler}, {Palacin},
  {Palaversa}, {Parsons}, {Paulsen}, {Pecoraro}, {Pedrosa}, {Pentik{\"a}inen},
  {Pereira}, {Pichon}, {Piersimoni}, {Pineau}, {Plachy}, {Plum}, {Poujoulet},
  {Pr{\v{s}}a}, {Pulone}, {Ragaini}, {Rago}, {Rambaux}, {Ramos-Lerate},
  {Ranalli}, {Rauw}, {Read}, {Regibo}, {Renk}, {Reyl{\'e}}, {Ribeiro},
  {Rimoldini}, {Ripepi}, {Riva}, {Rixon}, {Roelens}, {Romero-G{\'o}mez},
  {Rowell}, {Royer}, {Rudolph}, {Ruiz-Dern}, {Sadowski}, {Sagrist{\`a}
  Sell{\'e}s}, {Sahlmann}, {Salgado}, {Salguero}, {Sarasso}, {Savietto},
  {Schnorhk}, {Schultheis}, {Sciacca}, {Segol}, {Segovia}, {Segransan},
  {Serpell}, {Shih}, {Smareglia}, {Smart}, {Smith}, {Solano}, {Solitro},
  {Sordo}, {Soria Nieto}, {Souchay}, {Spagna}, {Spoto}, {Stampa}, {Steele},
  {Steidelm{\"u}ller}, {Stephenson}, {Stoev}, {Suess}, {S{\"u}veges}, {Surdej},
  {Szabados}, {Szegedi-Elek}, {Tapiador}, {Taris}, {Tauran}, {Taylor},
  {Teixeira}, {Terrett}, {Tingley}, {Trager}, {Turon}, {Ulla}, {Utrilla},
  {Valentini}, {van Elteren}, {Van Hemelryck}, {van Leeuwen}, {Varadi},
  {Vecchiato}, {Veljanoski}, {Via}, {Vicente}, {Vogt}, {Voss}, {Votruba},
  {Voutsinas}, {Walmsley}, {Weiler}, {Weingrill}, {Werner}, {Wevers},
  {Whitehead}, {Wyrzykowski}, {Yoldas}, {{\v{Z}}erjal}, {Zucker}, {Zurbach},
  {Zwitter}, {Alecu}, {Allen}, {Allende Prieto}, {Amorim},
  {Anglada-Escud{\'e}}, {Arsenijevic}, {Azaz}, {Balm}, {Beck}, {Bernstein},
  {Bigot}, {Bijaoui}, {Blasco}, {Bonfigli}, {Bono}, {Boudreault}, {Bressan},
  {Brown}, {Brunet}, {Bunclark}, {Buonanno}, {Butkevich}, {Carret}, {Carrion},
  {Chemin}, {Ch{\'e}reau}, {Corcione}, {Darmigny}, {de Boer}, {de Teodoro}, {de
  Zeeuw}, {Delle Luche}, {Domingues}, {Dubath}, {Fodor}, {Fr{\'e}zouls},
  {Fries}, {Fustes}, {Fyfe}, {Gallardo}, {Gallegos}, {Gardiol}, {Gebran},
  {Gomboc}, {G{\'o}mez}, {Grux}, {Gueguen}, {Heyrovsky}, {Hoar}, {Iannicola},
  {Isasi Parache}, {Janotto}, {Joliet}, {Jonckheere}, {Keil}, {Kim},
  {Klagyivik}, {Klar}, {Knude}, {Kochukhov}, {Kolka}, {Kos}, {Kutka}, {Lainey},
  {LeBouquin}, {Liu}, {Loreggia}, {Makarov}, {Marseille}, {Martayan},
  {Martinez-Rubi}, {Massart}, {Meynadier}, {Mignot}, {Munari}, {Nguyen},
  {Nordlander}, {Ocvirk}, {O'Flaherty}, {Olias Sanz}, {Ortiz}, {Osorio},
  {Oszkiewicz}, {Ouzounis}, {Palmer}, {Park}, {Pasquato}, {Peltzer}, {Peralta},
  {P{\'e}turaud}, {Pieniluoma}, {Pigozzi}, {Poels}, {Prat}, {Prod'homme},
  {Raison}, {Rebordao}, {Risquez}, {Rocca-Volmerange}, {Rosen}, {Ruiz-Fuertes},
  {Russo}, {Sembay}, {Serraller Vizcaino}, {Short}, {Siebert}, {Silva},
  {Sinachopoulos}, {Slezak}, {Soffel}, {Sosnowska}, {Strai{\v{z}}ys}, {ter
  Linden}, {Terrell}, {Theil}, {Tiede}, {Troisi}, {Tsalmantza}, {Tur},
  {Vaccari}, {Vachier}, {Valles}, {Van Hamme}, {Veltz}, {Virtanen}, {Wallut},
  {Wichmann}, {Wilkinson}, {Ziaeepour}, \& {Zschocke}}]{Gaia2016}
{Gaia Collaboration}, {Prusti}, T., {de Bruijne}, J.~H.~J., {et~al.} 2016,
  \aap, 595, A1, \dodoi{10.1051/0004-6361/201629272}

\bibitem[{{Gaia Collaboration} {et~al.}(2023){Gaia Collaboration}, {Vallenari},
  {Brown}, {Prusti}, {de Bruijne}, {Arenou}, {Babusiaux}, {Biermann},
  {Creevey}, {Ducourant}, {Evans}, {Eyer}, {Guerra}, {Hutton}, {Jordi},
  {Klioner}, {Lammers}, {Lindegren}, {Luri}, {Mignard}, {Panem}, {Pourbaix},
  {Randich}, {Sartoretti}, {Soubiran}, {Tanga}, {Walton}, {Bailer-Jones},
  {Bastian}, {Drimmel}, {Jansen}, {Katz}, {Lattanzi}, {van Leeuwen}, {Bakker},
  {Cacciari}, {Casta{\~n}eda}, {De Angeli}, {Fabricius}, {Fouesneau},
  {Fr{\'e}mat}, {Galluccio}, {Guerrier}, {Heiter}, {Masana}, {Messineo},
  {Mowlavi}, {Nicolas}, {Nienartowicz}, {Pailler}, {Panuzzo}, {Riclet}, {Roux},
  {Seabroke}, {Sordo}, {Th{\'e}venin}, {Gracia-Abril}, {Portell}, {Teyssier},
  {Altmann}, {Andrae}, {Audard}, {Bellas-Velidis}, {Benson}, {Berthier},
  {Blomme}, {Burgess}, {Busonero}, {Busso}, {C{\'a}novas}, {Carry}, {Cellino},
  {Cheek}, {Clementini}, {Damerdji}, {Davidson}, {de Teodoro}, {Nu{\~n}ez
  Campos}, {Delchambre}, {Dell'Oro}, {Esquej}, {Fern{\'a}ndez-Hern{\'a}ndez},
  {Fraile}, {Garabato}, {Garc{\'\i}a-Lario}, {Gosset}, {Haigron}, {Halbwachs},
  {Hambly}, {Harrison}, {Hern{\'a}ndez}, {Hestroffer}, {Hodgkin}, {Holl},
  {Jan{\ss}en}, {Jevardat de Fombelle}, {Jordan}, {Krone-Martins}, {Lanzafame},
  {L{\"o}ffler}, {Marchal}, {Marrese}, {Moitinho}, {Muinonen}, {Osborne},
  {Pancino}, {Pauwels}, {Recio-Blanco}, {Reyl{\'e}}, {Riello}, {Rimoldini},
  {Roegiers}, {Rybizki}, {Sarro}, {Siopis}, {Smith}, {Sozzetti}, {Utrilla},
  {van Leeuwen}, {Abbas}, {{\'A}brah{\'a}m}, {Abreu Aramburu}, {Aerts},
  {Aguado}, {Ajaj}, {Aldea-Montero}, {Altavilla}, {{\'A}lvarez}, {Alves},
  {Anders}, {Anderson}, {Anglada Varela}, {Antoja}, {Baines}, {Baker},
  {Balaguer-N{\'u}{\~n}ez}, {Balbinot}, {Balog}, {Barache}, {Barbato},
  {Barros}, {Barstow}, {Bartolom{\'e}}, {Bassilana}, {Bauchet}, {Becciani},
  {Bellazzini}, {Berihuete}, {Bernet}, {Bertone}, {Bianchi}, {Binnenfeld},
  {Blanco-Cuaresma}, {Blazere}, {Boch}, {Bombrun}, {Bossini}, {Bouquillon},
  {Bragaglia}, {Bramante}, {Breedt}, {Bressan}, {Brouillet}, {Brugaletta},
  {Bucciarelli}, {Burlacu}, {Butkevich}, {Buzzi}, {Caffau}, {Cancelliere},
  {Cantat-Gaudin}, {Carballo}, {Carlucci}, {Carnerero}, {Carrasco},
  {Casamiquela}, {Castellani}, {Castro-Ginard}, {Chaoul}, {Charlot}, {Chemin},
  {Chiaramida}, {Chiavassa}, {Chornay}, {Comoretto}, {Contursi}, {Cooper},
  {Cornez}, {Cowell}, {Crifo}, {Cropper}, {Crosta}, {Crowley}, {Dafonte},
  {Dapergolas}, {David}, {David}, {de Laverny}, {De Luise}, {De March}, {De
  Ridder}, {de Souza}, {de Torres}, {del Peloso}, {del Pozo}, {Delbo},
  {Delgado}, {Delisle}, {Demouchy}, {Dharmawardena}, {Di Matteo}, {Diakite},
  {Diener}, {Distefano}, {Dolding}, {Edvardsson}, {Enke}, {Fabre}, {Fabrizio},
  {Faigler}, {Fedorets}, {Fernique}, {Fienga}, {Figueras}, {Fournier},
  {Fouron}, {Fragkoudi}, {Gai}, {Garcia-Gutierrez}, {Garcia-Reinaldos},
  {Garc{\'\i}a-Torres}, {Garofalo}, {Gavel}, {Gavras}, {Gerlach}, {Geyer},
  {Giacobbe}, {Gilmore}, {Girona}, {Giuffrida}, {Gomel}, {Gomez},
  {Gonz{\'a}lez-N{\'u}{\~n}ez}, {Gonz{\'a}lez-Santamar{\'\i}a},
  {Gonz{\'a}lez-Vidal}, {Granvik}, {Guillout}, {Guiraud},
  {Guti{\'e}rrez-S{\'a}nchez}, {Guy}, {Hatzidimitriou}, {Hauser}, {Haywood},
  {Helmer}, {Helmi}, {Sarmiento}, {Hidalgo}, {Hilger}, {H{\l}adczuk}, {Hobbs},
  {Holland}, {Huckle}, {Jardine}, {Jasniewicz}, {Jean-Antoine Piccolo},
  {Jim{\'e}nez-Arranz}, {Jorissen}, {Juaristi Campillo}, {Julbe}, {Karbevska},
  {Kervella}, {Khanna}, {Kontizas}, {Kordopatis}, {Korn}, {K{\'o}sp{\'a}l},
  {Kostrzewa-Rutkowska}, {Kruszy{\'n}ska}, {Kun}, {Laizeau}, {Lambert},
  {Lanza}, {Lasne}, {Le Campion}, {Lebreton}, {Lebzelter}, {Leccia}, {Leclerc},
  {Lecoeur-Taibi}, {Liao}, {Licata}, {Lindstr{\o}m}, {Lister}, {Livanou},
  {Lobel}, {Lorca}, {Loup}, {Madrero Pardo}, {Magdaleno Romeo}, {Managau},
  {Mann}, {Manteiga}, {Marchant}, {Marconi}, {Marcos}, {Marcos Santos},
  {Mar{\'\i}n Pina}, {Marinoni}, {Marocco}, {Marshall}, {Martin Polo},
  {Mart{\'\i}n-Fleitas}, {Marton}, {Mary}, {Masip}, {Massari},
  {Mastrobuono-Battisti}, {Mazeh}, {McMillan}, {Messina}, {Michalik}, {Millar},
  {Mints}, {Molina}, {Molinaro}, {Moln{\'a}r}, {Monari}, {Mongui{\'o}},
  {Montegriffo}, {Montero}, {Mor}, {Mora}, {Morbidelli}, {Morel}, {Morris},
  {Muraveva}, {Murphy}, {Musella}, {Nagy}, {Noval}, {Oca{\~n}a}, {Ogden},
  {Ordenovic}, {Osinde}, {Pagani}, {Pagano}, {Palaversa}, {Palicio},
  {Pallas-Quintela}, {Panahi}, {Payne-Wardenaar}, {Pe{\~n}alosa Esteller},
  {Penttil{\"a}}, {Pichon}, {Piersimoni}, {Pineau}, {Plachy}, {Plum}, {Poggio},
  {Pr{\v{s}}a}, {Pulone}, {Racero}, {Ragaini}, {Rainer}, {Raiteri}, {Rambaux},
  {Ramos}, {Ramos-Lerate}, {Re Fiorentin}, {Regibo}, {Richards}, {Rios Diaz},
  {Ripepi}, {Riva}, {Rix}, {Rixon}, {Robichon}, {Robin}, {Robin}, {Roelens},
  {Rogues}, {Rohrbasser}, {Romero-G{\'o}mez}, {Rowell}, {Royer}, {Ruz Mieres},
  {Rybicki}, {Sadowski}, {S{\'a}ez N{\'u}{\~n}ez}, {Sagrist{\`a} Sell{\'e}s},
  {Sahlmann}, {Salguero}, {Samaras}, {Sanchez Gimenez}, {Sanna},
  {Santove{\~n}a}, {Sarasso}, {Schultheis}, {Sciacca}, {Segol}, {Segovia},
  {S{\'e}gransan}, {Semeux}, {Shahaf}, {Siddiqui}, {Siebert}, {Siltala},
  {Silvelo}, {Slezak}, {Slezak}, {Smart}, {Snaith}, {Solano}, {Solitro},
  {Souami}, {Souchay}, {Spagna}, {Spina}, {Spoto}, {Steele},
  {Steidelm{\"u}ller}, {Stephenson}, {S{\"u}veges}, {Surdej}, {Szabados},
  {Szegedi-Elek}, {Taris}, {Taylor}, {Teixeira}, {Tolomei}, {Tonello}, {Torra},
  {Torra}, {Torralba Elipe}, {Trabucchi}, {Tsounis}, {Turon}, {Ulla}, {Unger},
  {Vaillant}, {van Dillen}, {van Reeven}, {Vanel}, {Vecchiato}, {Viala},
  {Vicente}, {Voutsinas}, {Weiler}, {Wevers}, {Wyrzykowski}, {Yoldas}, {Yvard},
  {Zhao}, {Zorec}, {Zucker}, \& {Zwitter}}]{Gaia2023}
{Gaia Collaboration}, {Vallenari}, A., {Brown}, A.~G.~A., {et~al.} 2023, \aap,
  674, A1, \dodoi{10.1051/0004-6361/202243940}

\bibitem[{{Greisen}(2003)}]{Greisen2003}
{Greisen}, E.~W. 2003, in Astrophysics and Space Science Library, Vol. 285,
  Information Handling in Astronomy - Historical Vistas, ed. A.~{Heck}, 109,
  \dodoi{10.1007/0-306-48080-8_7}

\bibitem[{{Haas} {et~al.}(2003){Haas}, {Klaas}, {M{\"u}ller}, {Bertoldi},
  {Camenzind}, {Chini}, {Krause}, {Lemke}, {Meisenheimer}, {Richards}, \&
  {Wilkes}}]{Haas2003}
{Haas}, M., {Klaas}, U., {M{\"u}ller}, S.~A.~H., {et~al.} 2003, \aap, 402, 87,
  \dodoi{10.1051/0004-6361:20030110}

\bibitem[{{Helfand} {et~al.}(2015){Helfand}, {White}, \&
  {Becker}}]{Helfand2015}
{Helfand}, D.~J., {White}, R.~L., \& {Becker}, R.~H. 2015, \apj, 801, 26,
  \dodoi{10.1088/0004-637X/801/1/26}

\bibitem[{{Kaspi} {et~al.}(2005){Kaspi}, {Maoz}, {Netzer}, {Peterson},
  {Vestergaard}, \& {Jannuzi}}]{Kaspi2005}
{Kaspi}, S., {Maoz}, D., {Netzer}, H., {et~al.} 2005, \apj, 629, 61,
  \dodoi{10.1086/431275}

\bibitem[{{Kellermann} {et~al.}(1989){Kellermann}, {Sramek}, {Schmidt},
  {Shaffer}, \& {Green}}]{Kellermann1989}
{Kellermann}, K.~I., {Sramek}, R., {Schmidt}, M., {Shaffer}, D.~B., \& {Green},
  R. 1989, \aj, 98, 1195, \dodoi{10.1086/115207}

\bibitem[{{Kellermann} {et~al.}(1994){Kellermann}, {Sramek}, {Schmidt},
  {Green}, \& {Shaffer}}]{Kellermann1994}
{Kellermann}, K.~I., {Sramek}, R.~A., {Schmidt}, M., {Green}, R.~F., \&
  {Shaffer}, D.~B. 1994, \aj, 108, 1163, \dodoi{10.1086/117145}

\bibitem[{{Kimball} {et~al.}(2011){Kimball}, {Kellermann}, {Condon},
  {Ivezi{\'c}}, \& {Perley}}]{Kimball2011}
{Kimball}, A.~E., {Kellermann}, K.~I., {Condon}, J.~J., {Ivezi{\'c}}, {\v{Z}}.,
  \& {Perley}, R.~A. 2011, \apjl, 739, L29, \dodoi{10.1088/2041-8205/739/1/L29}

\bibitem[{{King} {et~al.}(2017){King}, {Lohfink}, \& {Kara}}]{King2017}
{King}, A.~L., {Lohfink}, A., \& {Kara}, E. 2017, \apj, 835, 226,
  \dodoi{10.3847/1538-4357/835/2/226}

\bibitem[{{Kukula} {et~al.}(1998){Kukula}, {Dunlop}, {Hughes}, \&
  {Rawlings}}]{Kukula1998}
{Kukula}, M.~J., {Dunlop}, J.~S., {Hughes}, D.~H., \& {Rawlings}, S. 1998,
  \mnras, 297, 366, \dodoi{10.1046/j.1365-8711.1998.01481.x}

\bibitem[{{Lacy} {et~al.}(2020){Lacy}, {Baum}, {Chandler}, {Chatterjee},
  {Clarke}, {Deustua}, {English}, {Farnes}, {Gaensler}, {Gugliucci},
  {Hallinan}, {Kent}, {Kimball}, {Law}, {Lazio}, {Marvil}, {Mao}, {Medlin},
  {Mooley}, {Murphy}, {Myers}, {Osten}, {Richards}, {Rosolowsky}, {Rudnick},
  {Schinzel}, {Sivakoff}, {Sjouwerman}, {Taylor}, {White}, {Wrobel},
  {Andernach}, {Beasley}, {Berger}, {Bhatnager}, {Birkinshaw}, {Bower},
  {Brandt}, {Brown}, {Burke-Spolaor}, {Butler}, {Comerford}, {Demorest}, {Fu},
  {Giacintucci}, {Golap}, {G{\"u}th}, {Hales}, {Hiriart}, {Hodge}, {Horesh},
  {Ivezi{\'c}}, {Jarvis}, {Kamble}, {Kassim}, {Liu}, {Loinard}, {Lyons},
  {Masters}, {Mezcua}, {Moellenbrock}, {Mroczkowski}, {Nyland}, {O'Dea},
  {O'Sullivan}, {Peters}, {Radford}, {Rao}, {Robnett}, {Salcido}, {Shen},
  {Sobotka}, {Witz}, {Vaccari}, {van Weeren}, {Vargas}, {Williams}, \&
  {Yoon}}]{Lacy2020}
{Lacy}, M., {Baum}, S.~A., {Chandler}, C.~J., {et~al.} 2020, \pasp, 132,
  035001, \dodoi{10.1088/1538-3873/ab63eb}

\bibitem[{{Laor}(2000)}]{Laor2000}
{Laor}, A. 2000, \apjl, 543, L111, \dodoi{10.1086/317280}

\bibitem[{{Laor} {et~al.}(2019){Laor}, {Baldi}, \& {Behar}}]{Laor2019}
{Laor}, A., {Baldi}, R.~D., \& {Behar}, E. 2019, \mnras, 482, 5513,
  \dodoi{10.1093/mnras/sty3098}

\bibitem[{{Laor} \& {Behar}(2008)}]{Laor2008}
{Laor}, A., \& {Behar}, E. 2008, \mnras, 390, 847,
  \dodoi{10.1111/j.1365-2966.2008.13806.x}

\bibitem[{{Laor} \& {Netzer}(1989)}]{Laor1989}
{Laor}, A., \& {Netzer}, H. 1989, \mnras, 238, 897,
  \dodoi{10.1093/mnras/238.3.897}

\bibitem[{{Leipski} {et~al.}(2006){Leipski}, {Falcke}, {Bennert}, \&
  {H{\"u}ttemeister}}]{Leipski2006}
{Leipski}, C., {Falcke}, H., {Bennert}, N., \& {H{\"u}ttemeister}, S. 2006,
  \aap, 455, 161, \dodoi{10.1051/0004-6361:20054311}

\bibitem[{{Lusso} {et~al.}(2012){Lusso}, {Comastri}, {Simmons}, {Mignoli},
  {Zamorani}, {Vignali}, {Brusa}, {Shankar}, {Lutz}, {Trump}, {Maiolino},
  {Gilli}, {Bolzonella}, {Puccetti}, {Salvato}, {Impey}, {Civano}, {Elvis},
  {Mainieri}, {Silverman}, {Koekemoer}, {Bongiorno}, {Merloni}, {Berta}, {Le
  Floc'h}, {Magnelli}, {Pozzi}, \& {Riguccini}}]{Lusso2012}
{Lusso}, E., {Comastri}, A., {Simmons}, B.~D., {et~al.} 2012, \mnras, 425, 623,
  \dodoi{10.1111/j.1365-2966.2012.21513.x}

\bibitem[{{Merloni} \& {Fabian}(2002)}]{Merloni2002}
{Merloni}, A., \& {Fabian}, A.~C. 2002, \mnras, 332, 165,
  \dodoi{10.1046/j.1365-8711.2002.05288.x}

\bibitem[{{Miller} {et~al.}(2011){Miller}, {Brandt}, {Schneider}, {Gibson},
  {Steffen}, \& {Wu}}]{Miller2011}
{Miller}, B.~P., {Brandt}, W.~N., {Schneider}, D.~P., {et~al.} 2011, \apj, 726,
  20, \dodoi{10.1088/0004-637X/726/1/20}

\bibitem[{{Miller} {et~al.}(1993){Miller}, {Rawlings}, \&
  {Saunders}}]{Miller1993}
{Miller}, P., {Rawlings}, S., \& {Saunders}, R. 1993, \mnras, 263, 425,
  \dodoi{10.1093/mnras/263.2.425}

\bibitem[{{Neugebauer} {et~al.}(1987){Neugebauer}, {Green}, {Matthews},
  {Schmidt}, {Soifer}, \& {Bennett}}]{Neugebauer1987}
{Neugebauer}, G., {Green}, R.~F., {Matthews}, K., {et~al.} 1987, \apjs, 63,
  615, \dodoi{10.1086/191175}

\bibitem[{{Njeri} {et~al.}(2024){Njeri}, {Deane}, {Radcliffe}, {Beswick},
  {Thomson}, {Muxlow}, {Garrett}, \& {Harrison}}]{Njeri2024}
{Njeri}, A., {Deane}, R.~P., {Radcliffe}, J.~F., {et~al.} 2024, \mnras, 528,
  6141, \dodoi{10.1093/mnras/stae381}

\bibitem[{{Panessa} {et~al.}(2019){Panessa}, {Baldi}, {Laor}, {Padovani},
  {Behar}, \& {McHardy}}]{Panessa2019}
{Panessa}, F., {Baldi}, R.~D., {Laor}, A., {et~al.} 2019, Nature Astronomy, 3,
  387, \dodoi{10.1038/s41550-019-0765-4}

\bibitem[{{Panessa} {et~al.}(2007){Panessa}, {Barcons}, {Bassani}, {Cappi},
  {Carrera}, {Ho}, \& {Pellegrini}}]{Panessa2007}
{Panessa}, F., {Barcons}, X., {Bassani}, L., {et~al.} 2007, \aap, 467, 519,
  \dodoi{10.1051/0004-6361:20066943}

\bibitem[{{Panessa} {et~al.}(2022){Panessa}, {P{\'e}rez-Torres},
  {Hern{\'a}ndez-Garc{\'\i}a}, {Casella}, {Giroletti}, {Orienti}, {Baldi},
  {Bassani}, {Fiocchi}, {La Franca}, {Malizia}, {McHardy}, {Nicastro}, {Piro},
  {Vincentelli}, {Williams}, \& {Ubertini}}]{Panessa2022a}
{Panessa}, F., {P{\'e}rez-Torres}, M., {Hern{\'a}ndez-Garc{\'\i}a}, L.,
  {et~al.} 2022, \mnras, 510, 718, \dodoi{10.1093/mnras/stab3426}

\bibitem[{{Petric} {et~al.}(2015){Petric}, {Ho}, {Flagey}, \&
  {Scoville}}]{Petric2015}
{Petric}, A.~O., {Ho}, L.~C., {Flagey}, N. J.~M., \& {Scoville}, N.~Z. 2015,
  \apjs, 219, 22, \dodoi{10.1088/0067-0049/219/2/22}

\bibitem[{{Petrov} \& {Kovalev}(2017)}]{Petrov2017}
{Petrov}, L., \& {Kovalev}, Y.~Y. 2017, \mnras, 467, L71,
  \dodoi{10.1093/mnrasl/slx001}

\bibitem[{{Raginski} \& {Laor}(2016)}]{Raginski2016}
{Raginski}, I., \& {Laor}, A. 2016, \mnras, 459, 2082,
  \dodoi{10.1093/mnras/stw772}

\bibitem[{{Richards} {et~al.}(2006){Richards}, {Lacy}, {Storrie-Lombardi},
  {Hall}, {Gallagher}, {Hines}, {Fan}, {Papovich}, {Vanden Berk}, {Trammell},
  {Schneider}, {Vestergaard}, {York}, {Jester}, {Anderson}, {Budav{\'a}ri}, \&
  {Szalay}}]{Richards2006}
{Richards}, G.~T., {Lacy}, M., {Storrie-Lombardi}, L.~J., {et~al.} 2006, \apjs,
  166, 470, \dodoi{10.1086/506525}

\bibitem[{{Runnoe} {et~al.}(2018){Runnoe}, {G{\"u}ltekin}, \&
  {Rupke}}]{Runnoe2018}
{Runnoe}, J.~C., {G{\"u}ltekin}, K., \& {Rupke}, D. S.~N. 2018, \apj, 852, 8,
  \dodoi{10.3847/1538-4357/aa9934}

\bibitem[{{Rybicki} \& {Lightman}(1986)}]{Rybicki1986}
{Rybicki}, G.~B., \& {Lightman}, A.~P. 1986, {Radiative Processes in
  Astrophysics}

\bibitem[{{Sanders} {et~al.}(1989){Sanders}, {Phinney}, {Neugebauer}, {Soifer},
  \& {Matthews}}]{Sanders1989}
{Sanders}, D.~B., {Phinney}, E.~S., {Neugebauer}, G., {Soifer}, B.~T., \&
  {Matthews}, K. 1989, \apj, 347, 29, \dodoi{10.1086/168094}

\bibitem[{{Schmidt} \& {Green}(1983)}]{Schmidt1983}
{Schmidt}, M., \& {Green}, R.~F. 1983, \apj, 269, 352, \dodoi{10.1086/161048}

\bibitem[{{Schmitt} {et~al.}(2001){Schmitt}, {Ulvestad}, {Antonucci}, \&
  {Kinney}}]{Schmitt2001a}
{Schmitt}, H.~R., {Ulvestad}, J.~S., {Antonucci}, R.~R.~J., \& {Kinney}, A.~L.
  2001, \apjs, 132, 199, \dodoi{10.1086/318957}

\bibitem[{{Shablovinskaya} {et~al.}(2024){Shablovinskaya}, {Ricci}, {Chang},
  {Tortosa}, {del Palacio}, {Kawamuro}, {Aalto}, {Arzoumanian}, {Balokovic},
  {Bauer}, {Gendreau}, {Ho}, {Kakkad}, {Kara}, {Koss}, {Liu}, {Loewenstein},
  {Mushotzky}, {Paltani}, {Privon}, {Smith}, {Tombesi}, \&
  {Trakhtenbrot}}]{Shablovinskaya2024}
{Shablovinskaya}, E., {Ricci}, C., {Chang}, C.-S., {et~al.} 2024, arXiv
  e-prints, arXiv:2403.19524, \dodoi{10.48550/arXiv.2403.19524}

\bibitem[{{Shepherd} {et~al.}(1994){Shepherd}, {Pearson}, \&
  {Taylor}}]{Shepherd1994}
{Shepherd}, M.~C., {Pearson}, T.~J., \& {Taylor}, G.~B. 1994, in Bulletin of
  the American Astronomical Society, Vol.~26, 987--989

\bibitem[{{Shi} {et~al.}(2014){Shi}, {Rieke}, {Ogle}, {Su}, \&
  {Balog}}]{Shi2014}
{Shi}, Y., {Rieke}, G.~H., {Ogle}, P.~M., {Su}, K.~Y.~L., \& {Balog}, Z. 2014,
  \apjs, 214, 23, \dodoi{10.1088/0067-0049/214/2/23}

\bibitem[{{Silpa} {et~al.}(2022){Silpa}, {Kharb}, {Harrison}, {Girdhar},
  {Mukherjee}, {Mainieri}, \& {Jarvis}}]{Silpa2022}
{Silpa}, S., {Kharb}, P., {Harrison}, C.~M., {et~al.} 2022, \mnras, 513, 4208,
  \dodoi{10.1093/mnras/stac1044}

\bibitem[{{Soderberg} {et~al.}(2006){Soderberg}, {Chevalier}, {Kulkarni}, \&
  {Frail}}]{Soderberg2006}
{Soderberg}, A.~M., {Chevalier}, R.~A., {Kulkarni}, S.~R., \& {Frail}, D.~A.
  2006, \apj, 651, 1005, \dodoi{10.1086/507571}

\bibitem[{{Ulvestad} {et~al.}(2005){Ulvestad}, {Antonucci}, \&
  {Barvainis}}]{Ulvestad2005}
{Ulvestad}, J.~S., {Antonucci}, R.~R.~J., \& {Barvainis}, R. 2005, \apj, 621,
  123, \dodoi{10.1086/427426}

\bibitem[{{Vasudevan} \& {Fabian}(2007)}]{Vasudevan2007}
{Vasudevan}, R.~V., \& {Fabian}, A.~C. 2007, \mnras, 381, 1235,
  \dodoi{10.1111/j.1365-2966.2007.12328.x}

\bibitem[{{Wang} {et~al.}(2023{\natexlab{a}}){Wang}, {An}, {Cheng}, {Ho},
  {Kellermann}, {Baan}, {Yang}, \& {Zhang}}]{Wang2023a}
{Wang}, A., {An}, T., {Cheng}, X., {et~al.} 2023{\natexlab{a}}, \mnras, 518,
  39, \dodoi{10.1093/mnras/stac3091}

\bibitem[{{Wang} {et~al.}(2023{\natexlab{b}}){Wang}, {An}, {Zhang}, {Cheng},
  {Ho}, {Kellermann}, \& {Baan}}]{Wang2023c}
{Wang}, A., {An}, T., {Zhang}, Y., {et~al.} 2023{\natexlab{b}}, \mnras, 525,
  6064, \dodoi{10.1093/mnras/stad2651}

\bibitem[{{Wang} {et~al.}(2023{\natexlab{c}}){Wang}, {An}, {Guo}, {Ho}, {Baan},
  {Braun}, {Chen}, {Cheng}, {Hartley}, {Yang}, \& {Zhang}}]{Wang2023b}
{Wang}, A., {An}, T., {Guo}, S., {et~al.} 2023{\natexlab{c}}, \mnras, 523, L30,
  \dodoi{10.1093/mnrasl/slad051}

\bibitem[{{Yang} {et~al.}(2012){Yang}, {Wu}, {Paragi}, \& {An}}]{Yang2012}
{Yang}, J., {Wu}, F., {Paragi}, Z., \& {An}, T. 2012, \mnras, 419, L74,
  \dodoi{10.1111/j.1745-3933.2011.01182.x}

\bibitem[{{Young} {et~al.}(2007){Young}, {Axon}, {Robinson}, {Hough}, \&
  {Smith}}]{Young2007}
{Young}, S., {Axon}, D.~J., {Robinson}, A., {Hough}, J.~H., \& {Smith}, J.~E.
  2007, \nat, 450, 74, \dodoi{10.1038/nature06319}

\bibitem[{{Zakamska} {et~al.}(2016){Zakamska}, {Lampayan}, {Petric}, {Dicken},
  {Greene}, {Heckman}, {Hickox}, {Ho}, {Krolik}, {Nesvadba}, {Strauss},
  {Geach}, {Oguri}, \& {Strateva}}]{Zakamska2016}
{Zakamska}, N.~L., {Lampayan}, K., {Petric}, A., {et~al.} 2016, \mnras, 455,
  4191, \dodoi{10.1093/mnras/stv2571}

\end{thebibliography}
\bibliographystyle{aasjournal}


\begin{table*}[ht!]
\caption{The VLBA coordinates, the separations from the {\it Gaia} positions, and the source sizes in the VLBA 8.4 and 23.6 GHz observations.}
\label{size}
\centering
\footnotesize
\begin{tabular}{llllllllll}
\hline\hline
Name & $z$ & Scale & $\nu$ & R.A. & Dec. & $\Delta$ & $\theta_{\rm maj}$ & $\theta_{\rm min}$ & PA \\
& & (pc\,mas$^{-1}$) & (GHz) & (hh:mm:ss) & (dd:mm:ss) & (mas) & (mas) & (mas) & (degree) \\
(1) & (2) & (3) & (4) & (5) & (6) & (7) & (8) & (9) & (10) \\
\hline
\multirow{2}{*}{PG\,0026+129} & \multirow{2}{*}{0.145} & \multirow{2}{*}{2.61}
& 8.4 & 00:29:13.7012 & +13:16:03.9478 & 0.6 & 1.36 & $<$0.42 & $-$32.21 \\
& & & 23.6 & 00:29:13.7012 & +13:16:03.9478 & 0.5 & $<$0.50 & $<$0.20 & $-$29.25 \\
\hline
PG\,0050+124 & 0.061 & 1.20 & & & & & & & \\
\hline
\multirow{2}{*}{PG\,0052+251} & \multirow{2}{*}{0.154} & \multirow{2}{*}{2.75}
& 8.4 & 00:54:52.1182 & +25:25:38.9846 & 0.8 & $<$1.22 & $<$0.43 & 2.25 \\
& & & 23.6 & 00:54:52.1182 & +25:25:38.9847 & 0.8 & 1.27 & 0.20 & $-$2.97 \\
\hline
PG\,0157+001 & 0.163 & 2.88 & & & & & & & \\
\hline
\multirow{2}{*}{PG\,0921+525} & \multirow{2}{*}{0.035} & \multirow{2}{*}{0.74}
& 8.4 C1 & 09:25:12.8479 & +52:17:10.3874 & 0.5 & $<$1.40 & $<$0.45 & 17.33 \\
& & & 8.4 C2 & 09:25:12.8478 & +52:17:10.3855 & 1.5 & 3.87 & $<$0.45 & $-$41.06 \\
\hline
PG\,0923+129 & 0.029 & 0.62 & 8.4 & 09:26:03.2696 & +12:44:03.7299 & 1.0 & $<$1.11 & $<$0.37 & 77.77 \\
\hline
\multirow{3}{*}{PG\,1149$-$110} & \multirow{3}{*}{0.049} & \multirow{3}{*}{1.01}
& 8.4 C1 & 11:52:03.5504 & $-$11:22:24.0934 & 2.0 & $<$1.14 & $<$0.38 & 12.52 \\
& & & 8.4 C2 & 11:52:03.5505 & $-$11:22:24.0922 & 1.4 & $<$1.14 & $<$0.38 & $-$16.34 \\
& & & 23.6 & 11:52:03.5504 & $-$11:22:24.0936 & 2.1 & 1.23 & $<$0.23 & $-$13.54 \\
\hline
\multirow{2}{*}{PG\,1216+069} & \multirow{2}{*}{0.331} & \multirow{2}{*}{4.92}
& 8.4 & 12:19:20.9317 & +06:38:38.4662 & 1.4 & $<$1.28 & $<$0.37 & $-$0.94 \\
& & & 23.6 & 12:19:20.9317 & +06:38:38.4661 & 1.4 & 0.90 & $<$0.18 & $-$27.85 \\
\hline
\multirow{2}{*}{PG\,1351+640} & \multirow{2}{*}{0.088} & \multirow{2}{*}{1.71}
& 8.4 C1 & 13:53:15.8312 & +63:45:45.6828 & 0.1 & 1.58 & 0.53 & 14.24 \\
& & & 8.4 C2 & 13:53:15.8305 & +63:45:45.6861 & 5.5 & 1.54 & 0.90 & $-$49.13 \\
\hline
\multirow{2}{*}{PG\,1501+106} & \multirow{2}{*}{0.036} & \multirow{2}{*}{0.76}
& 8.4 & 15:04:01.1935 & +10:26:15.7822 & 0.9 & 1.22 & $<$0.40 & 7.41 \\
& & & 23.6 & 15:04:01.1935 & +10:26:15.7821 & 0.9 & 0.71 & $<$0.18 & $-$17.40 \\
\hline
PG\,1534+580 & 0.030 & 0.63 & 8.4 & 15:35:52.4031 & +57:54:09.5141 & 0.5 & 3.81 & $<$0.43 & 5.60 \\
\hline
PG\,1700+518 & 0.289 & 4.47 & 8.4 & 17:01:24.8266 & +51:49:20.4461 & 1.8 & 2.73 & $<$0.45 & 16.09 \\
\hline
\multirow{2}{*}{PG\,2304+042} & \multirow{2}{*}{0.043} & \multirow{2}{*}{0.85}
& 8.4 C1 & 23:07:02.9147 & +04:32:57.1027 & 0.6 & 1.12 & 0.57 & 29.80 \\
& & & 8.4 C2 & 23:07:02.9148 & +04:32:57.1090 & 7.0 & 2.90 & $<$0.44 & $-$22.15 \\
\hline
\end{tabular}
\begin{flushleft}
\vspace{-0.3cm}
\tablecomments{Columns:
(1) the name,
(2) the redshift,
(3) the physical scale,
(4) the frequency,
(5) the right ascension of the centroid determined by the MODELFIT in DIFMAP,
(6) the declination of the centroid determined by the MODELFIT in DIFMAP,
(7) the separation between the VLBA and the {\it Gaia} positions,
(8) the deconvolved major FWHM of the source in unit of mas,
(9) the deconvolved minor FWHM of the source in unit of mas,
(10) the deconvolved position angle of the source in unit of degree.}
\end{flushleft}
\end{table*}

\begin{turnpage}
\begin{table*}[ht!]
\caption{The $uv$-coverage, the total and core flux densities, and the background noise in the full-array maps and the tapered maps in the VLBA 8.4 and 23.6 GHz observations.}
\label{flux_XK}
\centering
\footnotesize
\begin{tabular}{lllllllllll}
\hline\hline
Name & Frequency & \multicolumn{5}{c}{Full-array maps} & \multicolumn{4}{c}{Tapered maps} \\
& $\nu$ & $uv$-range & $F_{\rm total}$ & $F_{\rm core}$ & $\frac{F_{\rm core}}{F_{\rm total}}$ & RMS & $uv$-range & $F_{\rm total}$ & $F_{\rm core}$ & RMS \\
& (GHz) & (M$\lambda$) & (mJy) & (mJy\,beam$^{-1}$) & & (mJy\,beam$^{-1}$) & (M$\lambda$) & (mJy) & (mJy\,beam$^{-1}$) & (mJy\,beam$^{-1}$) \\
(1) & (2) & (3) & (4) & (5) & (6) & (7) & (8) & (9) & (10) & (11) \\
\hline
\multirow{2}{*}{PG\,0026+129}
& 8.4 & 5--246 & 0.76 $\pm$ 0.03 & 0.55 & 0.7 & 0.023 & 15--246 & 0.76 $\pm$ 0.04 & 0.54 & 0.024 \\
& 23.6 & 15--683 & 0.70 $\pm$ 0.05 & 0.57 & 0.8 & 0.037 & 15--246 & 0.78 $\pm$ 0.06 & 0.67 & 0.043 \\
\hline
\multirow{2}{*}{PG\,0050+124}
& 8.4 & 5--246 & & $<$ 0.10 & & 0.020 & 15--246 & & $<$ 0.10 & 0.021 \\
& 23.6 & 15--683 & & $<$ 0.25 & & 0.049 & 15--246 & & $<$ 0.31 & 0.063 \\
\hline
\multirow{2}{*}{PG\,0052+251}
& 8.4 & 6--246 & 0.19 $\pm$ 0.04 & 0.17 & 0.9 & 0.029 & 17--246 & 0.18 $\pm$ 0.04 & 0.17 & 0.030 \\
& 23.6 & 17--595 & 0.49 $\pm$ 0.13 & 0.26 & 0.5 & 0.055 & 17--246 & 0.61 $\pm$ 0.19 & 0.25 & 0.072 \\
\hline
\multirow{2}{*}{PG\,0157+001}
& 8.4 & 4--246 & & $<$ 0.14 & & 0.029 & 12--176 & & $<$ 0.14 & 0.028 \\
& 23.6 & 12--683 & & $<$ 0.22 & & 0.043 & 12--176 & & $<$ 0.25 & 0.051 \\
\hline
\multirow{3}{*}{PG\,0921+525}
& 8.4 C1 & 5--246 & 1.23 $\pm$ 0.07 & 1.22 & \multirow{2}{*}{0.7} & 0.035 & 14--246 & 1.30 $\pm$ 0.07 & 1.22 & 0.036 \\
& 8.4 C2 & 5--246 & 0.49 $\pm$ 0.10 & & & 0.035 & 14--246 & 0.47 $\pm$ 0.14 & & 0.036 \\
& 23.6 & 14--672 & & $<$ 0.52 & & 0.104 & 14--246 & & $<$ 0.64 & 0.129 \\
\hline
\multirow{2}{*}{PG\,0923+129}
& 8.4 & 5--246 & 0.13 $\pm$ 0.03 & 0.10 & 0.8 & 0.021 & 15--176 & 0.13 $\pm$ 0.11 & 0.13 & 0.022 \\
& 23.6 & 15--683 & & $<$ 0.21 & & 0.042 & 15--176 & & $<$ 0.26 & 0.051 \\
\hline
\multirow{4}{*}{PG\,1149$-$110}
& 8.4 C1 & 4--246 & 0.53 $\pm$ 0.06 & 0.48 & \multirow{2}{*}{0.6} & 0.030 & 11--120 & 0.55 $\pm$ 0.05 & 0.51 & 0.037 \\
& 8.4 C2 & 4--246 & 0.28 $\pm$ 0.06 & & & 0.030 & 11--120 & 0.29 $\pm$ 0.05 & & 0.037 \\
& 23.6 C1 & 11--678 & 0.45 $\pm$ 0.10 & 0.40 & 0.9 & 0.076 & 11--120 & 0.47 $\pm$ 0.19 & 0.40 & 0.086 \\
& 23.6 C2 & 11--678 & & $<$ 0.38 & & 0.076 & 11--120 & & $<$ 0.43 & 0.086 \\
\hline
\multirow{2}{*}{PG\,1216+069}
& 8.4 & 5--246 & 1.80 $\pm$ 0.05 & 1.53 & 0.9 & 0.036 & 15--246 & 1.79 $\pm$ 0.06 & 1.48 & 0.038 \\
& 23.6 & 15--679 & 0.41 $\pm$ 0.10 & 0.32 & 0.8 & 0.060 & 15--246 & 0.54 $\pm$ 0.14 & 0.40 & 0.077 \\
\hline
\multirow{3}{*}{PG\,1351+640}
& 8.4 C1 & 4--246 & 1.60 $\pm$ 0.03 & 1.10 & \multirow{2}{*}{0.4} & 0.027 & 13--246 & 1.50 $\pm$ 0.04 & 1.05 & 0.028 \\
& 8.4 C2 & 4--246 & 0.85 $\pm$ 0.04 & & & 0.027 & 13--246 & 0.75 $\pm$ 0.05 & & 0.028 \\
& 23.6 & 13--681 & & $<$ 0.21 & & 0.043 & 13--246 & & $<$ 0.26 & 0.053 \\
\hline
\multirow{2}{*}{PG\,1501+106}
& 8.4 & 5--246 & 0.34 $\pm$ 0.03 & 0.30 & 0.9 & 0.023 & 15--246 & 0.34 $\pm$ 0.04 & 0.30 & 0.023 \\
& 23.6 & 15--683 & 0.27 $\pm$ 0.06 & 0.22 & 0.8 & 0.038 & 15--246 & 0.29 $\pm$ 0.10 & 0.21 & 0.046 \\
\hline
\multirow{2}{*}{PG\,1534+580}
& 8.4 & 5--245 & 0.28 $\pm$ 0.05 & 0.13 & 0.5 & 0.029 & 14--245 & 0.25 $\pm$ 0.06 & 0.12 & 0.029 \\
& 23.6 & 14--683 & & $<$ 0.22 & & 0.044 & 14--245 & & $<$ 0.26 & 0.053 \\
\hline
\multirow{2}{*}{PG\,1700+518}
& 8.4 & 5--246 & 0.18 $\pm$ 0.05 & 0.14 & 0.8 & 0.033 & 14--246 & 0.15 $\pm$ 0.05 & 0.13 & 0.034 \\
& 23.6 & 14--683 & & $<$ 0.31 & & 0.061 & 14--246 & & $<$ 0.37 & 0.075 \\
\hline
\multirow{3}{*}{PG\,2304+042}
& 8.4 C1 & 5--246 & 0.70 $\pm$ 0.04 & 0.49 & \multirow{2}{*}{0.6} & 0.028 & 14--176 & 0.71 $\pm$ 0.05 & 0.49 & 0.030 \\
& 8.4 C2 & 5--246 & 0.18 $\pm$ 0.08 & & & 0.028 & 14--176 & 0.22 $\pm$ 0.09 & & 0.030 \\
& 23.6 & 14--683 & & $<$ 0.26 & & 0.051 & 14--176 & & $<$ 0.33 & 0.065 \\
\hline
\end{tabular}
\begin{flushleft}
\vspace{-0.3cm}
\tablecomments{Columns:
(1) the name,
(2) the frequency,
(3) the $uv$-range of the full-array map,
(4) the total flux density of the full-array map,
(5) the core flux density of the full-array map,
(6) the core to total flux ratio of the full-array map,
(7) the background noise of the full-array map,
(8) the $uv$-range of the tapered map,
(9) the total flux density of the tapered map,
(10) the core flux density of the tapered map,
(11) the background noise of the tapered map.}
\end{flushleft}
\end{table*}
\end{turnpage}

\begin{table*}[ht!]
\caption{The VLBA flux densities of the core component at 1.5 and 5.0 GHz in the full-array maps and the tapered (3--50\,M$\lambda$) maps adapted from \citet{Alhosani2022} and \citet{Chen2023}.}
\label{flux_LC}
\centering
\footnotesize
\begin{tabular}{lllll}
\hline\hline
Name & \multicolumn{2}{c}{Full-array maps} & \multicolumn{2}{c}{Tapered (3--50\,M$\lambda$) maps} \\
& $F_{\rm 1.5}$ & $F_{\rm 5.0}$ & $F_{\rm 1.5}$ & $F_{\rm 5.0}$ \\
& (mJy) & (mJy) & (mJy) & (mJy) \\
(1) & (2) & (3) & (4) & (5) \\
\hline
PG\,0026+129 & $<$ 0.10 & 0.31 $\pm$ 0.04 & $<$ 0.11 & 0.28 $\pm$ 0.04 \\
PG\,0050+124 C1 & 0.72 $\pm$ 0.12 & 0.24 $\pm$ 0.05 & 0.24 $\pm$ 0.07 & \\
PG\,0050+124 C2 & 2.66 $\pm$ 0.15 & 0.31 $\pm$ 0.07 & 1.42 $\pm$ 0.10 & \\
PG\,0052+251 & 0.42 $\pm$ 0.07 & 0.30 $\pm$ 0.03 & 0.35 $\pm$ 0.06 & \\
PG\,0157+001 & 6.10 $\pm$ 0.28 & 1.79 $\pm$ 0.30 & 2.60 $\pm$ 0.17 & 0.95 $\pm$ 0.16 \\
PG\,0921+525 & 1.24 $\pm$ 0.05 & 1.21 $\pm$ 0.04 & 1.19 $\pm$ 0.06 & 1.02 $\pm$ 0.03 \\
PG\,0923+129 & 0.91 $\pm$ 0.09 & 0.21 $\pm$ 0.05 & 0.29 $\pm$ 0.04 & 0.17 $\pm$ 0.02 \\
PG\,1149$-$110 & 1.06 $\pm$ 0.11 & 0.59 $\pm$ 0.04 & 0.86 $\pm$ 0.11 & \\
PG\,1216+069 & 0.47 $\pm$ 0.03 & 6.91 $\pm$ 0.04 & 0.60 $\pm$ 0.07 & 6.89 $\pm$ 0.05 \\
PG\,1351+640 C1 & 0.59 $\pm$ 0.07 & 2.30 $\pm$ 0.04 & 0.78 $\pm$ 0.03 & 2.42 $\pm$ 0.05 \\
PG\,1351+640 C2 & 8.13 $\pm$ 0.05 & 2.67 $\pm$ 0.04 & 7.29 $\pm$ 0.04 & 2.58 $\pm$ 0.05 \\
PG\,1501+106 & 0.37 $\pm$ 0.07 & 0.62 $\pm$ 0.08 & 0.37 $\pm$ 0.07 & 0.40 $\pm$ 0.05 \\
PG\,1534+580 & 0.63 $\pm$ 0.09 & 0.14 $\pm$ 0.02 & 0.20 $\pm$ 0.06 & 0.14 $\pm$ 0.02 \\
PG\,1700+518 & 1.46 $\pm$ 0.13 & 0.92 $\pm$ 0.04 & 1.36 $\pm$ 0.13 & 1.00 $\pm$ 0.05 \\
PG\,2304+042 & 0.53 $\pm$ 0.08 & 0.49 $\pm$ 0.04 & 0.55 $\pm$ 0.07 & \\
\hline
\end{tabular}
\begin{flushleft}
\vspace{-0.3cm}
\tablecomments{Columns:
(1) the name,
(2) the 1.5~GHz total flux density of the full-array map,
(3) the 5.0~GHz total flux density of the full-array map,
(4) the 1.5~GHz total flux density of the tapered map,
(5) the 5.0~GHz total flux density of the tapered map.
The tapered maps have a $uv$-range of $\sim$ 3--50\,M$\lambda$ in both bands.}
\end{flushleft}
\end{table*}

\begin{table*}[ht!]
\caption{The VLA 1.4, 3.0, 5.0, 8.5, 15, and 45~GHz flux densities with the A/B/C configurations from literature.}
\label{vla}
\centering
\footnotesize
\begin{tabular}{lllllll}
\hline\hline
Name & $F_{1.4}$ & $F_{3.0}$ & $F_{5.0}$ & $F_{8.5}$ & $F_{15}$ & $F_{45}$ \\
& (mJy) & (mJy) & (mJy) & (mJy) & (mJy) & (mJy) \\
(1) & (2) & (3) & (4) & (5) & (6) & (7) \\
\hline
PG\,0026+129 & 2.40 $\pm$ 0.70 $^a$ & 1.81 $\pm$ 0.14 & 0.20 $\pm$ 0.06 $^b$ & 0.17 $\pm$ 0.02 $^c$ & & 0.21 $\pm$ 0.04 $^d$ \\
PG\,0050+124 & 5.10 $\pm$ 0.40 $^a$ & 2.87 $\pm$ 0.15 & 1.80 $\pm$ 0.06 $^b$ & 0.90 $\pm$ 0.20 $^a$ & 1.06 $\pm$ 0.32 $^e$ & 0.30 $\pm$ 0.08 $^d$ \\
PG\,0052+251 & 1.10 $\pm$ 0.50 $^a$ & 0.95 $\pm$ 0.14 & 0.58 $\pm$ 0.01 $^c$ & 0.70 $\pm$ 0.10 $^a$ & & 0.37 $\pm$ 0.04 $^d$ \\
PG\,0157+001 & 24.70 $\pm$ 1.20 $^a$ & 9.43 $\pm$ 0.19 & 5.58 $\pm$ 0.06 $^b$ & 2.98 $\pm$ 0.01 $^c$ & 2.04 $\pm$ 0.40 $^e$ & 0.48 $\pm$ 0.06 $^d$ \\
PG\,0921+525 & 5.80 $\pm$ 0.60 $^a$ & 2.46 $\pm$ 0.19 & 1.87 $\pm$ 0.06 $^b$ & 1.70 $\pm$ 0.10 $^a$ & & \\
PG\,0923+129 & 7.45 $\pm$ 0.15 $^f$ & 4.15 $\pm$ 0.11 & 2.82 $\pm$ 0.02 $^g$ & 2.00 $\pm$ 0.05 $^h$ & & \\
PG\,1149$-$110 & 3.10 $\pm$ 0.80 $^a$ & 1.96 $\pm$ 0.13 & 1.08 $\pm$ 0.02 $^c$ & 1.30 $\pm$ 0.10 $^a$ & & 0.36 $\pm$ 0.03 $^d$ \\
PG\,1216+069 & 2.05 $\pm$ 0.13 $^f$ & 1.58 $\pm$ 0.10 & 4.95 $\pm$ 0.06 $^b$ & 12.06 $\pm$ 0.65 $^e$ & 8.36 $\pm$ 1.20 $^e$ & \\
PG\,1351+640 & 25.32 $\pm$ 0.15 $^f$ & 8.93 $\pm$ 0.11 & 7.31 $\pm$ 0.01 $^c$ & & 3.45 $\pm$ 0.56 $^e$ & \\
PG\,1501+106 & & 1.29 $\pm$ 0.17 & 0.50 $\pm$ 0.06 $^b$ & 0.55 $\pm$ 0.08 $^a$ & & \\
PG\,1534+580 & 5.11 $\pm$ 0.15 $^f$ & 2.29 $\pm$ 0.14 & 2.00 $\pm$ 0.02 $^c$ & & & \\
PG\,1700+518 & 19.20 $\pm$ 0.14 $^f$ & 8.88 $\pm$ 0.16 & 2.05 $\pm$ 0.06 $^b$ & 2.60 $\pm$ 0.20 $^a$ & 1.90 $\pm$ 0.39 $^e$ & \\
PG\,2304+042 & & 0.69 $\pm$ 0.05 & 1.10 $\pm$ 0.06 $^b$ & 1.00 $\pm$ 0.10 $^i$ & & 1.03 $\pm$ 0.05 $^j$ \\
\hline
\end{tabular}
\begin{flushleft}
\vspace{-0.3cm}
\tablecomments{Columns:
(1) the name,
(2) the 1.4~GHz flux densities from literature,
(3) the 3.0~GHz flux densities from the VLA Sky Survey \citep[VLASS,][]{Lacy2020},
(4) the 5.0~GHz flux densities from literature,
(5) the 8.5~GHz flux densities from literature,
(6) the 15~GHz flux densities from literature,
(7) the 45~GHz flux densities from literature.
Reference:
(a) \citet{Kukula1998},
(b) \citet{Kellermann1989},
(c) \citet{Leipski2006},
(d) \citet{Baldi2022},
(e) \citet{Barvainis1996},
(f) the Faint Images of the Radio Sky at Twenty-Centimeters \citep[FIRST,][]{Helfand2015},
(g) \citet{Berton2018},
(h) \citet{Schmitt2001a},
(i) \citet{Barvainis2005},
(j) data reduced from a VLA 45~GHz survery (Program ID: VLA/20A-169).}
\end{flushleft}
\end{table*}

\begin{table*}[ht!]
\caption{The Eddington ratio, the brightness temperature at 8.4~GHz, the 1.5--5.0~GHz and the 8.4--23.6~GHz spectral slopes in the VLBA observations.}
\label{slope}
\centering
\footnotesize
\begin{tabular}{llllllll}
\hline\hline
Name & $\log \frac{M_{\rm BH}}{M_{\odot}}$ & $\log L_{\rm bol}$ & $\log \frac{L}{L_{\rm Edd}}$ & $\log T_{\rm B}$ & $\alpha_{1.5-5.0}$ & $\alpha_{8.4-23.6}$ & Spectrum \\
& & (erg\,s$^{-1}$) & & (K) & & & \\
(1) & (2) & (3) & (4) & (5) & (6) & (7) & (8) \\
\hline
PG\,0026+129 & 7.74 & 46.15 & 0.30 & $>$ 7.59 & $>$ 0.89 $\pm$ 0.14 & 0.03 $\pm$ 0.09 & \multirow{5}{*}{Flat-Flat} \\
\cline{1-7}
PG\,0052+251 & 8.64 & 46.06 & $-$0.68 & $>$ 7.03 & $-$0.13 $\pm$ 0.16 & 1.18 $\pm$ 0.37 & \\
\cline{1-7}
\multirow{2}{*}{PG\,1149$-$110} & \multirow{2}{*}{7.34} & \multirow{2}{*}{44.75} & \multirow{2}{*}{$-$0.70} & C1 $>$ 7.52 & \multirow{2}{*}{$-$0.13 $\pm$ 0.12} & C1 $-$0.15 $\pm$ 0.40 & \\
& & & & C2 $>$ 7.24 & & C2 $<$ 0.38 $\pm$ 0.17 & \\
\cline{1-7}
PG\,1501+106 & 8.11 & 44.90 & $-$1.32 & $>$ 7.27 & 0.08 $\pm$ 0.21 & $-$0.15 $\pm$ 0.35 & \\
\hline
\multirow{2}{*}{PG\,0921+525} & \multirow{2}{*}{6.87} & \multirow{2}{*}{44.47} & \multirow{2}{*}{$-$0.51} & C1 $>$ 7.71 & \multirow{2}{*}{$-$0.14 $\pm$ 0.05} & C1 $<$ $-$0.69 $\pm$ 0.05 & \multirow{7}{*}{Flat-Steep} \\
& & & & C2 $>$ 6.87 & & C2 $<$ 0.30 $\pm$ 0.29 & \\
\cline{1-7}
PG\,1216+069 & 9.06 & 46.61 & $-$0.56 & $>$ 8.11 & 2.18 $\pm$ 0.10 & $-$1.16 $\pm$ 0.25 & \\
\cline{1-7}
\multirow{2}{*}{PG\,1351+640} & \multirow{2}{*}{8.49} & \multirow{2}{*}{45.31} & \multirow{2}{*}{$-$1.29} & C1 7.72 & C1 1.01 $\pm$ 0.04 & C1 $<$ $-$1.70 $\pm$ 0.03 & \\
& & & & C2 7.23 & C2 $-$0.93 $\pm$ 0.02 & C2 $<$ $-$1.03 $\pm$ 0.06 & \\
\cline{1-7}
\multirow{2}{*}{PG\,2304+042} & \multirow{2}{*}{7.91} & \multirow{2}{*}{44.49} & \multirow{2}{*}{$-$1.52} & C1 7.46 & \multirow{2}{*}{$-$0.09 $\pm$ 0.12} & C1 $<$ $-$0.74 $\pm$ 0.07 & \\
& & & & C2 $>$ 6.57 & & C2 $<$ 0.39 $\pm$ 0.40 & \\
\hline
PG\,0923+129 & 6.82 & 44.53 & $-$0.40 & $>$ 6.92 & $-$0.48 $\pm$ 0.19 & $<$ 0.67 $\pm$ 0.82 & \multirow{6}{*}{Indeterminate} \\
\cline{1-7}
PG\,1534+580 & 7.71 & 44.49 & $-$1.33 & $>$ 6.65 & $-$0.36 $\pm$ 0.29 & $<$ 0.04 $\pm$ 0.23 & \\
\cline{1-7}
PG\,1700+518 & 8.59 & 46.63 & $-$0.07 & $>$ 6.68 & $-$0.28 $\pm$ 0.10 & $<$ 0.87 $\pm$ 0.32 & \\
\cline{1-7}
\multirow{2}{*}{PG\,0050+124} & \multirow{2}{*}{6.99} & \multirow{2}{*}{45.12} & \multirow{2}{*}{0.03} & & 0.00 $\pm$ 0.29 & & \\
& & & & & $-$1.21 $\pm$ 0.19 & & \\
\cline{1-7}
PG\,0157+001 & 8.00 & 45.93 & $-$0.18 & & $-$0.90 $\pm$ 0.16 & & \\
\hline
\end{tabular}
\begin{flushleft}
\vspace{-0.3cm}
\tablecomments{Columns:
(1) the name,
(2) the logarithm scale of the black hole mass,
(3) the logarithm scale of the bolometric luminosity in units of erg\,s$^{-1}$,
(4) the logarithm scale of the Eddington ratio,
(5) the logarithm scale of the brightness temperature calculated at 8.4~GHz,
(6) the 1.5--5.0~GHz spectral slopes adapted from \citet{Alhosani2022} and \citet{Chen2023},
(7) the 8.4--23.6~GHz spectral slopes from the VLBA 8.4 and 23.6 GHz observations,
(8) the 1--24~GHz spectral type, in which the ``flat-flat'' refers to flat slopes at both 1.5--5.0~GHz and 8.4--23.6~GHz, the ``flat-steep'' refers to a flat slope at 1.5--5.0~GHz and a steep slope at 8.4--23.6~GHz, and the ``Indeterminate'' refers to an uncertain 8.4--23.6~GHz slope.
The black hole mass, the bolometric luminosity, and the Eddington ratio are from \citet{Davis2011} or \citet{Laor2019}.}
\end{flushleft}
\end{table*}

\begin{table*}[ht!]
\caption{The spatial extents of the radio core emission.}
\label{region}
\centering
\footnotesize
\begin{tabular}{lllllllll}
\hline\hline
Name & Spectrum & $\nu_0 (R_{\rm out})$ & $\nu_0 (R_{\rm in})$ & $R_{\rm out}$ & $R_{\rm in}$ & $R_{\rm BLR}$ & $\frac{R_{\rm out}}{R_{\rm BLR}}$ & $\frac{R_{\rm in}}{R_{\rm BLR}}$ \\
& & (GHz) & (GHz) & (pc) & (pc) & (pc) & & \\
(1) & (2) & (3) & (4) & (5) & (6) & (7) & (8) & (9) \\
\hline
PG\,0026+129 & \multirow{4}{*}{Flat-Flat} & 5.0--8.4 & $>$23.6 & 0.041--0.048 & $<$0.014 & 0.119 & 0.348--0.408 & $<$0.120 \\
PG\,0052+251 & & $<$1.5 & $>$23.6 & $>$0.188 & $<$0.013 & 0.107 & $>$1.758 & $<$0.119 \\
PG\,1149$-$110 & & $<$1.5 & $>$23.6 & $>$0.076 & $<$0.003 & 0.024 & $>$3.218 & $<$0.145 \\
PG\,1501+106 & & $<$1.5 & $>$23.6 & $>$0.040 & $<$0.002 & 0.028 & $>$1.427 & $<$0.080 \\
\hline
PG\,0921+525 & \multirow{4}{*}{Flat-Steep} & $<$1.5 & 5.0--8.4 & $>$0.058 & 0.010--0.017 & 0.017 & $>$3.360 & 0.598--0.998 \\
PG\,1216+069 & & 5.0--8.4 & 5.0--8.4 & 0.136--0.392 & 0.136--0.392 & 0.202 & 0.675--1.943 & 0.675--1.943 \\
PG\,1351+640 & & 5.0--8.4 & 5.0--8.4 & 0.030--0.058 & 0.030--0.058 & 0.045 & 0.660--1.283 & 0.660--1.283 \\
PG\,2304+042 & & $<$1.5 & 5.0--8.4 & $>$0.049 & 0.010--0.014 & 0.018 & $>$2.781 & 0.555--0.809 \\
\hline
PG\,0923+129 & \multirow{5}{*}{Indeterminate} & $<$1.5 & $>$5.0 & $>$0.044 & $<$0.007 & 0.018 & $>$2.408 & $<$0.402 \\
PG\,1534+580 & & $<$1.5 & $>$5.0 & $>$0.039 & $<$0.006 & 0.018 & $>$2.218 & $<$0.365 \\
PG\,1700+518 & & $<$1.5 & $>$5.0 & $>$0.622 & $<$0.155 & 0.207 & $>$3.012 & $<$0.751 \\
PG\,0050+124 & & $<$1.5 & $>$5.0 & $>$0.085 & $<$0.017 & 0.036 & $>$2.352 & $<$0.455 \\
PG\,0157+001 & & & $<$1.5 & & $>$0.560 & 0.092 & & $>$6.075 \\
\hline
\end{tabular}
\begin{flushleft}
\vspace{-0.3cm}
\tablecomments{Columns:
(1) the name,
(2) the 1--24~GHz spectral type which is the same as Table~\ref{slope},
(3) the turnover frequency where the slope changes from inverted to flat,
(4) the turnover frequency where the slope changes from flat to steep,
(5) the outer emission radius,
(6) the inner emission radius,
(7) the radius of the BLR,
(8) the outer emission radius in units of the BLR radius,
(9) the inner emission radius in units of the BLR radius.}
\end{flushleft}
\end{table*}

\end{document}